\documentclass[prl,twocolumn,superscriptaddress,amsmath,showkeys,amssymb]{revtex4}

\usepackage{graphicx}
\usepackage{indentfirst}
\usepackage{psfrag}
\usepackage{epsfig}
\usepackage{bm}
\usepackage{amsmath}
\usepackage{amssymb}
\usepackage{bm}
\usepackage{mathrsfs}
\newcommand{\be}{\begin{eqnarray}}
\newcommand{\ee}{\end{eqnarray}}

\newcommand{\hel}{\mathscr{ H}}

\begin{document}

\title{Fundamentals and model of  resonance helicity  (RHELT) and energy (RET) transfer between two magnetoelectric chiral  particles}
\author{Manuel Nieto-Vesperinas }
\affiliation{Instituto de Ciencia de Materiales de Madrid, Consejo Superior de
Investigaciones Cient\'{i}ficas\\
 Campus de Cantoblanco, Madrid 28049, Spain.\\ www.icmm.csic.es/mnv; 
mnieto@icmm.csic.es }



\begin{abstract}
We establish a classical electrodynamic theory for the non-radiative transfer of field helicity (RHELT) and energy (RET) between a donor and an acceptor, both being dipolar, magnetoelectric and bi-isotropic, chiral in particular, with rotating excited dipoles. We introduce  orientational factors that control this process. Also,  a RHELT and RET interaction radius  is put forward.   The detection of RHELT adds a wealth of information contained in the helicity of the transferred fields, never used or established to date. The nature of these dipolar  magnetoelectric bi-isotropic particles and/or molecules with induced dipoles possessing angular momentum,  enriches the number of variables and associated effects. Hence  the landscape  involved in this transfer phenomenon, never explored before, is  significantly broader than in  conventional FRET.  In this  way, chiral interacting objects convey  terms in the equations of   transfer rate of  helicity and energy that are discriminatory, so that   one can extract information on their structural  chirality  handedness and polarization rotation.  As such, not only the rate of electromagnetic helicity transfer, but also that of energy transfer may be negative, which for the latter means an enhanced emission from the donor  in  pressence of acceptor,  a phenomenon which does not exist in conventional FRET. Importantly, both  the RHELT  and RET rates, as well as the RHELT interaction radius, are very sensitive to changes in the helicity, or  state of polarization,  of the illumination, as well as to  the  polarization of the excited  electric and magnetic dipole moments of  donor and acceptor. Finally, we introduce the observable quanties in terms of which one can obtain the transfer rates and interaction radii.
\end{abstract}
\maketitle

\section{Introduction}
Electromagnetic wavefields with rotation of their polarization vectors and wavefronts \cite{allen1,andrews1,yao}  are subjects of increasing interest for their larger number of degrees of freedom as communication channels \cite{andrews2,boyd}, and for providing new capibilities to probing and manipulating both chiral and achiral structures at the micro and nanoscale in light-matter interactions \cite{schellman,richard,vuong,bliokh1,schukov1,brasse,bliokh2}. In this latter respect, the interplay between the structural symmetry  of the light probe and that of matter \cite{andrews3,nieto1,bliokh3,kivshar} is of upmost importance, since the latter governs the metabolism of living organisms and is becoming of increasing relevance for nanophotonic devices. Nevertheless, our knowledge in this regard is yet incomplete; therefore new probe techniques for chiral  nanostructures with rotating dipoles (and multipoles) are still necessary.

Progress in characterizing light chirality and its interaction with matter, on the one hand,  \cite{tang1,bliokh4,bliokh5,cameron1,cameron2,nieto2,nieto3,gutsche1,gutsche2,gutsche3,corbato1, zambra1,gutsche4}, and on  designing structured wavefields that enhance the usually weak  interaction with chiral molecules and nanoparticles   \cite{tang2,tang3,choi,klimov,dionne1,schafer1,giessen1,schafer2,dionne2,carminati,wang,hu},  has advanced together with  methods to increase the energy transfer between nanostructures \cite{kivshar,carminati}, like e.g.  F\"{o}rster energy tranfer (FRET) between molecules and/or particles \cite{fret1,fret2,circpollibro1}. Although this latter phenomenon constitutes nowadays an  established technique   in nanoscience and biology \cite{circpolemission,kagan}, and it has a well developed theory \cite{fret1,clegg,novotny}; and in spite of theoretical studies on  shifts and transfer of  energy between two chiral molecules based on their dipole and quadrupole interactions \cite{craig1,salam1,craig2,salam2},  aparently and as far as we know, there are not yet techniques based on characterizing helicity states of the detected light in FRET. This involves  field helicity transfer between particles and/or molecules, let them be chiral or achiral; but  FRET is based on detecting only   omnidirectional intensities emitted by the fluorophores, which can be hindered by  limitations depending on the molecule nature and environment configuration, and specially by the low signal-to-noise ratio \cite{silas}.

In previous work  \cite{nieto2} we established the law ruling the extinction of electromagnetic helicity of a twisted illuminating wavefield on scattering and absorption by a  wide sense \cite{g-etxarri,nieto4} (i.e. non-Rayleigh) dipolar  bi-isotropic particle, chiral in particular. Also we put  forward the helicity enhancement factor \cite{nieto3} when the particle is  in an inhomogeneous environment \cite{madrazo,garcia}. Such a quantity plays a role analogous to that of the Purcell factor for the energy. This extinction law encompasses a variety of new  processes involving the helicity of electromagnetic fields which are emitted, absorbed and/or scattered in presence of other objects \cite{nieto3, gutsche1,gutsche2}. 

Extending studies to  circularly or elliptically polarized dipoles opens a new landscape on interactions  of light and matter, in particular  between particles. This  has applications in new nanophotonic systems and techniques. Specifically, like electromagnetic theory formulates FRET  as the extinction (most frequently, the absorption) by the acceptor A of the energy emitted by the donor D, {\it  one may ask on  the existence of an analogous phenomenon between two dipoles D and A for the wavefield helicity on illumination of D by rotating  light, therefore possessing angular momentum which is conveyed to the donor; specifically when D and A are magnetoelectric and bi-isotropic elliptically, or circularly, polarized dipoles.  Addressing this question is the main aim of this paper} and, as matter of fact, we gave a hint  (cf. Eq.(36) of   \cite{nieto2}) on  employing our  helicity extinction  formulation  in modelling its transfer between two nanoscale magnetoelectric dipolar bodies. {\it The quantities involved in such a kind of phenomenon  go beyond those of standard FRET}. For instance, {\it we shall put forward  the  existence of several orientational factors}, instead of just one, $\kappa^2$,  of standard FRET.

Therefore, in this  work we establish a classical electromagnetic theory of {\it resonance helicity transfer, (which we shall abridge as RHELT), between two dipolar particles D and A}, (by "particle" we shall mean either a quantum dot, a molecule,  a synthesized material polarizable nanoparticle, or a hybrid between both),  {\it both being dipolar and magnetoelectric, bi-isotropic, chiral in particular}, and located in the near-field (i.e. non-radiative) region   of each other. This brings {\it  additional information to estimate their relative distance and orientations}, as well as to know donor and acceptor {\it constitutive parameters}, and  their generally elliptic polarization; thus making it possible {\it to characterize them according to their effects on the electromagnetic helicity}, both trasferred from D to A, as well as emitted by the acceptor,  (in this connection see e.g. \cite{nieto2,gutsche4}).
Besides, this configuration adapts to illumination and/or emission of twisting light, like in circularly polarized luminescence \cite{circpollibro1,circpolemission,kumar}.

 In parallel to RHELT, {\it we will also study   resonance  energy transfer  (RET) between  these dipolar magnetoelectric bi-isotropic (or chiral)  rotating donors and acceptors}. We   shall address  time-harmonic fields at optical frequencies; except when  we consider the emission and absorption of donors and acceptors  over a range of frequencies, a case in which  the fields are taken at a generic frequency of their spectrum, and the polarizabilities are given by their effective values, expressed as overlapping integrals of their respective emission and absorption - or extinction - spectra.  

Moreover, due to our lack of  data on energy and  helicity  transfer  between magnetoelectric rotating chiral particles, and incomplete knowledge on values of their polarizabilities, and  given the  progress in the last years  in devising and  building nanoparticles with a large magnetic response to the field of light \cite{nietoJOSA,g-etxarri,geffrin,kuznetsov,staude, kivshar_reviews}, (which leads to  phenomena stemming from  the interplay between the particle  induced electric and magnetic dipoles), we may envisage a near future of both theretical and experimental research leading to techniques with magnetoelectric conjugates of  molecules and  nanoparticles, (or even of bulky magnetoelectric  molecules) with angular momentum gained on illumination with twisted light.  Therefore, {\it we shall address rather  large rotating magnetoelectric chiral particles with a diameter of a few tens of $nm$}, (see e.g. \cite{pyramids, jaque,dionne,chinos}), whether they actually are molecules, dielectric or metallic nanoparticles,  or conjugates of  them both; and hence  {\it possessing   large magnetic and cross electric-magnetic  polarizabilities, besides the electric one}. In this  way,  {\it the  interaction distance that our theory yields with these larger particles gets values considerably greater than the typical $5-10 nm$  F\"{o}rster radii of FRET. }

Our results will  therefore be qualitative as we shall not address specific material parameters beyond certain models of emission and absorption distributions. {\it For both RHELT and RET we obtain  transfer rates that include}, besides the electric polarizability term like in conventional FRET,  {\it additional contributions of both the cross electric-magnetic and the magnetic polarizabilities  of the donor and acceptor} and that, therefore, depend on their chirality handedness.

However, we shall show that while the structural chirality,  and polarization helicity, of the donor dipoles is implicitely contained in its electric and magnetic dipole moments,  included in the  transfer rate  and interaction radii equations, {\it  the chirality  of the acceptor  appears explicitely as its  cross electric-magnetic polarizability  in some terms of  these RHELT and RET equations.  As such, these terms are  discriminatory, and thus uniquely characterize the symmetry handedness of A}.       

Our electromagnetic theory is  different from the quantum-mechanical one previously established for transfer  and shifts of energy between molecules \cite{salam1,salam2,craig1,craig2},  as we deal not only with absorption, but also with electromagnetic scattering. Therefore the extinction (rather than just the absorption) of energy and helicity is considered. Namely, these absorbed plus scattered  quantities \cite{nieto1,nieto2,nieto3} are addressed in this work. The $r^{-6}$ interdistance dependence is recovered  for RET, as well as for RHELT, as a consequence of D and A being assumed dipoles in the near-field of each other.

Also, while in conventional FRET, the axcitation of A by D conveys a decrease in the energy  emitted by D,  there being an increase in the energy emitted by A,  we shall see  that {\it  for  chiral A and D  the energy transfer rate may be negative, which will indicate that the donor emission is enhanced, rather than inhibited,  in  presence of the acceptor}; and, hence, {\it the proximity of A increases the spontaneous decay rate of D}.  However, a negative helicity rate  does not necessarily mean an enhancement of helicity extinction in the donor due to the presence of the acceptor, since {\it  the extinction of  helicity, at difference  with that of  energy, is expected to have either negative or  positive values, even for purely dielectric interacting particles, depending on the  sense of rotation of the emited wavefield}. 

In the following sections we develop the details of these new phenomena. Finally, we shall also introduce the observables linked to these effects, discussing their behavior and their relationships with the main quantities involved in these transfer processes.

\section{Dipolar excitation of  helicity and energy}
In standard FRET, a  molecule D excited on illumination  emits  falling to its ground state, and part of this emitted light is absorbed by  a molecule A which is  in the near-field zone of  D.  This process  requires that  the  emission spectrum of D and the absorption spectrum of  A  have a certain overlap \cite{fret1,clegg, novotny}, so that such a resonance (i.e. non-radiative)  transfer of energy from D to A may take place.  

As stated in the introduction, we assume A and D being "particles" in  general, by which we mean either quantum dots,  synthesized material nanoparticles,  molecules, or conjugates of both of them. We  address the spatial parts ${\bf E}({\bf r})$ and ${\bf B}({\bf r})$  of the electric and magnetic vectors of  time-harmonic electromagnetic  fields in their complex representation. [If A and D emit and extinguish, or absorb, light over a range of frequencies, these fields are understood at a generic frequency of their spectrum: ${\bf E}({\bf r},\omega)$ and ${\bf B}({\bf r,\omega})$]. Their interaction in a medium of refractive index $n =\sqrt{ \epsilon\mu}$ with a {\it magnetoelectric,  bi-isotropic } and {\it dipolar} "particle",  is  given through its  electric, magnetic, and cross electric-magnetic polarizability tensors: $\underline{\bm \alpha}_{e}$,  $\underline{\bm\alpha}_{m}$,  $\underline{\bm\alpha}_{em}$ and $\underline{\bm\alpha}_{me}$ .    $k=n\omega/c=2\pi n/\lambda$. Hence, the electric and magnetic dipole moments, $ {\bf p}$ and  ${\bf m}$, induced in the particle by this  field are given by the constitutive relations:  ${\bf p}=\underline{\bm\alpha}_{e}{\bf E}+\underline{\bm\alpha}_{em}{\bf B}$, ${\bf m}=\underline{\bm\alpha}_{me}{\bf E}+\underline{\bm\alpha}_{m}{\bf B}$.

In this work we shall  consider the bi-isotropic particle being {\it chiral } reciprocal, so that $\underline{\bm\alpha}_{em}=-\underline{\bm\alpha}_{me}^{^{\dag}}$. The sign $\dag$ standing for conjugate transpose.

Using a Gaussian system of units,  the time-averaged (written as $<\cdot>)$ electromagnetic {\it  helicity} density  of the wavefield: $\hel({\bf r})= <\hel({\bf r})>=\frac{1}{2k}\sqrt{\frac{\epsilon}{\mu}} \mbox{Im} [{\bf E}({\bf r})\cdot {\bf B}^*({\bf r})]$ is a conserved quantity fulfilling the continuity equation \cite{nieto2,bliokh4}: $\nabla \cdot {\cal F} =- \mathscr{ P}$, where ${\cal F}$ is the helicity density {\it flow}   which for these fields coincides with their spin angular momentum \cite{nieto2, bliokh4}, and $\mathscr{ P}$ denotes the {\it conversion}  of helicity, i.e. its decrease or increase by absorption and/or scattering of the wavefield by the particle \cite{nieto2,gutsche2, gutsche4}. Henceforth, Re and Im stand for real and imaginary part, respectively;  and $^*$ denotes complex conjugate.  Most optical wavefields can be decomposed into the sum of a field ${\bf E}^{+}({\bf r})$ with all plane wave components  being left circularly polarized (LCP), plus  a field ${\bf E}^{-}({\bf r})$ whose components are right circularly polarized (RCP), so that the above  helicity density may be expressed as \cite{nieto1,nieto3}:
${\hel}({\bf r})= (\epsilon/2k)[|{\bf E}^{+}({\bf r})|^2 -  |{\bf E}^{-}({\bf r})|^2]$;  while its time-averged energy density reads: $<{w}({\bf r})>= (\epsilon/8\pi)[|{\bf E}^{+}({\bf r})|^2 + |{\bf E}^{-}({\bf r})|^2]$.

With reference to a Cartesian framework $OXYZ$, (cf. Fig. 1), {\it our analysis  is based on the extinction of electromagnetic  helicity and energy  by the electric and magnetic  dipoles ${\bf p}_A$ and ${\bf m}_A$ of the  acceptor particle  A, placed at a point of position vector ${\bf r}_A$, on interaction with the field  ${\bf E}_D({\bf r}_A)$, ${\bf B}_D({\bf r}_A)$ emitted by the rotating electric and magnetic dipoles ${\bf p}_D$ and ${\bf m}_D$,  induced on illumination of the donor particle  D by  generally  twisted  light, propagating along a main direction defined by the ${\bf s}_i$ vector}, (see Fig. 1). In its most general form stemming from the   optical theorem, this reads  for the transfer of energy from D to A:
\be
 {\cal W}^{DA}= \frac{\omega}{2} \mbox{Im}[ {\bf p}_{A}  \cdot {\bf E}_{D}^{*}({\bf r}_{A})   +{\bf m}_{A}\cdot  {\bf B}_{D}^{*}({\bf r}_A)] . \label{topefret}
\ee
Whereas the extinction of  electromagnetic  helicity - or wavefield chirality  \cite{nieto1,tang1,gutsche4} - in A of the emission from D is, according to  the helicity optical theorem, (cf. Eq. (36) of \cite{nieto2}):
\be
{\cal W}_{\hel}^{DA}=2\pi c \mbox{Re} \{ -\frac{1}{n^2} {\bf p}_{A}  \cdot {\bf B}_{D}^{*}({\bf r}_A) + {\bf m}_{A}\cdot  {\bf E}_{D}^{*}({\bf r}_A) \}. \label{tohelfret}
\ee
Eq.  (\ref{tohelfret}) determines the total transfer of  helicity from D to A.
\begin{figure}[htbp]
\centerline{\includegraphics[width=1.15\columnwidth]{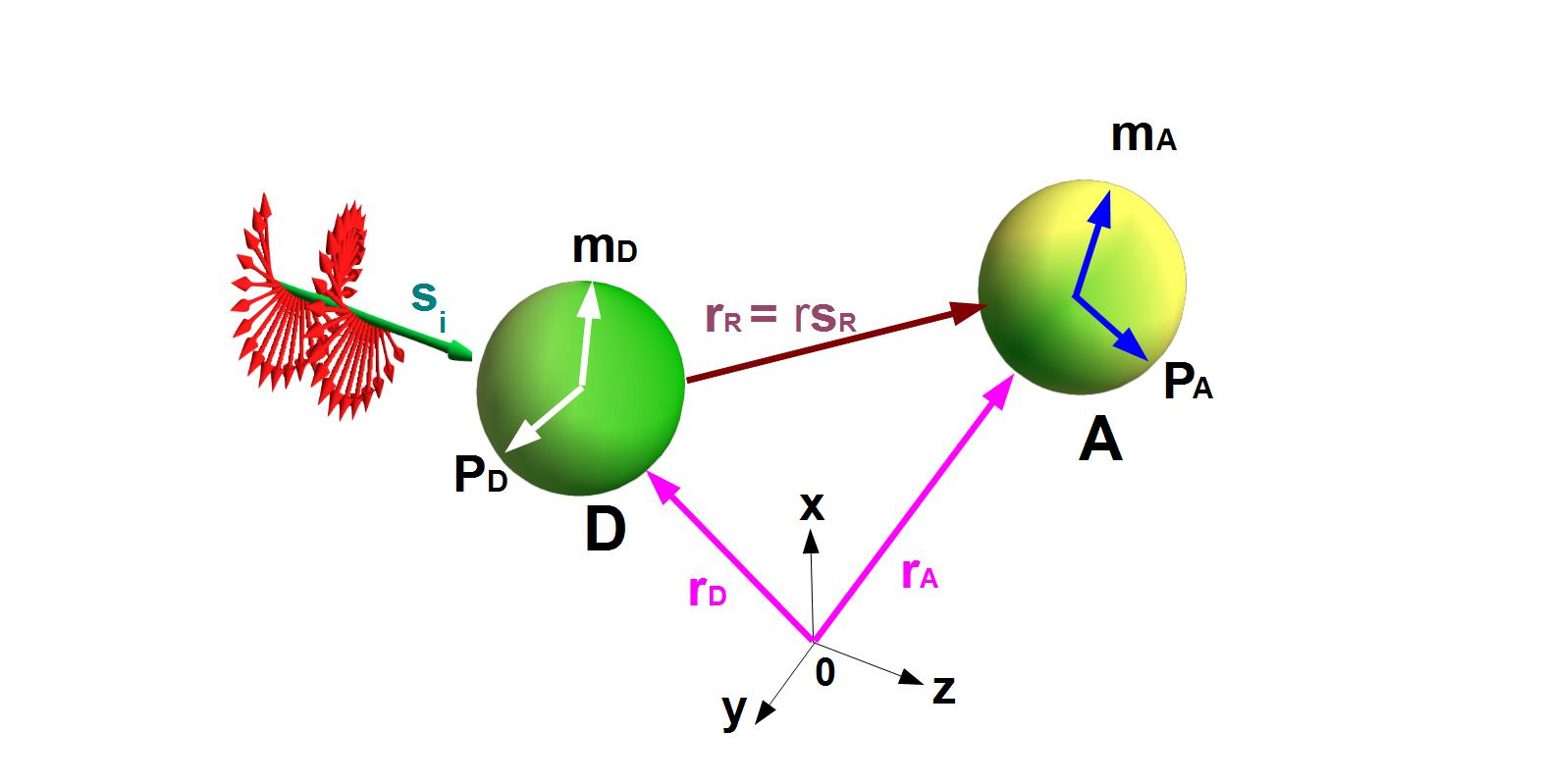}}
\caption{(Color online). A twisted wave, propagating along ${\bf s}_i$,  incides on a  polarizable donor particle D, inducing   dipole moments ${\bf p}_D$  and ${\bf m}_D$ which emit light, inducing dipoles  ${\bf p}_A$  and ${\bf m}_A$ in the acceptor  A. }
\end{figure}
Expressing the vector pointing from D to A as ${\bf r}_R = r {\bf s}_R$, $ r=|{\bf r}_R|$, $|{\bf s}_R|=1$,  $r$ being the distance between the centers of D and A, (see Fig.1),  we write the near-fields ${\bf E}_{D }$ and ${\bf B}_{D}$ emitted by D at the center of A at distance $r$ from the center of D: ${\bf r}_R={\bf r}_A -{\bf r}_D $:
\be
{\bf E}_{D}({\bf r}_A)= \frac{1}{\epsilon r^3} [3({\bf p}_D \cdot {\bf s}_R )-{\bf p}_D]; \,\,\,\,\,\,\,\,\, \nonumber \\
{\bf B}_{D}({\bf r}_A)= \frac{\mu}{ r^3} [3({\bf m}_D\cdot {\bf s}_R)-{\bf m}_D].\,\,
  \,\,\,\,\,\,\label{nfields}
\ee
A weak coupling regime between D and A is assumed, so that there is no scattering feedback between them.
\section{RHELT and RET between chiral particles}
\subsection{Bi-isotropic particles}
We shall first consider the donor D and acceptor A magnetoelectric and generally  bi-isotropic;  subsequently we shall particularize them being chiral.  We  assume  the  electric and magnetic dipoles induced  in A by the wavefields (\ref{nfields})  emitted by D,  (transition dipoles if A is a molecule),   polarized in   directions given by the complex unit vectors ${\bf s}_{A}^{p}$ and ${\bf s}_{A}^{m}$, thus presenting orientational photoselection with respect to the illuminating field from D \cite{circpollibro1,circpolemission} :
\be
{p}_{ A\, i}=\alpha_{e}^{A}s_{A\, i}^{p}s_{A\, j}^{p\,*} {E}_{D \, j}+ \alpha_{em}^{A}s_{A\,i}^{p}s_{A\, j}^{m\,*} {B}_{D \, j}=p_A s_{A\,i}^{p} . \,\,\,\, \,\,\,\,\nonumber \\
{m}_{ A\, i}=\alpha_{me}^{A}s_{A\, i}^{m}s_{A\, j}^{p\,*} {E}_{D \, j}+ \alpha_{m}^{A}s_{A\, i}^{m}s_{A\, j}^{m\,*} {B}_{D \, j}=m_A s_{A\,i}^{m}. \,\,\,\, \,\,\,\,\nonumber \\
p_A=\alpha_{e}^{A}s_{A\, j}^{p\,*} {E}_{D \, j}+\alpha_{em}^{A}s_{A\, j}^{m\,*} {B}_{D \, j}.\,\, \,\,\,    \nonumber \\
m_A=\alpha_{me}^{A}s_{A\, j}^{p\,*} {E}_{D \, j}+ \alpha_{m}^{A}s_{A\, j}^{m\,*} {B}_{D \, j}. \,\,\,\,( i,j=1,2,3). \,\,\,\, \,\,\,\, \label{constifret}
\ee
Having used the notation of summing over all repeated indices.   If  also  D were orientationally photoselective,    we would consider  its electric and magnetic  dipoles, induced by an incident  field ${\bf E}^{(i)}, {\bf B}^{(i)}$, also polarized in   directions given by the complex unit vectors ${\bf s}_{D}^{p}$ and ${\bf s}_{D}^{m}$ as:
\be
{p}_{ D\, i}=\alpha_{e}^{D}s_{D\, i}^{p}s_{D\, j}^{p\,*} {E}_{j}^{(i)}+ \alpha_{em}^{D}s_{D\,i}^{p}s_{D\, j}^{m\,*} {B}_{ j}^{(i)}=p_D s_{D\,i}^{p} . \,\,\,\, \,\,\,\,\nonumber \\
{m}_{ D\, i}=\alpha_{me}^{D}s_{D\, i}^{m}s_{D\, j}^{p\,*} {E}_{j}^{(i)}+ \alpha_{m}^{D}s_{D\, i}^{m}s_{D\, j}^{m\,*} {B}_{j}^{(i)}=m_D s_{D\,i}^{m}. \,\,\,\, \,\,\,\,\nonumber \\
p_D=\alpha_{e}^{D}s_{D\, j}^{p\,*} {E}_{j}^{(i)}+\alpha_{em}^{D}s_{D\, j}^{m\,*} {B}_{ j}^{(i)}.\,\, \,\,\,  \nonumber \\
 m_D=\alpha_{me}^{D}s_{D\, j}^{p\,*} {E}_{j}^{(i)}+ \alpha_{m}^{D}s_{D\, j}^{m\,*} {B}_{j}^{(i)}. \,\,\,\,( i,j=1,2,3). \,\,\,\, \,\,\,\, \label{constifret1} 
\ee 

 In (\ref{constifret}) and (\ref{constifret1}) the  corresponding  eight  polarizability tensors of the induced dipoles  have been expressed in terms of the unit 3-D complex vectors ${\bf s}_{A}^{p}$ and  ${\bf s}_{A}^{m}$, and ${\bf s}_{D}^{p}$ and  ${\bf s}_{D}^{m}$, with components that we write in a condensed manner as: $(\underline{\bm\alpha}_{e,m,em,me}^{A,D})_{ij}=\alpha_{e,m,em,me}^{A,D}s_{A,D\, i}^{p,m}s_{A,D\, j}^{p,m\,*}$.  $( i,j=1,2,3)$. Of course all  polarizabilities are functions of frequency as explicitely shown in Eqs. (A2-6) - (A2-9) of Appendix 2.
 
Taking into account that according to the right side of (\ref{constifret}) and (\ref{constifret1}), $p_A$, $m_A$, $p_D$ and  $m_D$ are complex scalars,  and    ${\bf s}_D^{p}$, ${\bf s}_D^{m}$, ${\bf s}_A^{p}$ and $ {\bf s}_A^{m}$ are complex unit vectors, we obtain by introducing  (\ref{constifret}) and  (\ref{constifret1}) into (\ref{nfields}),   (\ref{topefret}) and   (\ref{tohelfret}), the extinction of  helicity 
${\cal W}_{\hel}^{DA}$ and energy  ${\cal W}^{DA}$ in A of the field emitted by D, (see the derivation in Appendix 1). Namely,
\be
{\cal W}_{\hel}^{DA}=\frac{2\pi c}{\epsilon r^6} \{-\frac{1}{\epsilon} \mbox{Re} [\alpha_{e}^{A} p_{D} m_{D}^{*} {\cal K} ^{(1)}_{\cal H}] 
+\mu  \mbox{Re} [\alpha_{m}^{A}  p_{D}^{*} m_{D}  {\cal K} ^{(2)}_{\cal H}] \nonumber \\
- \mu | m_{D}|^{2} \mbox{Re} [\alpha_{em}^{A}  {\cal K} ^{(3)}_{\cal H}]
+ \frac{| p_{D}|^{2}}{\epsilon} \mbox{Re} [\alpha_{me}^{A}   {\cal K} ^{(4)}_{\cal H}]\}, \,\,\,\,\,\,\,\,\,\,  \label{whfr}
\ee
and
\be
{\cal W}^{DA}=\frac{\omega}{2   r^{6}}  \{ \frac{\mbox{Im} [\alpha_{e}^{A}]}{\epsilon^{2}} | p_{D}|^{2}{\cal K} ^{(1)}  + \mu ^{2} |m_{D}|^{2} \mbox{Im} [\alpha_{m}^{A}]  {\cal K} ^{(2)}\nonumber \\
+\frac{\mu}{\epsilon} \mbox{Im} [\alpha_{em}^{A}  p_{D}^{*} m_{D}{\cal K} ^{(3)}+\alpha_{me}^{A}  p_{D} m_{D}^{*} {\cal K} ^{(3) \, *}] 
 \}  .  \,\,\,\, \label{wwfr}
\ee
{\it Eq. (\ref{whfr}) governs RHELT as the transfer of helicity from D to A} on extinction in A of the wavefield  helicity emitted by D. {\it This  equation constitutes a new law} to be considered  together with  Eq. (\ref{wwfr})  for  resonance energy transfer (RET) between magnetoelectric bi-isotropic particles.  Both transfer laws have an $r^{-6}$ dependence as a consequence of dealing with dipolar near fields,  [cf. Eq. (\ref{nfields})]. 

\subsection{Orientational factors}
In addition to this interaction distance $r$ between D and A, the  orientational ${\cal K}$-factors, shown in Eqs.(\ref{K1H})-(\ref{K3}) below,  determine the transfer of field helicity and energy. The RHELT orientational factors read, (see their derivation in Section A1.a of Appendix 1):
\be
{\cal K} ^{(1)}_{\cal H}=[ 3({\bf s}_{A}^{p}\cdot {\bf s}_{R}) ({\bf s}_{R}\cdot {\bf s}_{D}^{m })-  ({\bf s}_{A}^{p}\cdot {\bf s}_{D}^{m})] 
\,\,\,\,\,\,\,\,\,\,\,\,\nonumber \\ \times   
[3({\bf s}_{R}\cdot {\bf s}_{A}^{p}) ({\bf s}_{D}^{p}\cdot {\bf s}_{R})-  ({\bf s}_{A}^{p\,*}\cdot {\bf s}_{D}^{p\,*})].  \,\,\,\,\,\,\,\,\,\,\,\,\,\, \label{K1H} \\
 {\cal K} ^{(2)}_{\cal H}= [3({\bf s}_{A}^{m}\cdot {\bf s}_{R}) ({\bf s}_{R}\cdot {\bf s}_{D}^{p})-  ({\bf s}_{A}^{m}\cdot {\bf s}_{D}^{p})]
\,\,\,\,\,\,\,\,\,\,\,\,\,\,\,\,\,\,\,\nonumber \\ \times 
 [3({\bf s}_{R}\cdot {\bf s}_{A}^{m}) ({\bf s}_{D}^{m }\cdot {\bf s}_{R})-  ({\bf s}_{A}^{m\,*}\cdot {\bf s}_{D}^{m\,*})] \}.\,\,\,\,\,\,\, \label{K2H}\\
{\cal K} ^{(3)}_{\cal H} = [3({\bf s}_{A}^{p}\cdot{\bf s}_{R} ) ( {\bf s}_{R}\cdot {\bf s}_{D}^{m})-  ({\bf s}_{A}^{p}\cdot {\bf s}_{D}^{m})] 
\,\,\,\,\,\,\,\,\,\,\,\,\,\,\,\,\nonumber \\  \times
[3({\bf s}_{R}\cdot {\bf s}_{A}^{m})   ({\bf s}_{D}^{m}\cdot {\bf s}_{R})-({\bf s}_{A}^{m\,*}\cdot {\bf s}_{D}^{m\,*})] .\,\,\,\,\,\,
\label{K3H} \\
{\cal K} ^{(4)}_{\cal H} =[3({\bf s}_{A}^{m}\cdot {\bf s}_{R}) ({\bf s}_{R}\cdot {\bf s}_{D}^{p})-  ({\bf s}_{A}^{m}\cdot {\bf s}_{D}^{p})] 
\,\,\,\,\,\,\,\,\,\,\,\,\,\,\,\, \nonumber \\
\times 
 3({\bf s}_{R}\cdot {\bf s}_{A}^{p}) ({\bf s}_{D}^{p}\cdot {\bf s}_R) -  ({\bf s}_{A}^{p\,*}\cdot {\bf s}_{D}^{p\,*})]. \label{K4H}\,\,\,\,\,\,\,\,\,\,
\ee
While those of RET are (cf. Section A1.b of Appendix 1):
\be
 {\cal K} ^{(1)}=|3 ({\bf s}_{R}\cdot {\bf s}_{A}^{p}) ({\bf s}_{D}^{p}\cdot {\bf s}_{R})-({\bf s}_{D}^{p}\cdot {\bf s}_{A}^{p})] |^2 .\,\,\,\,\,\,\,\,\,\,\,\,\,\,\,\,\label{K1}\,\,\,\,
 \\
{\cal K} ^{(2)}= |3({\bf s}_{R}\cdot {\bf s}_{A}^{m}) ({\bf s}_{D}^{m}\cdot {\bf s}_{R}) - ({\bf s}_{D}^{m}\cdot {\bf s}_{A}^{m})]|^2.\,\,\,\,\,\,\,\,\,\,\,\,\,\,\,\,\label{K2}\,\,\,\, \\
{\cal K} ^{(3)}=  [3({\bf s}_{A}^{p}\cdot {\bf s}_{R}) ({\bf s}_{R}\cdot {\bf s}_{D}^{p}) - ({\bf s}_{A}^{p}\cdot {\bf s}_{D}^{p})] 
\,\,\,\,\,\,\,\,\,\,\,\,\,\,\,\,\, \,\,\,\,\,\,\,\,\,\,\, \,\,\,\,\,\,\,\,\nonumber \\\times 
[3 ( {\bf s}_{R}\cdot {\bf s}_{A}^{m}) ({\bf s}_{D}^{m}\cdot {\bf s}_{R} )- ({\bf s}_{D}^{m}\cdot{\bf s}_{A}^{m})]. \,\,\,\,\,\,\,\,\,\,\,\,\,\,\,\,\, \,\,\,\,\,\,\,\,\,\,\, \label{K3} 
\ee
We have employed the notation of scalar product in the Hilbert space of complex vectors:  ${\bf a} \cdot {\bf b}= a_i b_{i}^*$, $(i=1,2,3)$. {\it Eqs. (\ref{whfr}) and (\ref{wwfr}), with the orientation factors  (\ref{K1H}) - (\ref{K3}), are the main result of this paper}.

 Eq.(\ref{wwfr}) for  the transfer of energy,  ${\cal W}^{DA}$, as well as ${\cal K} ^{(1)}$,   reduce to those well-known of conventional FRET with orientation factor $\kappa^2$ when only linear  electric dipole moments ${\bf p}_D$ and ${\bf p}_A$ are excited in  D and  A, respectively. Obviously in this case there is no transfer of helicity, and Eq. (\ref{whfr}) yields  ${\cal W}_{\hel}^{DA}=0$. 

However,  Eqs. (\ref{whfr}) and (\ref{wwfr}) involve a rich variety of configurations and associated physical phenomena. This is seen on comparing, for instance, the full  Eqs. (\ref{whfr}) and (\ref{wwfr}) with their form in the particular case in which  only D were magnetoelectric and  A  were not bi-isotropic, so that only the   dipole moments ${\bf p}_D$, ${\bf m}_D$ and ${\bf p}_A$ would be excited. Then  (\ref{wwfr})  would resemble  that of  FRET  with {\it  circularly polarized donor emission}, while (\ref{whfr}) shows that in this case  there would also exist a helicity transfer proportional to $-{\epsilon}^{-1} \mbox{Re} [\alpha_{e}^{A} p_{D} m_{D}^{*} {\cal K} ^{(1)}_{\cal H}]$.

When   both A and D are bi-isotropic,   ${\cal W}_{\hel}^{DA}$ and ${\cal W}^{DA}$ contain additional information via the magnetic and the cross electric-magnetic polarizabilities of A, which appear linked with the factors $|p_{D}|^2$,  $| m_{D}|^2$,  and those of electric-magnetic dipole interference: $ p_{D}^{*} m_{D}$ and   $ p_{D} m_{D}^{*}$, in a reciprocal way in (\ref{whfr}) and (\ref{wwfr}).  Notice also that {\it the terms of  (\ref{whfr}) and (\ref{wwfr}) are discriminatory as they depend on the  chirality handedness of  both D and A}, namely,  on the sign of  the cross electric-magnetic polarizabilities of D and A.
\subsection{Chiral particles}
Next we address both  D and A  being {\it chiral reciprocal} \cite{gutsche4,shivola}, i.e.  $\alpha^{A,D}_{em}=-\alpha^{A,D}_{me}$. Then Eqs. (\ref{whfr}) and  (\ref{wwfr}) become
\be
{\cal W}_{\hel}^{DA}=\frac{2\pi c}{\epsilon r^6} \{-\frac{1}{\epsilon} \mbox{Re} (\alpha_{e}^{A} p_{D} m_{D}^{*} {\cal K} ^{(1)}_{\cal H})  \nonumber \\
+\mu  \mbox{Re} (\alpha_{m}^{A}  p_{D}^{*} m_{D}  {\cal K} ^{(2)}_{\cal H})+ \mu | m_{D}|^{2} \mbox{Re} (\alpha_{me}^{A}  {\cal K} ^{(3)}_{\cal H}) \nonumber \\
+ \frac{| p_{D}|^{2}}{\epsilon} \mbox{Re} ( \alpha_{me}^{A}   {\cal K} ^{(4)}_{\cal H}) \}, \,\,\,\,\,\,\,\,\,\,  \label{whhfr}
\ee
and
\be
{\cal W}^{DA}=\frac{\omega}{2   r^{6}}  \{ \frac{\mbox{Im} (\alpha_{e}^{A})}{\epsilon^{2}} | p_{D}|^{2}{\cal K} ^{(1)}  + \mu ^{2} |m_{D}|^{2} \mbox{Im} (\alpha_{m}^{A})  {\cal K} ^{(2)}
\nonumber \\
-2\frac{\mu}{\epsilon} \mbox{Re} (\alpha_{me}^{A}) \mbox{Im} (  p_{D}^{*} m_{D}{\cal K} ^{(3)}
) \}  .   \,\,\,\,\,\,\,\,\,\, 
\label{wwwfr}  \,\,\,\,\,\,\,\,\,\, 
\ee 
 As shown by (\ref{whhfr}), either the chirality of the acceptor A, or the excitation of both electric and  magnetic dipoles in the donor D, gives rise to a non-zero transfer rate of field helicity between D and A. This new  equation may be employed together with   Eq.(\ref{wwwfr}) of  field energy transfer.  Again, {\it  (\ref{whhfr}) and  (\ref{wwwfr}) contain   discriminatory terms which depend on both $\alpha_{me}^{D}$, (which influences on $p_D$ and $m_D$), and   $\alpha_{me}^{A}$}.  On the other hand,  Eqs. (\ref{whfr}) and  (\ref{whhfr}) allows us to introduce the new concept of    {\it   RHELT radius}, as shown next,  which  differs from the standard   F\"orster  radius: $ \frac{3}{2\epsilon k^{3}}  {\kappa}^{2}\,  \mbox{Im} \{ \alpha_{e}^{A} \}$  \cite{clegg, novotny}.

Also we notice in (\ref{wwfr}) and  (\ref{wwwfr}) that while the chirality of D  affects $p_D$ and $m_D$,  the chirality of A introduces terms with $\alpha_{em}^{D}$ and/or $\alpha_{me}^{D}$, which, as we shall see,  in some configurations  may be  larger than the sum of the first two terms of these equations, so that  {\it the energy transfer would be negative}. {\em I.e. the emission of energy from D in presence of A may be enhanced}, rather than reduced as in standard FRET \cite{berney}, {\em on account of the chirality  of A}, which contributes  to this effect through  {\it a sufficiently large} $\alpha_{me}^{A}$. This new  phenomenon is analysed in more detail later in this paper, in the section:  {\it Observables. Donor emission and decay rates}.

The discriminatory third term of   (\ref{wwwfr}) due to the {\it interference of  the electric and magnetic  dipoles ${\bf p}_D$ and ${\bf m_D}$, induced  in D,   makes the energy transfer to distinguish between an acceptor particle and its enantiomer}. I.e.  for a given factor $\mbox{Im} [  p_{D}^{*} m_{D}{\cal K} ^{(3)}]$, the sign of  $\mbox{Re}\{\alpha_{me}^{A}\}$ determines that of this third term of  (\ref{wwwfr}). Moreover, for large enough $\mbox{Re}\{\alpha_{me}^{A}\}$   {\it a chiral acceptor might give rise to ${\cal W}^{DA}<0$ while its enanatiomer, with opposite sign of  Re$\{\alpha_{me}^{A}\}$,  would produce  ${\cal W}^{DA}>0$}.  This is a remarkable effect, ruled out in conventional FRET.

{\it The same discriminatory phenomenon is seen in the transfer of  helicity}, Eq. (\ref{whhfr}).  Of course the helicity transfer rate   may be either positive or  negative, and the $\alpha_{me}^{A}$ discriminatory terms of  (\ref{whhfr}) influence the magnitude of ${\cal W}_{\hel}^{DA}$.

The condition of large  $ \mbox{Re}\{ \alpha_{me}^{A}\}$  may be investigated  at wavelengths where this cross electric-magnetic polarizability of  A  has strong  resonances, like in e.g. hybrids of nanostructures and large molecules  \cite{pyramids, dionne1, jaque, cana}.

\section{RHELT and RET radii}
Taking into account the right sides of the energy and helicity optical theorems  (\ref{topefret}) and (\ref{tohelfret}),  the helicity ${\cal W} _{\hel}^{0}$ and energy   ${\cal W}^{0}$ of the wavefield emitted by D  in absence of A, are \cite{nieto2}:
\be
 {\cal W} _{\hel}^{0}= \frac{8\pi c k^3}{3\epsilon}\mbox{Im} [ p_{D} m_{D}^{*}({\bf s}_{D}^{p}\cdot{\bf s}_{D}^{m \, *})].  \,\,\,\,\,\,\,\,\ \nonumber \\
 {\cal W}^{0}=\frac{ck^4}{3n}(\frac{|p_{D}|^{2}}{\epsilon}
+\mu |m_{D}|^{2}). \,\,\, \label{W0}
\ee
Having written as in (\ref{constifret1}): ${\bf p}_{D}={p}_{D}{\bf s}_{D}^{p}$, \,\, ${\bf m}_{D}={m}_{D}{\bf s}_{D}^{m}$.

For the normalized transfer rates of helicity and energy we get
\be
\frac{\gamma _{\hel}^{DA}}{ \gamma _{\hel}^{0}} =\frac{{\cal W} _{\hel}^{DA}}{ {\cal W} _{\hel}^{0}},\,\,\,\,\,\,\,\, \,\,\,\,\,\,\,\,\frac{\gamma^{DA}}{ \gamma^{0}} =\frac{{\cal W}^{DA}}{ {\cal W}^{0}} . \label{fFRHg}
\ee
Where $\gamma _{\hel}^{DA}$ and  $\gamma^{DA}$ are the  rates of helicity and energy transfer from  donor to  acceptor, respectively;  while  $\gamma _{\hel}^{0}$ and $\gamma ^{0}$ represent  the helicity and energy spontaneous decay rates from the donor when there is no acceptor.

From the above ratios we may introduce  the {\it  RHELT} and {\it RET  interaction radii}, $R _{\hel}$ and  $R _{\cal E}$, respectively, between D and A:
\be
\frac{|{\cal W} _{\hel}^{DA}|}{| {\cal W} _{\hel}^{0}|} =  [\frac{R _{\hel}}{r}]^6. \,\,\,\,\,\,\,\,\,\,\,\,\,\,\,\,
\frac{|{\cal W}^{DA}|} {{\cal W}^{0}}= [\frac{R _{\cal E}}{r}]^6. \label{fFRH}
\ee
Notice that in order to obtain a distance, in (\ref{fFRH})  we have written modulus of the transferred and emitted quantities that may be negative.

 From Eqs. (\ref{whhfr}),  (\ref{wwwfr}) and (\ref{fFRH}) we obtain $R _{\hel}$ and $R _{\cal E}$ 
expressed as
\be
   R _{\hel}^{6}= \frac{3}{4 k^{3}}\,|  -\frac{1}{\epsilon} \mbox{Re} (\alpha_{e}^{A} p_{D} m_{D}^{*} {\cal K} ^{(1)}_{\cal H})  \nonumber \\
+\mu 
 \mbox{Re} (\alpha_{m}^{A}  p_{D}^{*} m_{D}  {\cal K} ^{(2)}_{\cal H})+ \mu | m_{D}|^{2} \mbox{Re} (\alpha_{me}^{A}  {\cal K} ^{(3)}_{\cal H}) \nonumber \\
+ \frac{| p_{D}|^{2}}{\epsilon} \mbox{Re} (\alpha_{me}^{A}   {\cal K} ^{(4)}_{\cal H}) |/|\mbox{Im} [ p_{D} m_{D}^{*}({\bf s}_{D}^{p} \cdot {\bf s}_{D}^{m\, *})]|. \,\,\, \,\,\label{FRH}
\ee
\be
R _{\cal E}^{6}=  \frac{3}{2 k^{3}} \{ \frac{\mbox{Im} (\alpha_{e}^{A})}{\epsilon^{2}} | p_{D}|^{2}{\cal K} ^{(1)}  + \mu ^{2} |m_{D}|^{2} \mbox{Im} (\alpha_{m}^{A})  {\cal K} ^{(2)}  \nonumber \\
-2\frac{\mu}{\epsilon} \mbox{Re} (\alpha_{me}^{A}) \mbox{Im} (  p_{D}^{*} m_{D}{\cal K} ^{(3)}
) \}/(\frac{|p_{D}|^{2}}{\epsilon}
+\mu| m_{D}|^{2}). \,\,\,\,\,\label{FRE}
\ee
Notice that {\it the radii} introduced in Eqs. (\ref{FRH}) and (\ref{FRE}),  {\it  are functions of}  $\lambda$ and, hence, their bandwidth is limited by that of the emission and absorption (or extinction) spectra of D and A, respectively. 

On the other hand, we know from FRET theory that the interaction radius conveys an overlap integration of D and A spectra.  Therefore,  considering the range of wavelengths at which the donor and acceptor emits and absorbs, respectively, one should  substitute in the above RHELT and RET equations  the    acceptor polarizabilities $\alpha^A(\omega)$'s by  their {\it effective} values  $\alpha^{A\, eff}$, expressed in terms of the overlapping integrals of the donor emission spectra $f_e^{D}(\lambda)$, $f_m^{D}(\lambda)$, and  $f_{me\,A}^{CD}(\lambda)$ and the acceptor   cross-sections $\sigma_{e}^{a}(\lambda)$, $\sigma_{m}^{a}(\lambda)$, and $\sigma_{me}^{CD}(\lambda)$,  (cf. Eqs. (A2-19) - (A2-21) of Appendix 2):
\be
\mbox{Im}\{\alpha_{e}^{A\,eff}\}=\frac{3c}{4\pi}    \int_{0}^{\infty}d\lambda \, \frac{\epsilon(\lambda) f_{e}^{D} (\lambda)\sigma_{e\,A}^{a}(\lambda)}{ n(\lambda) \lambda}. \,\,\,\,\, \, \label{sigmaso1a}
\\
\mbox{Im}\{\alpha_{m}^{A\, eff}\}=\frac{3c}{4\pi}    \int_{0}^{\infty}d\lambda \, \frac{ f_{m}^{D}(\lambda)\sigma_{m\,A}^{a}(\lambda)}{ n(\lambda)\mu(\lambda) \lambda}
. \,\, \, \,\, \,\,\, \,\label{sigmaso22} \\
\mbox{Re}\{\alpha_{me}^{A\,eff}\}=\frac{3c}{16\pi}    \int_{0}^{\infty}d\lambda\, \sqrt{\frac{\epsilon(\lambda)}{\mu(\lambda)}}\frac{f_{me\,D}^{CD}(\lambda)\sigma_{me\,A}^{CD}(\lambda)} {n(\lambda) \lambda}. \,\,\, \, \,\, \,\,\,   \label{sigmaso3a} 
\ee
If scattering were strong in A, extinction rather than absorption should be considered in the acceptor particle.

Since, however,  our aim  is to understand  the helicity and energy transfer in terms of the polarizability spectra, (which of course depend on the emission and extinction - or absorption -  spectra of D and A), in the numerical examples to show later, {\it instead of determining through (\ref{sigmaso1a})-(\ref{sigmaso3a}) just one number for the value of   $ R _{\cal E}$  and  $R _{\hel}$ corresponding to a concrete donor and acceptor with experimentally  determined polarizabilities}, (which are scarce as far as we know),  {\it we will make use of  (\ref{whhfr}), (\ref{wwwfr}), (\ref{FRH}) and (\ref{FRE})   to establishing the values acquired by ${\cal W} _{\hel}^{DA}(\lambda)$, ${\cal W}^{DA}(\lambda)$,  $ R _{\cal E}(\lambda)$  and  $R _{\hel}(\lambda)$ in a range of wavelengths}.

 Of course one may envisage this point of view as if the donor emitted at frequencies $\omega'$ with the distribution $f_{e,m, me}^{D}(\omega')\delta(\omega'-\omega)$, also $\omega$ being   variable, since evidently $\int_{0}^{\infty}d\omega' f_{e,m, me}^{D,CD}(\omega')\delta(\omega'-\omega)\sigma_{e,m,me\,A}^{a, \,CD}(\omega')= f_{e,m, me}^{D}(\omega)\sigma_{e,m,me\,A}^{a, \,CD}(\omega)$.

Appendix 3  contains a {\it test and calibration} of our formulation with existing data. This confirms the  validity of our equations for the RHELT and RET   rates and interaction radii.

\section{Free orientation of donor dipoles with incident polarization. Both donor and acceptor being  chiral}
Let a  time-harmonic, elliptically polarized, plane wave with ${\bf E}_i={\bf e}_{i} e^{ik({\bf s}_{i}\cdot {\bf r})}$,  ${\bf B}_i={\bf b}_{i} e^{ik({\bf s}_{i}\cdot {\bf r})}$ be incident on the donor chiral generic  particle D, (cf. Fig.2). ${\bf b}_{i}=n {\bf s}_{i} \times  {\bf e}_{i}$, ${\bf e}_{i} \cdot {\bf s}_{i}= {\bf b}_{i} \cdot {\bf s}_{i}=0$.

We consider    ${\bf s}_i$ along $OZ$, (see Fig. 2); expressing ${\bf E}_i$ and ${\bf B}_i$  in the helicity basis:   ${\bm \epsilon}^{\pm}=(1/\sqrt{2})(\hat{\bf x} \pm i \hat{\bf y})$   as the sum of a left-handed (LCP)  and a right-handed  (RCP) circularly polarized (CPL)  plane wave,  so that
 \be
{\bf e}_{i}=( e_{i x},e_{i y},0)=e_{i}^{+}{\bm \epsilon}^{+}+ e_{i}^{-}{\bm \epsilon}^{-}. \,\,\,\,\,\,  {\bf  b}_{i}= ( b_{i x},b_{i y},0)  \nonumber \, \,\,\,\,\,\,\,  \,\,\,\,\,\,\,\\ 
=n ( -e_{i y},e_{i x},0)=b_{i}^{+}{\bm \epsilon}^{+}+ b_{i}^{-}{\bm \epsilon}^{-}  
=-ni(e_{i}^{+}{\bm \epsilon}^{+}- e_{i}^{-}{\bm \epsilon}^{-}). \,\,\,\,  \label{bheli1}
\ee
The  $+$ and $-$ superscripts standing for LCP (+) and RCP (-), respectively. In this representation, the incident helicity density  \cite{nieto2} reads:    
\be
{\hel^{i}}=(\epsilon/k)\mbox{Im}[e_{i x}^{*}e_{i y}]= (\epsilon/2k) S_{3}=(\epsilon/2k)[|e_{i}^{+}|^2-|e_{i}^{-}|^2].\, \,\,\,\label{helhel1}
\ee
which is the well-known expression of  ${\hel^{i}}$ as the difference between the LCP and RCP intensities of the field.   $S_3 = 2 \mbox{Im}[e_{i x}^{*}e_{i y}]= |e_{i}^{+}|^2-|e_{i}^{-}|^2$ is the 4th Stokes parameter \cite{born}. Also  $|e_{i}|^2= | e_{i x}|^2+|e_{i y}|^2=\frac{8\pi}{c}\sqrt{\frac{\mu}{\epsilon}} <S>=\frac{8\pi}{\epsilon} <w>$. $<S>$ and $<w>$ representing the incident field  time-averaged Poynting vector magnitude  and electromagnetic  energy density, respectively.  $<w>=<w_e>+<w_m>$ .  $<w_e>=(\epsilon/16 \pi)|{\bf E}_{i}|^{2}$, $<w_m>=(1/16 \pi  \mu)|{\bf B}_{i}|^{2}$.  

\subsection{Characterization of the donor dipole moments}
The angular momentum of the twisted  incident field is transferred to the donor, so that it induces dipoles ${\bf p}_D$ and ${\bf m}_D$  in D which are  free to orient and rotate with the polarization of the  illumination. Then Eqs. (\ref{constifret1}) hold with ${ s}_{D\, i}^{p} { s}_{D\, j}^{p\,*}={ s}_{D\, i}^{m} { s}_{D\, j}^{m\,*}=s_{D\,i}^{p}s_{D\, j}^{m\,*} =s_{D\, i}^{m}s_{D\, j}^{p\,*} = \delta_{ij}$, ($i,j=1,2,3$). 

In the helicity basis $ \{{\bm \epsilon}^{+}, {\bm \epsilon}^{-}\}$ Eqs. (\ref{constifret1}) yield for these induced dipoles
\be
{\bf p}_D=p_{D}^{+}{\bm \epsilon}^{+}+ p_{D}^{-}{\bm \epsilon}^{-}=p_{D}{\bf s}_{D}^{p}. \,\,\,\,\,  \,\,\,\,\,\,\,\,\,\   \,\,\,\, \nonumber \\
 p_{D}^{\pm}=(\alpha_{e}^{D} \pm n i \alpha_{me}^{D}) {e}_{i}^{\pm}.\,\,\,  
 p_{D}=|{\bf p}_D|=[|p_{D}^{+}|^{2}+ |p_{D}^{-}|^{2}]^{\frac{1}{2}};    \nonumber \\
\,\,\,\, {\bf s}_D^{p}=  p_{D}^{-1} [p_{D}^{+}{\bm \epsilon}^{+}+ p_{D}^{-}{\bm \epsilon}^{-}]; \,\,\,\,
|{\bf s}_D^{p}|^2 = s_{D \, i}^{p} \, {s}_{D\, i}^{p\, *} =1.  \,\,\,\,\,\,\, \label{p1}
\ee
\be
{\bf m}_D=m_{D}^{+}{\bm \epsilon}^{+}+ m_{D}^{-}{\bm \epsilon}^{-}=m_{D}{\bf s}_{D}^{m}. \,\,\,\,\,  \,\,\,\,\,\,\,\,\,\   \,\,\,\,\nonumber \\
 m_{D}^{\pm}=(\alpha_{me}^{D} \mp n i \alpha_{m}^{D}) {e}_{i}^{\pm}.\,\,\,  
 m_{D}=|{\bf m}_D|=[|m_{D}^{+}|^{2}+ |m_{D}^{-}|^{2}]^{\frac{1}{2}};    \nonumber \\
\,\,\,\, {\bf s}_D^{m}=  m_{D}^{-1} [m_{D}^{+}{\bm \epsilon}^{+}+ m_{D}^{-}{\bm \epsilon}^{-}] ;  \,\,
|{\bf s}_D^{m}|^2 ={ s}_{D\,i}^{m} \, {s}_{D\,i}^{m\, *} =1.\,\,\,\,\,  \,\,\label{m1}  
\ee
Notice that now the amplitudes $p_D$ and $m_D$ of the dipole moments, defined in (\ref{p1}) and (\ref{m1}), are real. This is in contrast with  previous sections where they were introduced as complex amplitudes [cf. Eq. (\ref{constifret1})]  which would coincide with either  $p_{D}^{\pm}$ and $m_{D}^{\pm}$ for incident circular polarization. In this latter case, since either   ${e}_{i}^{+}=0$  or ${e}_{i}^{-}=0$,   the near fields (\ref{nfields}) emitted by D would be circularly polarized along OZ \cite{nieto3}.  Then if D is a molecule, this emission of D would resemble the situation of {\it circularly polarized luminescence} \cite{circpollibro1,circpolemission}. 

According to Eqs. (\ref{K1H})-(\ref{K3}),  in order to evaluate the orientational factors we need the unitary vectors ${\bf s}_{D}^{p}$  and ${\bf s}_{D}^{m}$ of  ${\bf p}_D$ and ${\bf m}_D$. Without loss of generality,  the  coordinate framework of Fig. 2 is chosen such the $XY$-plane  contains  ${\bf s}_{D}^{p}$ and ${\bf s}_{D}^{m}$.  We write
\be
{\bf s}_D^{m}= \bm \xi  {\bf s}_D^{p},  \label{eym}
\ee
where $\bm \xi$ is the diagonal matrix: $\xi_{jk} =\xi_{j}\delta_{jk}$, ($j,k=1,2)$, which is orthogonal  since $|{\bf s}_D^{m}|=  |{\bf s}_D^{p}|=1$, and hence  $|\xi_{j}|= 1$, ($j=x,y$). Therefore

\be
 {\bf s}_D^{p}=  s_{D\, x}^{p} {\bf \hat {x}}+ s_{D\, y}^{p} {\bf \hat {y}},  \nonumber \\ \,\,\,\,\,\,\,\,\,\   \,\,\,\,
 {\bf s}_D^{m}=  s_{D\, x}^{m} {\bf \hat {x}}+ s_{D\, y}^{m} {\bf \hat {y}} =  \xi_x s_{D\, x}^{p} {\bf \hat {x}}+ \xi_y s_{D\, y}^{p} {\bf \hat {y}} . \label{carteym}
\ee

If the field impinging D is circularly polarized (CPL), then
\be
 {\bf s}_D^{p\, (\pm)}=e^{i\phi_e^{\pm}} {\bm \epsilon}^{\pm}, \,\,\,\,\,\,\,\,\,\,\,\,\,
 {\bf s}_D^{m\, (\pm)}=e^{i\phi_m^{\pm}} {\bm \epsilon}^{\pm}.  \label{circeym}
\ee
$\phi_e^{\pm}$ and $\phi_m^{\pm}$ are the arguments of the complex  $p_D^{\pm}$ and $m_D^{\pm}$, respectively. Thus (\ref{circeym}) conveys :
\be
 e^{-i\phi_e^{\pm}}{\bf s}_D^{p\, (\pm)}=  
e^{-i\phi_m^{\pm}} {\bf s}_D^{m\, (\pm)}, \,\,\,\,\,  {\bf s}_{D}^{m}=e^{i(\psi_{m}^{\pm}-\psi_{e}^{\pm})}{\bf s}_{D}^{p}. \label{cirsym}
\ee
Also  $\xi_{jk} =e^{i(\psi_{m}^{\pm}-\psi_{e}^{\pm})}\delta_{jk}$, ($j,k=1,2)$.

If the  illumination were CPL, a special situation would be when the donor D is a {\it dual} nanoparticle, $\epsilon^{-1}\alpha_D=\mu \alpha_m$ \cite{nieto2,corbato1,zambra1},  a case in which the donor dipoles illuminated by a field of  {\it well-defined helicity} (WDH), like an LCP or an RCP  plane wave  fulfill,  [cf. Eqs.(\ref{p1}) and (\ref{m1})]:  ${\bf p}_D^{ (\pm)}=\pm i n{\bf m}_D^{ (\pm)}$ \cite{nieto3,gutsche4,corbato1}, and hence  ${ p}_D^{ (\pm)}=\pm i n { m}_D^{ (\pm)}$. Several configurations pertaining to the relative orientations of  dipoles in D and A when a CPL plane wave illuminates D,  are  dicussed later.

Another interesting situation is that of {\it right} and {\it left} molecules \cite{klimov, dionne1}, corresponding to parallel and antiparallel ${\bf s}_D^{p}$ and ${\bf s}_D^{m}$: ${\bf s}_D^{m}=\pm {\bf s}_D^{p}$.

Recalling that $ p_D$ and $ m_D$ are real,  Eqs. (\ref{p1}) and (\ref{m1}) give
\be
 s_{D\, x}^{p}=\frac{1}{p_D\sqrt{2}} (p_{D}^{+} +p_D^-);\,\,\,\,s_{D\, y}^{p}=\frac{i}{p_D\sqrt{2}}(p_{D}^{+} -p_D^-).  \label{cartsDp}
\ee

And

\be
|{ s}_{D\,x,\,y}^{p}|^2=\frac{1}{2}\pm\frac{\mbox{Re}\{p_D^{+}p_D^{-\,*}\}}{p_D^2}\nonumber \\
=\frac{1}{2}(1\pm \mbox{Re} [e_{i}^{+} e_{i}^{-\,*}]\frac{|\alpha_e^{D}|^2-n^2 |\alpha_{me}^{D}|^2}{|\alpha_e^{D}|^2+n^2 |\alpha_{me}^{D}|^2}). \label{Sdperp}
\ee
where the upper and lower sign  in $\pm$   apply to ${ s}_{D\,x}^{p}$ and ${ s}_{D\,y}^{p}$,  respectively. Analogously, and with the same notation, we write
\be
|{ s}_{D\,x,\,y}^{m}|^2=\frac{1}{2}\pm\frac{\mbox{Re}\{m_D^{+}m_D^{-\,*}\}}{m_D^2}\nonumber \\
=\frac{1}{2}(1\mp \mbox{Re} [e_{i}^{+} e_{i}^{-\,*}]\frac{n^2|\alpha_m^{D}|-|\alpha_{me}^{D}|^2}{n^2|\alpha_m^{D}|^2 +  |\alpha_{me}^{D}|^2}). \label{Sdperpp}
\ee

\subsection{Characterization of the acceptor dipole moments}
In the $OXYZ$ framework of  Fig. 2,  using  polar and azimuthal angles $\alpha$ and $\beta$, we  write the unit vector ${\bf s}_R$ pointing from D to A  as:
\be
{\bf s}_R= \sin\alpha \cos\beta\hat{\bf x}+ \sin\alpha \sin\beta \hat{\bf y} + \cos\alpha  \hat{\bf z} \label{sR}
\ee
\begin{figure}[htbp]
\centerline{\includegraphics[width=1.15\columnwidth]{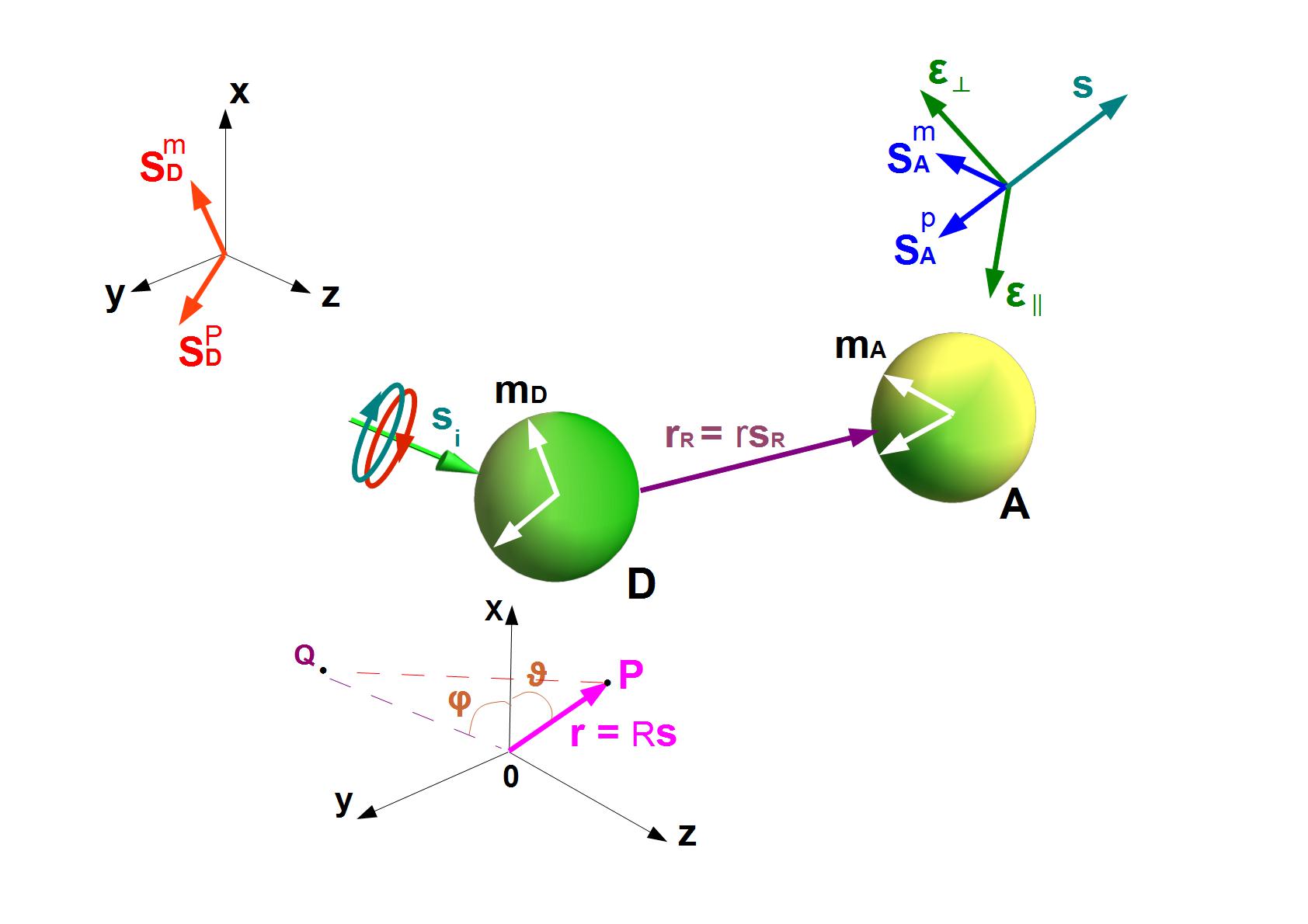}}
\caption{(Color online). In the  $OXYZ$  framework an elliptically polarized plane wave, propagating along ${\bf s}_i=\hat{\bm z}$,  incides on a   donor particle D. The fields emitted by the dipoles moments, oriented along ${\bf s}_{D}^{p}$ and  ${\bf s}_{D}^{m}$,  are evaluated at a generic point $P$: ${\bf r}=R \bf s$ , $|{\bf s}|=1$, of  coordinates   $(R, \theta, \phi)$ which eventually coincides with the center of the acceptor A with  dipoles moments oriented along ${\bf s}_{A}^{p}$ and  ${\bf s}_{A}^{m}$, thus becoming ${\bf r}={\bf r}_A$. The point  $Q$ is the projection of $P$ on the plane $ OXY$; the scattering plane being $OPQ$.  We  show the three orthonormal vectors:  ${\bf s}$,  ${\bm \epsilon}_{\parallel}$ (in the plane $OPQ$ and in the sense of rotation of $\theta$), and ${\bm \epsilon}_{\perp}$ (normal to $OPQ$). which form the helicity basis for the fields  $ {\bf E}_{D}({\bf r}_A), \,\, {\bf B}_{D}({\bf r}_A)$ in A:  $\{\hat{\bm \eta}^{+},\hat{\bm \eta}^{-}\}$, 
$\hat{\bm \eta}^{\pm}=\frac{1}{\sqrt{2}}(\hat{\bf e}_{\perp}\pm i \hat{\bf e}_{\parallel})$; while that  for the field incident on  D is  $\{\hat{\bm \epsilon}^{+},\hat{\bm \epsilon}^{-}\}$,  ${\bm \epsilon}^{\pm}=(1/\sqrt{2})(\hat{\bf x} \pm i \hat{\bf y})$. }
\end{figure}
As for the acceptor dipole moments  ${\bf p}_A=p_A {\bf s}_{A}^{p}$ and ${\bf m}_A=m_A {\bf s}_{A}^{m}$ , making the generic point $P$ to coincide with the center of the acceptor, i.e.  ${\bf r}={\bf r}_A$, we consider the origin of coordinates $O$, and therefore the orientation of the unit vector ${\bf s}={\bf r}_A/r_A$, such that without loss of generality both  ${\bf s}_{A}^{p}$ and ${\bf s}_{A}^{m}$ vary in the plane defined by the vectors  $\hat {\bf e}_{\perp}$ and $ \hat {\bf e}_{\parallel}$, normal to its position unit vector ${\bf s}=(\sin\theta\cos\phi, \sin\theta\sin\phi,\cos\theta)$, as shown in Fig. 2.    Notice that the orientations of  ${\bf s}$ and ${\bf s}_R$ are not linked to each other, which will later be important for calculating the orientational averages of the ${\cal K}$ - factors.  Therefore we write, (cf. Fig. 2):
\be
 {\bf p}_A=p_A {\bf s}_{A}^{p},\, p_A=|{\bf p}_A| , \,{\bf s}_{A}^{p}=  s_{A\, \perp}^{p}  \hat {\bf e}_{\perp}+ s_{A\, \parallel}^{p} \hat {\bf e}_{\parallel} ; \nonumber \\
 {\bf s}_A^{m}=  s_{A\, \perp}^{m}  \hat  {\bf e}_{\perp}+ s_{A\, \parallel}^{m}  \hat {\bf e}_{\parallel} =  \zeta_{\perp} s_{A\, \perp}^{p} \hat { \bf e}_{\perp}+ \zeta_{\parallel} s_{A\, \parallel}^{p}  \hat  {\bf e}_{\parallel}  . \,\,\,\,\, \label{Acarteym} 
\ee
\be
| s_{A\, \perp}^{m}|^2+ | s_{A\, \parallel}^{m}|^2=|\zeta_{\perp}|^2| s_{A\, \perp}^{p}|^2+ |\zeta_{\parallel}|^2| s_{A\, \parallel}^{p}|^2=1. \,\,\,\,\,\,\,\,\,\,\,\,\,\,\,\,\,\,\label{pAA}
\ee
 And the unitarity of  both ${\bf s}_A^{p}$ and ${\bf s}_A^{m}$  imply  $|\zeta_{\perp}|=|\zeta_{\parallel}|= 1$. 

Since
\be
 \hat {\bf e}_{\perp} =  \cos\theta \cos\phi\, \hat{\bf x}+\cos\theta \sin\phi \, \hat{\bf y}-\sin\theta \,\hat{\bf z},  \nonumber \\
 \hat {\bf e}_{\parallel}= \sin\phi \, \hat{\bf x}- \cos\phi \,\hat{\bf y}. \label{eparyeperp}
\ee
${\bf s}_{A}^{p}$ referred to the  $\hat{\bf x}, \hat{\bf y}, \hat{\bf z}$ basis is
\be
{\bf s}_{A}^{p}=
({ s}_{A\,\perp}^{p}\sin\phi+ { s}_{A\,\parallel}^{p} \cos\theta \cos\phi) \hat{\bf x}\nonumber \\
+ 
(-{ s}_{A\,\perp}^{p}\cos\phi+ { s}_{A\,\parallel}^{p} \cos\theta \sin\phi) \hat{\bf y} -  { s}_{A\,\parallel}^{p}\sin\theta \, \hat{\bf z}. \,\,\,\label{SAp}
\ee
And from (\ref{Acarteym}) we have for the acceptor  magnetic dipole  moment ${\bf m}_A=m_A {\bf s}_{A}^{m}$:
\be
{\bf s}_{A}^{m}=
(\zeta_{\perp}{ s}_{A\,\perp}^{p}\sin\phi+ \zeta_{\parallel}{ s}_{A\,\parallel}^{p} \cos\theta \cos\phi) \hat{\bf x} \nonumber \\
-(\zeta_{\perp}{ s}_{A\,\perp}^{p}\cos\phi- \zeta_{\parallel}{ s}_{A\,\parallel}^{p} \cos\theta \sin\phi) \hat{\bf y} -  \zeta_{\parallel}{ s}_{A\,\parallel}^{p}\sin\theta \, \hat{\bf z}.  \,\,\,\, \,,\label{SAm}
\ee
The components ${ s}_{A\,\perp}^{p}$ and ${ s}_{A\,\parallel}^{p}$  of the unit vector ${ \bf s}_{A}^{p}$ define the plane of reference of rotation of A which contain $\hat{\bf e}_{\perp}$ and $ \hat{\bf e}_{\parallel}$, [cf. Eqs. (\ref{Acarteym})], yielding the helicity basis $\{\hat{\bm \eta}^{+},\hat{\bm \eta}^{-}\}$:
$\hat{\bm \eta}^{\pm}=\frac{1}{\sqrt{2}}(\hat{\bf e}_{\perp}\pm i \hat{\bf e}_{\parallel})$,
 $({\bm \eta}^{\pm\,*}\cdot {\bm \eta}^{\mp}=0)$,  (see Fig. 2).
So that
 \be
{\bf p}_A=p_{A}^{+}\hat{\bm \eta}^{+}+ p_{A}^{-}\hat{\bm \eta}^{-}=p_{A}{\bf s}_{A}^{p}.\nonumber \\ 
{\bf m}_A=m_{A}^{+}\hat{\bm \eta}^{+}+ m_{A}^{-}\hat{\bm \eta}^{-}=m_{A}{\bf s}_{A}^{m}. \label{pAA11}
\ee
And then
\be
 p_{A}=[|p_{A}^{+}|^{2}+ |p_{A}^{-}|^{2}]^{\frac{1}{2}};    
\,\,\,\, {\bf s}_A^{p}=  p_{A}^{-1} [p_{A}^{+}\hat{\bm \eta}^{+}+ p_{A}^{-}\hat{\bm \eta}^{-}]; \nonumber \\
|{\bf s}_A^{p}|^2 ={s}_{A\,i}^{p}\,  { s}_{A\,i}^{p\, *} =1. \label{pA11}
\ee
\be
 m_{A}=[|m_{A}^{+}|^{2}+ |m_{A}^{-}|^{2}]^{\frac{1}{2}};   
\,\,\,\, {\bf s}_A^{m}=  m_{A}^{-1} [m_{A}^{+}\hat{\bm \eta}^{+}+ m_{A}^{-}\hat{\bm \eta}^{-}] ;  \nonumber \\
 |{\bf s}_A^{m}|^2 ={ s}_{A\,i}^{m}\,  s_{A\,i}^{m\, *} =1. \,\,\,\,\, \label{mA11}
\ee
Again we see that, like for the donor, the amplitudes $p_{A}$ and $m_{A}$ of the acceptor dipole moments  in this configuration are real rather than complex.

The $s_{A\, \perp}^{p}$ and $s_{A\, \parallel}^{p}$  components  are according to (\ref{pAA11})-(\ref{mA11}): 
\be
 s_{A\, \perp}^{p}=\frac{1}{p_A\sqrt{2}} (p_{A}^{+} +p_A^-);\,\,\,\,s_{A\, \parallel}^{p}=\frac{i}{p_A\sqrt{2}}(p_{A}^{+} -p_A^-).  \label{AcartsDp1}
\ee
With analogous expressions for $ s_{A\, \perp}^{m}$ and  $ s_{A\, \parallel}^{m}$ on substituting $p$ by $m$.

On the other hand, from  (\ref{pA11})-(\ref{AcartsDp1}):
\be
|{ s}_{A\,\perp,\,\parallel}^{p}|^2=\frac{1}{2}\pm\frac{\mbox{Re}\{p_A^{+}p_A^{-\,*}\}}{p_A^2}; 
\,\, \nonumber \\
|{ s}_{A\,\perp,\,\parallel}^{m}|^2=\frac{1}{2}\pm\frac{\mbox{Re}\{m_A^{+}m_A^{-\,*}\}}{m_A^2}.\,\,\,\, \label{Saperp1}
\ee
Where the upper and lower sign  in $\pm$   apply to $\perp$ and $\parallel$, respectively.

\subsection{Transfer of helicity and energy.  Interaction radii}
Finally, since according to (\ref{p1}) and (\ref{m1}) $p_D$ and $m_D$ are now real, and so are
 ${\cal K} ^{(1)}$ and ${\cal K} ^{(2)}$, the helicity and energy, transferred from D to A, [cf.  Eqs.(\ref{whhfr}) and (\ref{wwwfr})] become
\be
{\cal W}_{\hel}^{DA}=\frac{2\pi c}{\epsilon r^6}  [ p_{D}m_{D}\mbox{Re}  \{
-\frac{\alpha_{e}^{A}}{\epsilon}{\cal K} ^{(1)}_{\cal H}\,\,\,\,\,\,
\nonumber \\
+ \mu\alpha_{m}^{A}{\cal K} ^{(2)}_{\cal H}\} 
+ \mu m_{D}^{2}\mbox{Re} \{\alpha_{me}^{A}\, {\cal K} ^{(3)}_{\cal H}\}+ \frac{p_{D}^{2 }}{\epsilon}\mbox{Re} \{\alpha_{me}^{A}\, {\cal K} ^{(4)}_{\cal H}\}].  \,\,\,\,\,\, \label{tohelfret_2}
\ee
\be
 {\cal W}^{DA}=\frac{\omega}{2 r^{6}}  [ \frac{\mbox{Im} \{\alpha_{e}^{A}\}}{\epsilon^{2}} \, p_{D}^{2}\, {\cal K} ^{(1)}+\mu^{2} \mbox{Im} \{\alpha_{m}^{A}\}\, m_{D}^{2}\,{\cal K} ^{(2)} \nonumber \\
 -2\frac{\mu}{\epsilon} p_{D} m_{D} \mbox{Re}\{ \alpha_{me}^{A}\}\mbox{Im} \{{\cal K} ^{(3)}\} ]. \,\,\,\,\,\,\,\,\,\,\,\,\,\,\,\,\,\,
\label{topefret_2}
\ee

Once again, while Eq. (\ref{tohelfret_2}) accounts for a transfer of helicity between D and A, either positive or negative as a consequence of the handedness of  A and D, {\it  a negative transferred energy $ {\cal W}^{DA}$ may exist}, in contrast with conventional FRET, {\it  when the  last term of  (\ref {topefret_2})  is larger than the sum of the  first two terms}. 

The interaction radii $R _{\hel}$ and $R _{\cal E}$ now read
\be
   R _{\hel}^{6}= \frac{3}{4 k^{3}}\,   |  p_{D}m_{D}\mbox{Re}  \{
-\frac{\alpha_{e}^{A}}{\epsilon}{\cal K} ^{(1)}_{\cal H}
+ \mu\alpha_{m}^{A}{\cal K} ^{(2)}_{\cal H}\} \nonumber \\
+ \mu m_{D}^{2}
 \mbox{Re} \{\alpha_{me}^{A}\, {\cal K} ^{(3)}_{\cal H}\} 
  + \frac{p_{D}^{2 }}{\epsilon}\mbox{Re} \{\alpha_{me}^{A}\, {\cal K} ^{(4)}_{\cal H}\}| \nonumber \\
 /|\mbox{Im} ( p_{D}^{+} m_{D}^{+\,* } +  p_{D}^{-} m_{D\,*}^{- })|\,\,\,\,\, .\label{FR_H}
\ee
\be
R _{\cal E}^{6}=  \frac{3}{2 k^{3}}  | \frac{\mbox{Im} \{\alpha_{e}^{A}\}}{\epsilon^{2}} \, p_{D}^{2}\, {\cal K} ^{(1)}+\mu^{2} \mbox{Im} \{\alpha_{m}^{A}\}\, m_{D}^{2}\,{\cal K} ^{(2)}  \nonumber \\
 -2\frac{\mu}{\epsilon} p_{D} m_{D} \mbox{Re}\{ \alpha_{me}^{A}\}\mbox{Im} \{{\cal K} ^{(3)}\} |/ (\frac{p_{D}^{2}}{\epsilon}+\mu m_{D}^{2}).\,\,\,\,\,\,\,\label{FR_E}
\ee
\subsection{Orientational averages of the ${\cal K}$-factors}
 The orientation of the dipole moments  of D and A often randomly vary, so that the relative axes of rotation between D and A is unknown. In this case it is pertinent to work with the orientational averages of the ${\cal K}$-factors.

Using the form of the several unit vectors ${\bf s}_A$ and ${\bf s}_D$ shown above, these averages are 
\be
<{\cal K} ^{(1)}_{\cal H}>= \frac{1}{10}\{7|{ s}_{A\,\perp}^{p}|^2 +\frac{19}{3}|{ s}_{A\,\parallel}^{p}|^2\}
 \{\xi_{x}^{*}|s_{D\, x}^{p}|^2 \nonumber \\
 + \xi_{y}^{*}|s_{D\, y}^{p}|^{2}\}. \,\,\,\,\, \label{mkh1}
\ee
\be
 <{\cal K} ^{(2)}_{\cal H}>= \{\frac{7}{10}(|\zeta_{\perp}|^2|{ s}_{A\,\perp}^{p}|^2 +\frac{1}{3}|\zeta_{\parallel}|^2 |{ s}_{A\,\parallel}^{p}|^2)  \,\,\, \,\,\,\,\,\,\,\,\, \,\,\,\,\,\,\, \,\,\,\,\,\,\,\,\, \,\,\,\,  \nonumber \\
 +\frac{2}{5}|\zeta_{\parallel} |^2|{ s}_{A\,\parallel}^{p}|^{2} \}
\{\xi_x |s_{D\, x}^{p}|^2 +\xi_y |s_{D\, y}^{p}|^2\} =\frac{1}{10}\{7|{ s}_{A\,\perp}^{p}|^2 \,\,\nonumber \\
 +\frac{19}{3}|{ s}_{A\,\parallel}^{p}|^2 
 \}\{\xi_x |s_{D\, x}^{p}|^2 +\xi_y |s_{D\, y}^{p}|^2\}=<{\cal K} ^{(1)\,*}_{\cal H}> . \,\,\,\,\,\, \,\,\,\,\,\,\,\,\,\,\,\,\,\,\,\,\, \label{mkh2}
\ee
\be
 <{\cal K} ^{(3)}_{\cal H}> = \frac{1}{10} (7\zeta_{\perp}^{\,*}|{ s}_{A\,\perp}^{p}|^2 +\frac{19}{3}\zeta_{\parallel}^{\,*}|{ s}_{A\,\parallel}^{p}|^2). \,\,\label{mkh3} 
\ee
\be
<{\cal K} ^{(4)}_{\cal H}> = \frac{1}{10} (7\zeta_{\perp}|{ s}_{A\,\perp}^{p}|^2 +\frac{19}{3}\zeta_{\parallel}|{ s}_{A\,\parallel}^{p}|^2)= <{\cal K} ^{(3)\,*}_{\cal H}>. \,\,\, \,
 \label{mkh4}
\ee
\be
 <{\cal K} ^{(1)}>=  \frac{1}{10} (7|{ s}_{A\,\perp}^{p}|^2 +\frac{19}{3}|{ s}_{A\,\parallel}^{p}|^2). \label{mk1}
\ee
\be
<{\cal K} ^{(2)}>=\frac{7}{10}(|\zeta_{\perp}|^{2}|{ s}_{A\,\perp}^{p}|^2 +\frac{1}{3}|\zeta_{\parallel}|^{2}|{ s}_{A\,\parallel}^{p}|^2)\,\,\,\,\,\, \,\,\,\,\,\,\,\,\,\nonumber \\
+\frac{2}{5}|\zeta_{\parallel}|^{2} |{ s}_{A\,\parallel}^{p}|^{2}=
\frac{1}{10}(7|{ s}_{A\,\perp}^{p}|^2 +\frac{19}{3}|{ s}_{A\,\parallel}^{p}|^2)
=<{\cal K} ^{(1)}>. \, \,\,\,\, \label{mk2}
\ee
\be
<{\cal K} ^{(3)}>=  \frac{1}{10}\{7 \zeta_{\perp}^{\,*}|{ s}_{A \,\perp}^{p}|^2 +\frac{19}{3}\zeta_{\parallel}^{\,*}|{ s}_{A\,\parallel}^{p}|^2
\} \nonumber \\
\times\{\xi_x |s_{D\, x}^{p}|^2 +\xi_y |s_{D\, y}^{p}|^2\}. \, \,\,\,\,\,\,\  \label{mk3}
\ee
 The  procedure to calculate thes orientational averages  is illustrated in Appendix 4.

Notice that these  averages have been written in terms of the components of ${\bf  s}_{D}^{p}$
and  ${\bf  s}_{A}^{p}$, as well as of ${\bm \xi}$ and ${\bm  \zeta}$. The latter characterizing the respective orientation of ${\bf  s}_{D}^{m}$ and  ${\bf  s}_{A}^{m}$ relative to that of ${\bf  s}_{D}^{p}$
and  ${\bf  s}_{A}^{p}$, as shown before. 

We can further average the quantities $|s_{D\,x,\, y}^{p}|^2$ and $|{ s}_{A\,\perp,\,\parallel}^{p}|^2$ in the respective plane of rotation of   ${\bf  s}_{D}^{p}$ and  ${\bf  s}_{A}^{p}$. According to Eqs. (\ref{carteym}) and (\ref{Acarteym}), on expressing the two components of these unit vectors in polar coordinates as a cosine and a sine of the rotation angle, these  averages are:  $<|s_{D\,x,\, y}^{p}|^2>= <|{ s}_{A\,\perp,\,\parallel}^{p}|^2>=1/2$. Hence, with this additional averaging, the orientational mean  ${\cal K}$-factors finally become
\be
<{\cal K} ^{(1)}_{\cal H}>=\frac{1}{3} (\xi_{x}^{*}+ \xi_{y}^{*}),  \label{KK1} \\
<{\cal K} ^{(2)}_{\cal H}>=\frac{1}{3} (\xi_{x}+ \xi_{y})=<{\cal K} ^{(1)\,*}_{\cal H}>.\,\,\,\, \label{mmkh1}
\\
 <{\cal K} ^{(3)}_{\cal H}> = \frac{1}{20} (7\zeta_{\perp}^{\,*} +\frac{19}{3}\zeta_{\parallel}^{\,*}). \,\,\, \,\,\,\,\,\,\,\,\, \,\,\,\, \,\,\,\,\,\,  \label{mmkh2}\\
<{\cal K} ^{(4)}_{\cal H}> = \frac{1}{20} (7\zeta_{\perp} +\frac{19}{3}\zeta_{\parallel})= <{\cal K} ^{(3)\,*}_{\cal H}>. \, \,\,\,\,\,\, \,\,\,\,\,\,\
 \label{mmkh3}
\\
 <{\cal K} ^{(1)}>= \frac{2}{3}=
<{\cal K} ^{(2)}>,  \, \,\,\,\,\,\, \,\,\,\,\,\,\nonumber\\
<{\cal K} ^{(3)}>=  \frac{1}{40} (7 \zeta_{\perp}^{\,*}+\frac{19}{3}\zeta_{\parallel}^{\,*}
) (\xi_x+\xi_y ). \, \,\,\,\,\,\,\  \label{mmk1}
\ee
Therefore these averages are expressed in terms of the relative orientations of the magnetic moments  with respect to the electric ones, both in D and A.  For example, in the particular case in which $\xi_x=\xi_y=1$, $\zeta_{\perp} =\zeta_{\parallel}=1$, (i.e. both   D and A being linearly polarized {\it right} dipoles), each of these quantities  would reduce to  $2/3$. On the other hand, when ${\bf m}_D={\bf m}_A=0$,  the only  polarizability of the acceptor different from zero is $\alpha_{e}^{A}$;  hece  Eq. (\ref{tohelfret_2}) yields no transfer of helicity: ${\cal W}_{\hel}^{DA}=0$, while  Eq. (\ref{topefret_2})  reproduces the energy transfer $W^{DA}$ of standard FRET with $ <{\cal K} ^{(1)}>=2/3=\kappa^2$.


\subsection{Case in which donor and  acceptor have well defined helicity}
We  now address the RHELT and RET,  Eqs. (\ref{tohelfret_2}) and (\ref{topefret_2}), with deterministically oriented  dipole moments of D and A. Let us assume {\it well defined helicity  (WDH)} \cite{nieto3,corbato1,gutsche4}  in the illumination of  D, for example {\it let a CPL plane wave be incident on the donor}.  Then in (\ref{p1}) and (\ref{m1}) either $p_D^{+}=0$ or $p_D^{-}=0$ depending on whether this illumination is RCP or LCP, respectively.  Therefore the induced electric and magnetic dipole moments of D rotate in the $OXY$-plane, (cf. Fig. 2).  The near field along the $OZ$ axis emitted by D is circularly polarized \cite{nieto3}: ${\bf E}_{nf}(\hat{\bm z})=-(k^2/\epsilon r^3) p^{\pm}\hat{\bm \epsilon}^{\pm}$, \,\,  ${\bf B}_{nf}(\hat{\bm z})=\mp i n {\bf E}_{nf}(\hat{\bm z})$.

 As an instance. we  consider in the $OXYZ$ framework: ${\bf s}_R=(0,0,1)$, (see Fig. 2),  the  field incident on D being CPL. The dipoles ${\bf p}_{D}$ and ${\bf m}_{D}$ rotate  in the plane OXY, while ${\bf p}_{A}$ and ${\bf m}_{A}$ do it in a plane parallel to OXY at distance $|{\bf r}_R|=z=r$, ($|{\bf s}_R|=1$). I.e.  From (\ref{cirsym})  we have:  ${\bf s}_D^{p}= \exp(i\phi_{e}^{\pm}){\bm \epsilon}^{\pm} =\frac{\exp(i\phi_{e}^{\pm})}{\sqrt{2}}(1, \pm i,0)$,  ${\bf s}_D^{m} =\exp [i(\phi_{m}^{\pm}-\phi_{e}^{\pm})] {\bf s}_D^{p}$.  

On the other hand, from (\ref{Acarteym}), (\ref{pAA11}), (\ref{pA11}) and (\ref{mA11}) we have for the acceptor
\be
{\bf p}_A^{\pm}=p_{A}^{\pm}\hat{\bm \eta}^{\pm},\,{\bf m}_A^{\pm}=m_{A}^{\pm}\hat{\bm \eta}^{\pm}; \,
{\bf s}_{A}^{p}=e^{i\psi_{e}^{\pm}}\hat{\bm \eta}^{\pm},\,
{\bf s}_{A}^{m}=e^{i\psi_{m}^{\pm}}\hat{\bm \eta}^{\pm}. \,\,\,\label{SpAyM1}
\ee
$\psi_{e}^{\pm}$ and $\psi_{m}^{\pm}$ representing the arguments of the complex scalars $p_{A}^{\pm}$ and $m_{A}^{\pm}$, respectively. Hence
\be
{\bf s}_{A}^{m}=e^{i(\psi_{m}^{\pm}-\psi_{e}^{\pm})}{\bf s}_{A}^{p}. \label{spsa1}
\ee


Therefore the following values hold:
\be
{ s}_{A\,\perp}^{p,m}=\frac{e^{i\psi_{e,m}^{\pm}}}{\sqrt{2}},\,\, \,\,{ s}_{A\,\parallel}^{p,m}=\pm i \frac{e^{i\psi_{e,m}^{\pm}}}{\sqrt{2}}, \,\,\,\, \nonumber \\
|{ s}_{A\,\perp}^{p}|^2 = |{ s}_{A\,\parallel}^{p}|^2=|{ s}_{A\,\perp}^{m}|^2 = |{ s}_{A\,\parallel}^{m}|^2=\frac{1}{2}. \label{SpAyM2}
\ee

 Let  ${\bf s}_A^{p}$  be rotated  with respect to  ${\bf s}_D^{p}$ in the  $XY$-plane by an angle $\Theta$; namely, ${\bf s}_A^{p}=\exp(i\alpha) {\bf R}{\bf s}_D^{p}$. $\alpha$ being a phase shift and ${\bf R}$ denoting the rotation matrix:
\be
{ \bf R}=
  \left[ {\begin{array}{cc}
  \cos \Theta & - \sin \Theta \\
   \sin \Theta & \cos \Theta \\
 \end{array} } \right].  \label{rot}
\ee
According to (\ref{circeym}) and (\ref{SpAyM1}), both   ${\bf s}_D^{p}$ and ${\bf s}_A^{p}$ are helicity vectors aside from a phase factor, hence   ${\bf R}{\bf s}_D^{p}=\exp(\mp i \Theta){\bf s}_D^{p}$,  and   $|{\bf s}_A^{p}\cdot {\bf s}_D^{p}|^2 =0$.

Then for these magnetoelectric dipoles one has from  (\ref{K1H}) - (\ref{K3}):
\be
{\cal K}^{(1)}={\cal K}^{(2)}=|{\bf s}_D^{p}\cdot{\bf s}_A^{p} |^2, \,\,\,\,\,\,\,\,\nonumber \\
{\cal K}^{(3)}=|{\bf s}_D^{p}\cdot{\bf s}_A^{p} |^2
\exp\{i[(\phi_{m}^{\pm}-\phi_{e}^{\pm})-(\psi_{m}^{\pm}-\psi_{e}^{\pm})]\}, \,\,\,\, \nonumber \\
{\cal K}_{\cal H}^{(1)}=|{\bf s}_A^{p}\cdot {\bf s}_D^{p\,*}|^2\exp[-i(\phi_{m}^{\pm}-\phi_{e}^{\pm})], \,\,\,\,\,\,\,\,\,\,\,\,\,\,\,\,\nonumber \\
 {\cal K}_{\cal H}^{(2)}=|{\bf s}_D^{p}\cdot{\bf s}_A^{p} |^2\exp[i(\phi_{m}^{\pm}-\phi_{e}^{\pm})],\,\,\,\,\,\,\,\,\, \nonumber\\ 
{\cal K}_{\cal H}^{(3)}=|{\bf s}_D^{p}\cdot{\bf s}_A^{p} |^2\exp[-i(\psi_{m}^{\pm}-\psi_{e}^{\pm})],\,\,\,\,\,\,\,\,\nonumber \\
{\cal K}_{\cal H}^{(4)}=|{\bf s}_D^{p}\cdot{\bf s}_A^{p} |^2\exp[i(\psi_{m}^{\pm}-\psi_{e}^{\pm})]. \,\,\,\,\label{KnegatW}
\ee
But  $|{\bf s}_D^{p}\cdot{\bf s}_A^{p} |^2 =1$, hence (\ref{tohelfret_2}) and (\ref{topefret_2}) become
\be
{\cal W}_{\hel}^{DA}=\frac{2\pi c}{\epsilon r^6}  [ p_{D}m_{D}\mbox{Re}  \{
-\frac{\alpha_{e}^{A}}{\epsilon}e^{-i(\phi_{m}^{\pm}-\phi_{e}^{\pm})}
+ \mu\alpha_{m}^{A}e^{i(\phi_{m}^{\pm}-\phi_{e}^{\pm})}\} \nonumber \\
+ \mu m_{D}^{2}\mbox{Re} \{\alpha_{me}^{A}\, e^{-i(\psi_{m}^{\pm}-\psi_{e}^{\pm})}\}+ \frac{p_{D}^{2 }}{\epsilon}\mbox{Re} \{\alpha_{me}^{A}\, e^{i(\psi_{m}^{\pm}-\psi_{e}^{\pm})}\}].  \,\,\,\,\,\, \label{tohelfret_22}
\ee
\be
 {\cal W}^{DA}=\frac{\omega}{2r^{6}}  [ \frac{\mbox{Im} \{\alpha_{e}^{A}\}}{\epsilon^{2}} \, p_{D}^{2}\, +\mu^{2} \mbox{Im} \{\alpha_{m}^{A}\}\, m_{D}^{2}\,\,\,\,\,\,\, \nonumber \\
-2\frac{\mu}{\epsilon} p_{D} m_{D} \mbox{Re}\{ \alpha_{me}^{A}\}\sin[(\phi_{m}^{\pm}-\phi_{e}^{\pm})-(\psi_{m}^{\pm}-\psi_{e}^{\pm})]. \,\,\,\,\,\, \label{Wda4}
\label{topefret_22}
\ee
And again ${\cal W}^{DA}$ may become negative for large enough $\Re\{ \alpha_{me}^{A}\}$,
 and oscillate with the difference  of  phase-shifts  of ${\bf s}_D^{m}$  and ${\bf s}_D^{p}$, and  of ${\bf s}_A^{m}$ and  ${\bf s}_A^{p}$. The radii
 $R _{\hel}$ and $R _{\cal E}$ are now given by:
\be
   R _{\hel}^{6}= \frac{3}{4 k^{3}}\, |p_{D}m_{D}\mbox{Re}  \{
-\frac{\alpha_{e}^{A}}{\epsilon}e^{-i(\phi_{m}^{\pm}-\phi_{e}^{\pm})}\,\,\,\,\,\,\,\,\,\,\,\,\,\,\,\,\,\,\,\,\,\,\,\,\,\,\,\,\,\,\,\,\,  \nonumber \\
+ \mu\alpha_{m}^{A}e^{i(\phi_{m}^{\pm}-\phi_{e}^{\pm})}\}
+ \mu m_{D}^{2}\mbox{Re} \{\alpha_{me}^{A}\, e^{-i(\psi_{m}^{\pm}-\psi_{e}^{\pm})} \} \,\,\,\,\,\,\,\,\,\,\,\,\,\,\,\,\,\,\,\,  \nonumber \\
+ \frac{p_{D}^{2 }}{\epsilon}\mbox{Re}\{\alpha_{me}^{A} e^{i(\psi_{m}^{\pm}-\psi_{e}^{\pm})}\}|
 / p_{D} m_{D} |\sin(\phi_{e}^{\pm}-\phi_{m}^{\pm})|. \,\,\,\,\,\,\,\,\,\, \label{FR_HH}
\ee
\be
R _{\cal E}^{6}=  \frac{3}{2 k^{3}} \,\, |\frac{\mbox{Im} \{\alpha_{e}^{A}\}}{\epsilon^{2}} \, p_{D}^{2}\, +\mu^{2} \mbox{Im} \{\alpha_{m}^{A}\}\, m_{D}^{2}\,\,\,\,\,\,\, \nonumber \\
-2\frac{\mu}{\epsilon}p_{D} m_{D}\,\mbox{Re}\{ \alpha_{me}^{A}\}\sin[(\phi_{m}^{\pm}-\phi_{e}^{\pm})-(\psi_{m}^{\pm}-\psi_{e}^{\pm})] |\nonumber \\
/ (\frac{p_{D}^{2}}{\epsilon}+\mu m_{D}^{2}). \,\,\,\,\,\label{FR_EE}\,\,\,\,\,\,\,\,\,\,\,\,\,\,\,\,\,\,\,\,\,\,\,\,\,
\ee
Notice that while in conventional  FRET  if D and A have dipoles linearly polarized  in  planes normal to ${\bf s}_R$, there is no energy transfer if ${\bf s}_D^{p}$ and ${\bf s}_A^{p}$ are perpendicular to each other,  i.e. $ {\cal K}^{(1)}=0$,  this is not the case  in the configuration  of circularly rotating dipoles in D and A,  since, as seen above, in this latter case the unit vectors ${\bf s}_A^{p}$ and ${\bf s}_D^{p} $ are complex and $|{\bf s}_A^{p}\cdot {\bf s}_D^{p}|^2 =1$.

\subsection{Consequences when D and/or A is dual}
\subsubsection{1. a. The dipoles  induced in both   and A  are  dual} 
In this  case ${\epsilon}^{-1}\alpha_{e}^{A}= \mu\alpha_{m}^{A}$; ${\epsilon}^{-1}\alpha_{e}^{D}= \mu\alpha_{m}^{D}$. From (\ref{p1}), (\ref{m1}),
(\ref{pAA11}),  (\ref{pA11}) and  (\ref{mA11}) one has $p_D^{\pm}=\pm in m_D^{\pm}$,  and   $p_A^{\pm}=\pm in m_A^{\pm}$,  and  both D and A    emit fields of  well defined helicity \cite{nieto3,corbato1, gutsche4}, (e.g. CPL fields for incident CPL illumination).  Then ${\bf s}_D^{m} =\mp i{\bf s}_D^{p}$,  and ${\bf s}_A^{m} =\mp i {\bf s}_A^{p}$. I.e. $\phi_{m}^{\pm}-\phi_{e}^{\pm}$ and $\psi_{m}^{\pm}-\psi_{e}^{\pm}$ are either $ 3\pi/2$ for the upper (+) sign and $\pi/2$ for the lower (-) sign.  Hence   one has that $(\phi_{m}^{\pm}-\phi_{e}^{\pm})-(\psi_{m}^{\pm}-\psi_{e}^{\pm})$ is either   $0$ or $\pi$. Also  $p_D =n m_D$.

 Then $ {\cal W}^{DA}$,  given by (\ref{Wda4}) becomes
\be
{\cal W}^{DA}=\frac{\omega}{\epsilon^2 r^{6} } p_{D}^{2}\mbox{Im} \{\alpha_{e}^{A}\}. \label{Wbothdual}
\ee
 Also,  the sum of the third and fourth terms of  (\ref{tohelfret_22}) also vanishes since $\cos(\psi_{m}^{\pm}-\psi_{e}^{\pm})=0$, and therefore
\be
{\cal W}_{\hel}^{DA}=\pm\frac{4\pi c}{\epsilon^2 r^6}\,  p_{D}^{2}\, \frac{\mbox{Im}  \{\alpha_{e}^{A}\}}{n}. \,\,\,\,\,\,\,\,\,\,\,\,  \label{WHbothdual}
\ee
Therefore the RET and   RHELT  rates are equivalent apart from a constant, and are functionally similar  to the transfer of energy of a standard FRET with circularly polarized D and A   when one considers in (\ref{WHbothdual}) the effective polarizability given by Eq. (\ref{sigmaso1a}). This result is consistent with the fact that in this case the light emitted by both  D and A is circularly polarized and, therefore, there is an equivalence, apart from a constant factor, between the extinction of helicity and of energy \cite{cameron1,nieto2,nieto3}.

This is manifested by the radii:
\be
   R _{\hel}^{6}= \frac{3}{2\epsilon  k^{3}}\mbox{Im}  \{\alpha_{e}^{A}\} , \,\,\,\,\,\,\,\,\,\,\,\,\,\,\,\,\,\,\,\,\,\, R _{\cal E}^{6}=  \frac{3}{\epsilon k^{3}}\mbox{Im}  \{\alpha_{e}^{A}\}        .
\label{FR_HHh}\,\,\,\,\,\,\,\,\,\,\,\,\,\,\,\,\,\,\,\,\,\,\,\,
\ee
Which makes $R _{\hel}$ and $R _{\cal E}$ proportional to the radius of the acceptor, and  $R _{\cal E}=1.12 R _{\hel}$.

\subsubsection{1. b. Only A is dual}
In this  case  (\ref{tohelfret_22}) and (\ref{Wda4}) lead to
\be
{\cal W}_{\hel}^{DA}=
-\frac{4\pi c}{\epsilon^2 r^6}  p_D m_D \mbox{Im}  \{
\alpha_{e}^{A}\}\sin(\phi_{m}^{\pm}-\phi_{e}^{\pm}).   \label{tohelfret_3}
\ee
\be
 {\cal W}^{DA}=\frac{ck}{2\epsilon n r^{6}}  \{ (\frac{p_D^2}{\epsilon}+\mu m_D^2)\mbox{Im}\{\alpha_{e}^{A}\}\,\,\,\,\,\,\, \nonumber \\
\mp 2 \mu p_{D} m_D \mbox{Re}\{ \alpha_{me}^{A}\}\cos(\phi_{m}^{\pm}-\phi_{e}^{\pm})\}. \,\,\,\,\,\,\,\,\,\,\,\,\,\,
\label{topefret_3}
\ee
Since now  ${\cal W} _{\hel}^{0}=8\pi c k^3 p_{D}^2 \sin (\phi_{m}^{\pm}-\phi_{e}^{\pm}) /3\epsilon n$ and ${\cal W}^{0}=2ck^4 p_{D}^{2}$, Eqs.(\ref{FR_HH}) and (\ref{FR_EE}) yield for the interaction radii:
\be
R _{\hel}^6=\frac{3} {2k^3 \epsilon}  \mbox{Im}  \{\alpha_{e}^{A}\},   \label{RH3}
\ee
\be
 R _{\cal E}^6= \frac{3}{2\epsilon k^3}\{\mbox{Im}\{ \alpha_{e}^{A}\}\,\,\,\,\,\,\, \nonumber \\
\mp\frac{2 \mu p_D m_D}{\frac{p_D^2}{\epsilon}+\mu m_D^2} |\mbox{Re}\{ \alpha_{me}^{A}\}\cos(\phi_{m}^{\pm}-\phi_{e}^{\pm})|\}.  \,\,\,\,\label{RE3}
\ee

We see comparing (\ref{RH3}) and  (\ref{FR_HHh}) that 
 when A is dual, $R _{\hel}$ is not affected by whether D is dual or not. On the  other hand, (\ref{RE3}) is  consistent with (\ref{FR_HHh}) since if D is also dual, $\cos(\phi_{m}^{\pm}-\phi_{e}^{\pm})=0$.

\subsubsection{1.c. Only D  is dual}
Assuming for simplicity that  $\epsilon=\mu=n=1$, we have from (\ref{tohelfret_2}) and (\ref{Wda4}):
\be
{\cal W}_{\hel}^{DA}=\frac{2\pi c}{ r^6}p_{D}^{2} [ \mp(\mbox{Im}  \{
\alpha_{e}^{A}\}+\mbox{Im}\{\alpha_{m}^{A}\})\nonumber \\
+2\mbox{Re}\{\alpha_{me}^{A}\} \cos(\psi_{m}^{\pm}-\psi_{e}^{\pm})],   \label{tohelfret_222}
\ee
\be
 {\cal W}^{DA}=\frac{ck}{2r^{6}}p_D^2\{ \mbox{Im}  \{\alpha_{e}^{A }\}
+ \mbox{Im}  \{\alpha_{m}^{A }\} \nonumber \\ 
\pm \mbox{Re} \{ \alpha_{me}^{A}\}\cos(\psi_{m}^{\pm}-\psi_{e}^{\pm})\}.\,\,\,\, \label{Wda44}
\ee
\begin{figure*}
\begin{centering}
\includegraphics[width=18cm]{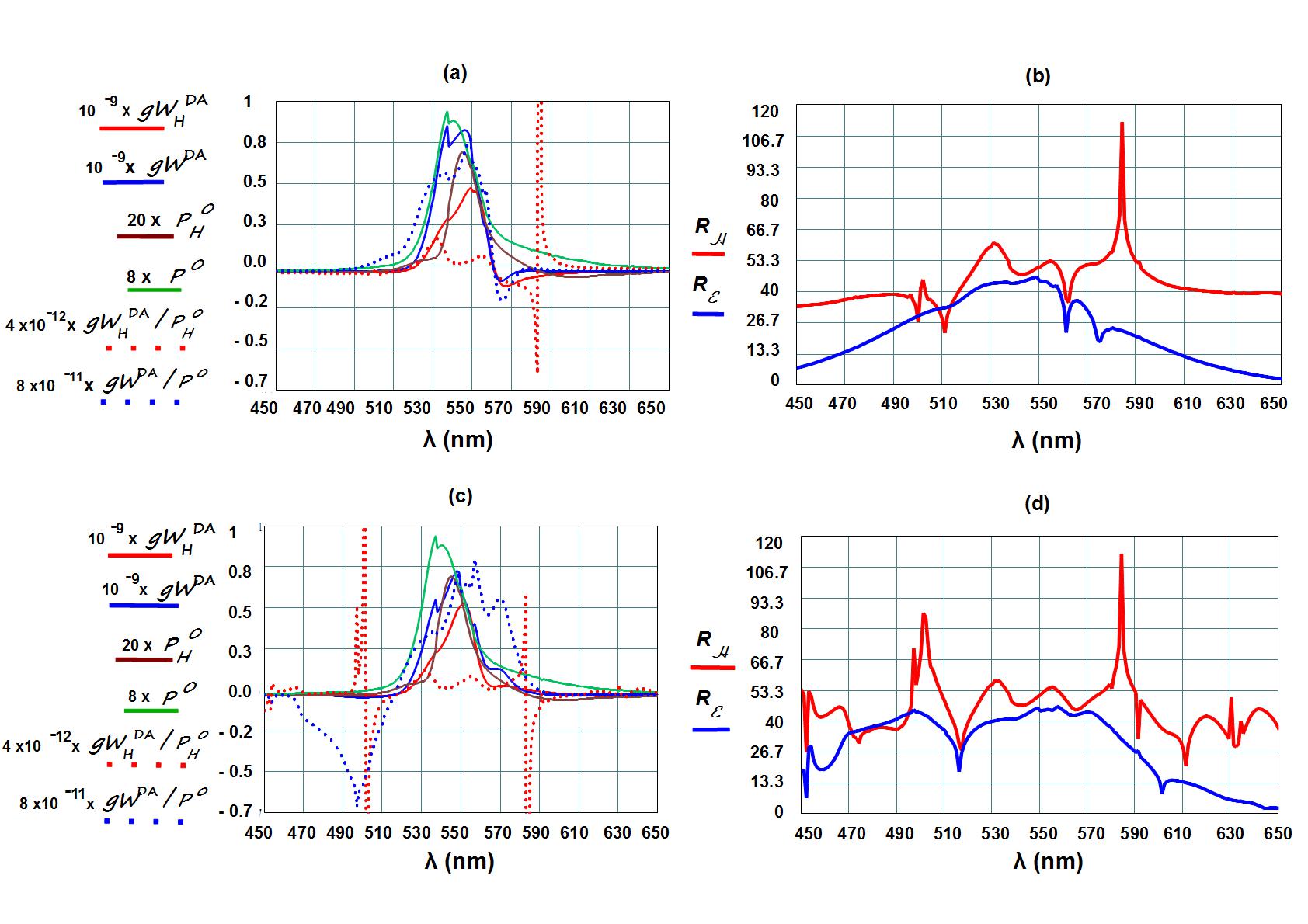}
\par\end{centering}
\caption{(Color online). Both D and A are chiral and magnetoelectric.  ${\bf p}_D$ and ${\bf m}_D$, are excited by elliptically polarized light.  ${\bf p}_A$ and ${\bf m}_A$ are also elliptically polarized, (see text). $g=\frac{3r^6}{4a^6 k^3(\lambda)}$.
 (a)  $g{\cal W}_{\hel}^{DA}(\lambda)$, ($k(\lambda)=n(\lambda)\frac{2\pi}{\lambda}$, $n(\lambda)=1$), and  $g{\cal W}^{DA}(\lambda)$, with $a=15 nm$ being the radius of D and A, and ${\cal W}_{\hel}^{DA}$ and ${\cal W}^{DA}$ given by (\ref{tohelfret_2}) and (\ref{topefret_2}), respectively. Also shown are ${\cal P}_{\hel}^{0}(\lambda)=\mbox{Im} [ p_{D}^{+} m_{D}^{+\,* } +  p_{D}^{-} m_{D}^{-\,* }]/a^6$  and ${\cal P}^{0}(\lambda)=(\frac{p_{D}^{2}}{\epsilon}+\mu m_{D}^{2})/a^6$, as well as $g{\cal W}_{\hel}^{DA}/{\cal  P}_{\hel}^{0}$ and  $g{\cal W}^{DA}/{\cal  P}^{0}$. These quantities are plotted in arbitrary units. (b)  $R _{\hel}(\lambda)$ and $R _{\cal E}(\lambda)$ in $nm$. (c) and (d) Same as (a) and (b), respectively,   with ${\bf p}_A$ and ${\bf m}_A$  randomly oriented with respect to  ${\bf p}_D$ and ${\bf m}_D$, and the ${\cal K}$-factors being  orientationally averaged. }
\end{figure*}
And
\be
R _{\hel}^6=\frac{3} {4k^3} |\pm(\mbox{Im}  \{
\alpha_{e}^{A}\}+\mbox{Im} \{\alpha_{m}^{A}\})\nonumber \\
+2\mbox{Re}\{\alpha_{me}^{A}\}\cos(\psi_{m}^{\pm}-\psi_{e}^{\pm})|, \, \,\,\,\label{RH4}
\ee
\be
 R _{\cal E}^6= \frac{3}{2 k^{3}}| \mbox{Im}  \{\alpha_{e}^{A }\} +\mbox{Im}  \{\alpha_{m}^{A }\}  \pm \mbox{Re}\{ \alpha_{me}^{A}\} \cos(\psi_{m}^{\pm}-\psi_{e}^{\pm})|, \,\,\,\label{RE4}
\ee
which show the cosinusoidal oscillation of these RHELT and RET quantities with amplitude $\mbox{Re}\{ \alpha_{me}^{A}\}$  about the level:  $\pm(\mbox{Im}  \{\alpha_{e}^{A }\} +\mbox{Im}  \{\alpha_{m}^{A }\})$.

\section{Examples:  RHELT and RET when both donor and acceptor are chiral and magnetoelectric} 
\subsection{Example A: Donor is  illuminated by an elliptically polarizated wavefield. The electric and magnetic dipole moments of the acceptor are elliptically polarized}

We  use Eqs. (\ref{tohelfret_2})-(\ref{FR_E}) on addressing chiral magnetoelectric  D and A. Let the magnetic and cross electric-magnetic polarizabilities dominate in A. We assume the radius of both particles to be: $a\simeq 15 nm$

Eqs. (A3-1) and (A3-2) of Appendix 3,  with the parameters of Appendix 5,  fit a model  of donor emission spectra and acceptor extinction cross-sections. They are plotted in Fig.A5-1 (a) of Appendix 5.  With these emission spectra,  from Eqs.(A2-6)-(A2-13) of Appendix 2 one gets the  donor and acceptor  polarizabilities, shown in Figs.A5-1 (b) and  (c) of Appendix 5.

We consider an elliptically polarized plane wave, propagating along $OZ$, incident on D, [cf. Fig. 2 and Eqs. (\ref{bheli1})], with $e_i^+=7$ and $e_i^-=3$ [in arbitrary units  (a.u.)]. The surrounding medium has $n(\lambda)=1$. Fig. A5-1(d) of Appendix 5 shows $p_D^2(\lambda)$, $m_D^2(\lambda)$ and $p_D(\lambda)m_D(\lambda)$,  obtained from Eqs. (\ref{p1}) and  (\ref{m1}) for this polarization. These latter equations show the transfer of incident angular momentum to the donor dipoles, and yield  ${\bf p}_D$ and ${\bf m}_D$, as well as their  respective unit vectors  ${\bf s}_D^p$ and ${\bf s}_D^m$. 

\begin{figure*}[t!]
\begin{centering}
\includegraphics[width=18cm]{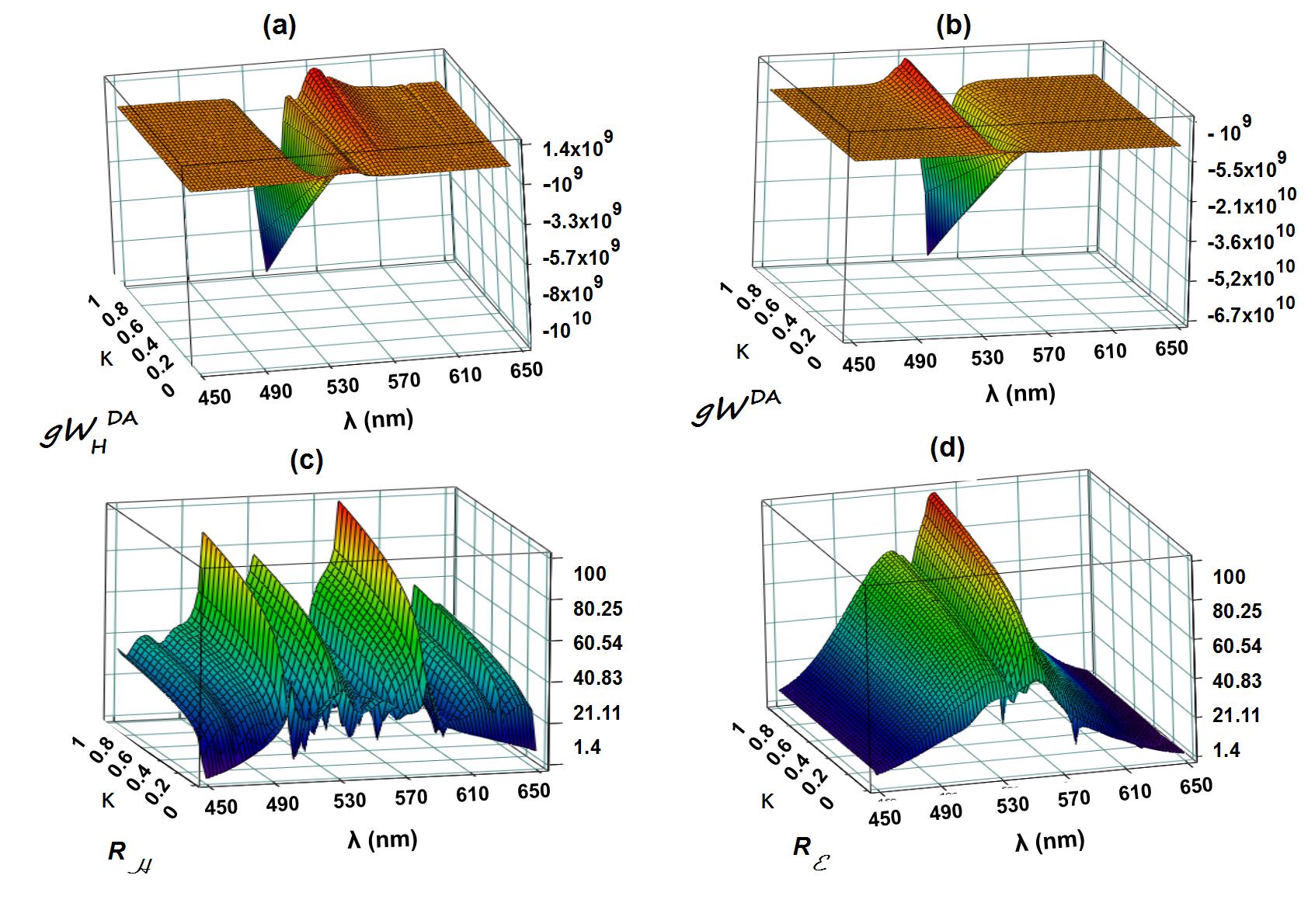}
\par\end{centering}
\caption{(Color online).  Effect of the chirality of a  magnetoelectric acceptor with elliptically polarized dipoles. Elliptic incident polarization. Same configuration as for Figs. 3 (a) and 3 (b).  (a)   $g{\cal W}_{\hel}^{DA}(\lambda, \kappa)$ in arbitrary units. (b)  $g{W}^{DA}(\lambda, \kappa)$ in arbitrary units. ($g=\frac{3r^6}{4a^6 k^3(\lambda)})$. (c)  $R _{\hel}(\lambda, \kappa)$  in $nm$. (d) $R _{\cal E}(\lambda, \kappa)$  in $nm$.  For $\kappa \in (0,-1]$, (not shown), the negative minima of   $g{\cal W}_{\hel}^{DA}/{\cal  P}_{\hel}^{0}$ and  $g{\cal W}^{DA}/{\cal  P}^{0}$ become positive maxima.  }
\end{figure*}
\begin{figure*}[t!]
\begin{centering}
\includegraphics[width=18cm]{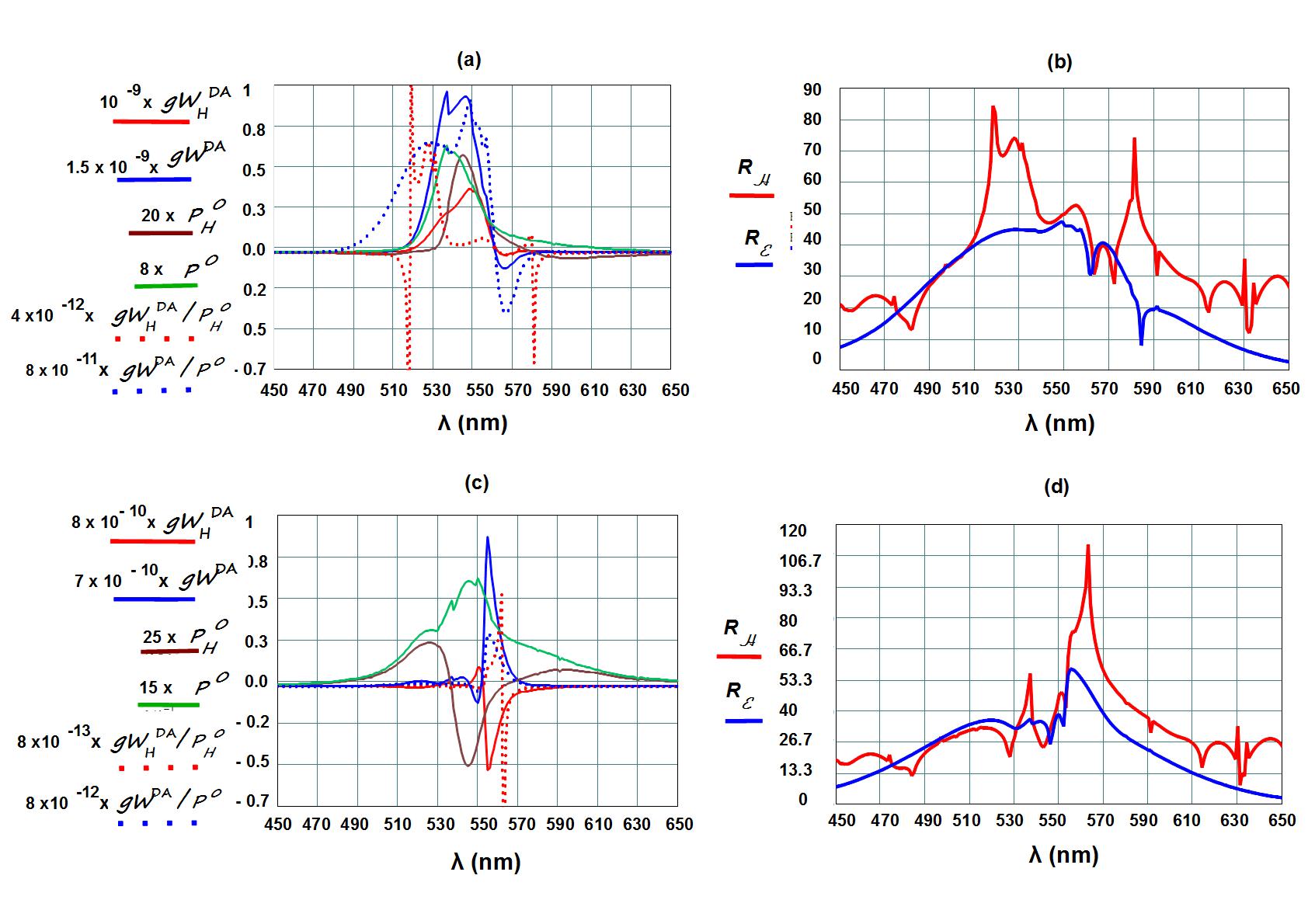}
\par\end{centering}
\caption{(Color online). The donor and the acceptor are chiral and magnetoelectric . Both ${\bf p}_D$ and ${\bf m}_D$, as well as ${\bf p}_A$ and ${\bf m}_A$, have well-defined helicity, (see text). (a) D and A have LCP  dipole moments.  $g{\cal W}_{\hel}^{DA}(\lambda)$,  and  $g{\cal W}^{DA}(\lambda)$, ($k(\lambda)=n(\lambda)\frac{2\pi}{\lambda}$, $n(\lambda)=1$, $g=\frac{3r^6}{4a^6 k^3(\lambda)}$), with $a=15 nm$ being the radius of D and A, and ${\cal W}_{\hel}^{DA}$ and ${\cal W}^{DA}$ given by (\ref{tohelfret_2}) and (\ref{topefret_2}), respectively. Also shown are  ${\cal P}_{\hel}^{0}(\lambda)=\mbox{Im} [ p_{D}^{+} m_{D}^{+\,* } +  p_{D}^{-} m_{D}^{-\,* }]/a^6$  and ${\cal P}^{0}(\lambda)=(\frac{p_{D}^{2}}{\epsilon}+\mu m_{D}^{2})/a^6$,   as well as $g{\cal W}_{\hel}^{DA}/{\cal  P}_{\hel}^{0}$ and  $g{\cal W}^{DA}/{\cal  P}^{0}$.  (b)  D and A with LCP moments. All these quantities are plotted in arbitrary units. $R _{\hel}(\lambda)$ and $R _{\cal E}(\lambda)$ in $nm$. (c) and (d) Same as (a) and (b), respectively, when both D and A have RCP dipole moments.}
\end{figure*}

The acceptor dipole moments ${\bf p}_A$ and ${\bf m}_A$, and their respective unit vectors  ${\bf s}_A^p$ and ${\bf s}_A^m$,     according to Eqs. (\ref{Acarteym}) - (\ref{Saperp1}) are set as $p_A^{\pm}=(\alpha_{e}^{A}\pm ni \alpha_{me}^{A})e^{\pm}$ and  $m_A^{\pm}=(\alpha_{me}^{A}\mp ni \alpha_{m}^{A})e^{\pm}$, choosing   $e^{+}=5 a.u.$ and $e^{-}=-2 a.u.$.

 The overlapping integrals (\ref{sigmaso1a})-(\ref{sigmaso3a})  for the effective polarizabilities: $\mbox{Im}\{\alpha_{e}^{A\, eff}\}$, $\mbox{Im}\{\alpha_{m}^{A\, eff}\}$ and $\mbox{Re}\{\alpha_{me}^{A\, eff}\}$, used in Eqs. (\ref{FR_H}) and (\ref{FR_E}) instead of their spectra:  $\mbox{I}m\{\alpha_{e}^{A}(\lambda)\}$, $\mbox{Im}\{\alpha_{m}^{A}(\lambda)\}$ and $\mbox{R}e\{\alpha_{me}^{A}(\lambda)\}$,  would yield a number for the RHELT and RET interaction radii. However, as mentioned before, we shall rather employ the polarizability spectra,  whose variation with  $\lambda$  gives us more information on the range of values of $R _{\hel}(\lambda)$ and $R _{\cal E}(\lambda)$ on comparison with the spectra of D and A. In fact,  the values of the polarizabilitiy spectra  of A are in the same range as their effective values. This is seen  on comparing Fig. A5-1(c) of Appendix 5  with these overlapping integrals (\ref{sigmaso1a})-(\ref{sigmaso3a})  that yield: $\mbox{Im}\{\alpha_{e}^{A\, eff}\}=3.68\times 10^3$ $nm^3$,  $\mbox{Im}\{\alpha_{m}^{A\, eff}\}=7.25\times 10^3$ $nm^3$,  $\mbox{Re}\{\alpha_{me}^{A\, eff}\}=2.13\times 10^4$ $nm^3$.

 Choosing ${\bf s}_R=(0,0,1)$, so that the basis $\{ \hat{\bf e}_{\perp}, \hat{\bf e}_{\parallel}, {\bf s}\}$ becomes $\{\hat{\bf x}, \hat{\bf y}, \hat{\bf z}\}$, (cf. Fig. 2), we show in  Fig. 3(a)  the spectra 
 ${\cal P}_{\hel}^{0}(\lambda)=\mbox{Im} [ p_{D}^{+} m_{D}^{+\,* } +  p_{D}^{-} m_{D}^{-\,* }]/a^6$  and ${\cal P}^0(\lambda)=(\frac{p_{D}^{2}}{\epsilon}+\mu m_{D}^{2})/a^6$, as well as 
 $g{\cal W}_{\hel}^{DA}(\lambda)$ and  $ g{\cal W}^{DA}(\lambda)$; with  ${\cal W}_{\hel}^{DA}$ and ${\cal W}^{DA}$ given by (\ref{tohelfret_2}) and (\ref{topefret_2}), respectively, ($g=\frac{3 r^6}{4a^6 k^3(\lambda)}$,  $k(\lambda)=n(\lambda)\frac{2\pi}{\lambda}$).  All these functions have  peaks at wavelengths near the  the donor and acceptor dipole resonances, as seen on comparing Fig. 4(a) with the lineshapes of Figs.  Figs. A5-1 (a) - (d) of Appendix 5. On the other hand,   $g{\cal W}_{\hel}^{DA}/{\cal  P}_{\hel}^{0}$ and  $g{W}^{DA}/{\cal P}^{0}$,   also  shown in Fig. 3 (a), {\it  are  less   influenced by the maxima of  the  polarizability peaks of both D and A. This indicates that  these latter quotients wash out to some extent  the effect of these D and A resonant values}. However $g{\cal W}_{\hel}^{DA}/{\cal  P}_{\hel}^{0}$ has sharp spikes at wavelengths where  ${\cal P}_{\hel}^{0}(\lambda)=0$ {\it by  changing sign, a feature not always shared by}  $g{W}^{DA}/{\cal P}^{0}$  since ${\cal P}^0(\lambda)$ cannot be negative and rarely has zeros within its support. 

 It is remarkable that, as mentioned before, [cf. Eq.(\ref{topefret_2})], and due to the strong chirality of A, manifested by its large cross electric-magnetic polarizability $\alpha_{me}^A$, {\it the transferred energy acquires negative values  between $560$ and $630$ $nm$}, [See.Fig. 3(a)],  {\it thus manifesting an increase, rather than a reduction, of the  energy emitted by the donor in presence of the acceptor}, (see next Section on Observables), a phenomenon ruled out in standard FRET and in RET between  achiral particles. 

Figures 3 (b) shows the  radii  $R _{\hel}(\lambda)$ and $R _{\cal E}(\lambda)$. The former has larger values than the latter, and both are in ranges above $25$ $nm$ in the wavelength region between $490$ and $580$ $nm$. The {\it large spikes in  $R _{\hel}(\lambda)$ coincide with those of  $g{\cal W}_{\hel}^{DA}/{\cal  P}_{\hel}^{0}(\lambda)$ again where  ${\cal  P}_{\hel}^{0}(\lambda)=0$}.

Finally, Figs. 3 (c) and 3 (d) show the same as Figs. 3 (a)  and 3(b), respectively, when  ${\bf p}_A$ and ${\bf m}_A$ are randomly oriented with respect to  ${\bf p}_D$ and ${\bf m}_D$, and so is ${\bf s}_R$. In this case we have taken the orientational averages of all ${\cal K}$-factors according to  Eqs. (\ref{KK1}) - (\ref{mmk1}). The ${\cal K}$-averaging has a noticeable effect on  both
 $R _{\cal E}(\lambda)$ and  $R _{\hel}(\lambda)$. As seen in the figures, the change is larger for the RHELT interaction radii.

 If we adopted the criterion employed in connection to the case dealt with in Fig. A3-1 (b) of Appendix 3, namely  choosig the wavelength at which the D and A lineshapes cross each other in Fig. A5-1 (a) of Appendix 5, we would get about  $540nm$ for this wavelength and an estimation: $R _{\hel}\simeq 48nm$ and $R _{\cal E}\simeq 44 nm$ for the case studied in Fig. 3 (b), while $R _{\hel}\simeq  R _{\cal E}\simeq 30 nm$ for the system addressed in Fig. 3 (d). As seen in Figs. 3(b) and 3(d), one can  choose configurations in which at certain $\lambda$'s: $R _{\hel}>>R _{\cal E}$; however, (and although not shown here for brevity), we have observed  $R _{\hel}<R _{\cal E}$ in some cases with  either ${\bf s}_R=(1,0,0)$ or  ${\bf s}_R=(1/\sqrt{2})(1,1,0)$. Also, on comparing with the standard FRET interaction radius: $5$ - $7.6$ $nm$ of  the interacting molecules dealt with in Figs. A3-1 of Appendix 3, we see in Figs. 3 (b) and 3 (d) that {\it larger particles convey greater RHELT and RET interaction radii}, as we have observed in all our studied cases.

\subsubsection{Influence of the chirality of A}
We choose the following variation of the real part of $\alpha_{me}^{A}(\omega)$, (cf. Eqs.  (A2-7) of Appendix 2):
\be
\alpha_{me}^{A\,R}(\omega)= \frac{2\pi}{k}\sqrt{\frac{\epsilon}{\mu}}\sigma_{me}^{A}(\omega) = \frac{\lambda}{n}\sqrt{\frac{\epsilon}{\mu}}\sigma_{me}^{A}(\omega) .  \,\,\, \omega=\frac{2\pi c}{\lambda}.\,\,\,\,\, \label{sigmasookk}
\ee
Where $\sigma_{me}^{A}(\omega) =\kappa \sigma_{me\,A}^{CD}(\omega).$

The  $\kappa$-factor  varies between $-1$ and $1$. In this regard, the polarizability  $\alpha_{me}^{A\,R}(\omega)$ employed in Figs. 3 and in Figs. 5 (below) corresponds to $\kappa=3\cdot 50/2(4\pi)^2 = 0.47$.

The surfaces of Fig. 4 show the variation with $\lambda$ and $\kappa$ of 
 $g{\cal W}_{\hel}^{DA}(\lambda,\kappa)$,  and $g{\cal W}^{DA}(\lambda,\kappa)$, as well as  $R _{\hel}(\lambda,\kappa)$ and $R _{\cal E}(\lambda,\kappa)$.  The configuration of  illumination and polarization of D and A is the same as for Figs. 3(a) and 3(b). We observe how the RHELT and RET radii increase with $\kappa$. Also, {\it  the negative values of  both helicity and energy transfers,  $g{\cal W}_{\hel}^{DA}$ and $g{\cal W}^{DA}$, around $550$ $nm$ grow with $\kappa$ due to the resulting  stronger chirality of the acceptor, manifested by larger values of  $\alpha_{me}^{A\,R}$}.

\subsection{Exaample B: Donor is  illuminated by light of well-defined helicity}
Let the same interacting  particles as in Example A be now illuminated by a wavefield of well-defined helicity, incident on D. Specifically we assume a circularly polarized (CPL)  plane wave, setting for both D and A: $e_i^+=6 a.u.$ and $e_i^-=0 a.u.$ for left circular polarization (LCP) and  $e_i^+=0 a.u.$ and $e_i^-=6 a.u.$ for right circular polarization (RCP). 

{\it Inverting the helicity of the incident field from LCP to RCP, and thus that of the induced dipoles in D and A, has a dramatic effect in all quantities shown in Fig. 5(a) and 5(c), both in their sign as on their shape, as well as on the range of wavelengths where the transferred energy  $g{\cal W}^{DA}(\lambda)$ acquires negative values}, thus signing the enhancement of the emission from D in presence of A, which is studied in more detail in the next section. We see, therefore, that {\it the incident helicity is discriminatory as it greatly influences both the RHELT and RET rates}.

{\it The effect of the incident polarization} is, however, not so heavy in $R _{\cal E}(\lambda)$, as a comparison of  Figs. 3(b), 3(d), 5(b) and 5(d) show; however, it {\it has a larger influence on   $R _{\hel}(\lambda)$}. This latter effect may be seen as a  kind of {\it  RHELT circular dichroism} on extinction in A of the rotating light emited by D.

\section{Observables. Donor emission and decay rates}

 Generally, the transfer rates of energy and  helicity, ${\cal W}^{DA}$ and  ${\cal W} _{\hel}^{DA}$, between donor and acceptor   would  not be directly accessible in experiments. However  due to their existence, both the intensity ${\cal W}^{D}$   and helicity  ${\cal W} _{\hel}^{D}$  of the light emitted by D in presence of A change with respect to those values of intensity ${\cal W}^{0}$ and helicity  ${\cal W} _{\hel}^{0}$  in absence of acceptor, and are amenable of being detected in experiments. 

Thus, after detection, one can consider relative values of these quantities, such as  ${\cal W}_{\hel}^{D}/{\cal W}_{\hel}^{0}$ and ${\cal W}^{D}/{\cal W}^{0}$ for polarizable particles whose emission is characterized by the intensity of their scattered field, or quivalently, by their extinction and scattering cross-sections, (see Appendix 2); or  one may address relative {\it decay rates}: ${\gamma _{\hel}^{D}}/{ \gamma _{\hel}^{0}}={\cal W}_{\hel}^{D}/{\cal W}_{\hel}^{0}$  and  ${\gamma^{D}}/{ \gamma^{0}}={\cal W}^{D}/{\cal W}^{0}$, like in Eqs. (\ref{fFRHg}), (see also \cite{novotny}), for quantum dots and molecules.

\subsection{Transfer of energy}

In the case of RET, one  may define the transfer efficiency $E_{\cal E}$, like in classical FRET \cite{clegg,novotny}, by means of the two alternative expressions in terms of either ${\cal W}^{D}$ or ${\cal W}^{DA}$
\be
E_{\cal E}=1-\frac{{\cal W}^{D}}{{\cal W}^{0}}\,\,;  \,\,\,\, E_{\cal E}=\frac{{\cal W}^{DA}}{{\cal W}^{0}+{\cal W}^{DA}}=
\frac{1}{1+ [\frac{r}{R_{\cal E}}]^6}, \label{effiE}
\ee
from which we obtain the important relationship
\be
{\cal W}^{DA}={\cal W}^{0}[\frac{{\cal W}^{0}}{{\cal W}^{D}}-1]. \label{relatDAyD}
\ee
{\it This equation  explicitely illustrates the  statement}, quoted above  in several parts of this paper, [cf. among others: Introduction, second paragraph below Eq. (\ref{wwwfr}), and Conclusions], namely that  {\it a negative value of the transfer rate ${\cal W}^{DA}$} due to its discriminatory terms arising from  the chirality of A, {\it conveys that the emitted energy ${\cal W}^{D}$ of D in presence of A is enhanced, rather than quenched, with respect to that ${\cal W}^{0}$ emitted by D in absence of A}.

Equation (\ref{relatDAyD}) yields {\it  the observable}  ${\cal W}^{D}$ in terms of  ${\cal W}^{DA}$:
\be
{\cal W}^{D}=\frac{{\cal W}^{0 \,2}}{{\cal W}^{DA}+{\cal W}^{0}}\,\,, \,\,\,\, {\cal W}^{D}\geq 0.\, \label{WD}
\ee
Notice that when  ${\cal W}^{DA}<0$ the energy ${\cal W}^{D}$ asymptotically increases as  $|{\cal W}^{DA}|$ approaches  ${\cal W}^{0}$. Now,  this grow will stop at a saturation of the excitation of D determined by its {\it  lifetime}. This also means that if   ${\cal W}^{DA}< 0$, since   ${\cal W}^{D}\geq 0$ then necessarily  $|{\cal W}^{DA}|<{\cal W}^{0}$, so that the denominator of (\ref{WD}) cannot be negative. This is a {\it  physical consequence} of the fact that {\it the donor cannot transfer more energy to the acceptor than that associated to its spontaneous decay rate in free-space}.

The RET efficiency definitions  (\ref{effiE})  are valid if  ${\cal W}^{DA}\geq 0$. However, when ${\cal W}^{DA}<0$ and hence ${\cal W}^{D}>{\cal W}^{0}$, the efficiencies compatible with Eq. (\ref{relatDAyD}) should read  in terms of either ${\cal W}^{D}$ or ${\cal W}^{DA}$:
\be
E_{\cal E}=\frac{1}{1+\frac{{\cal W}^{D}}{{\cal W}^{D}-{\cal W}^{0}}}\,\,;  \,\, E_{\cal E}=\frac{-{\cal W}^{DA}}{{\cal W}^{0}-{\cal W}^{DA}}=
\frac{1}{1+ [\frac{r}{R_{\cal E}}]^6}. \,\,\,\,\,\label{effiEneg}
\ee

As said above,  {\it  ${\cal W}^{0}$ and ${\cal W}^{D}$ are experimentally observable by measurements of the emission from D}, e.g. by fluorescence or scattering, either isolated and in presence of A, respectively; like in standard FRET. Thus from these two quantities we can derive both the RET rate ${\cal W}^{DA}$  from  (\ref{relatDAyD}), and its effciency $E_{\cal E}$  through  either (\ref{effiE}) or (\ref{effiEneg}), respectively. Equations (\ref{effiE}) and  (\ref{effiEneg}) also yield the RET radius $R_{\cal E}$ from these observables.

\subsection{Transfer of helicity}
Like  with energy detection, the measurement of helicity ${\cal W}_{\hel}^{D}$ of the wavefield emitted  by D is affected by the transfer rate of helicity ${\cal W} _{\hel}^{DA}$ to A.  That quantity ${\cal W}_{\hel}^{D}$ may be observed by ellipsometric measurements through the Stokes parameter $S_3$ of the field ${\bf E}_D$ emitted by D \cite{gutsche4, banzer}, which with reference to the paragraph below Eq. (\ref{helhel1}) and the notation of  Eqs. (\ref{nfields}),  is expressed as  $S_3^D 
 = 2 \mbox{Im} [E_{D\, x}^{*}E_{D\, y}]= |E_{D}^{+}|^2-|E_{D}^{-}|^2 =(2k/\epsilon){\cal W}_{\hel}^{D}$. We recall that the  $+$ and $-$ superscripts represent  LCP (+) and RCP (-) polarization, respectively.

Therefore, these  measurements  of field  helicity emitted by D yield either ${\cal W} _{\hel}^{0}$ and ${\cal W} _{\hel}^{D}$, according to whether they are performed  in absence or in presence of the acceptor A.

If  $|{\cal W} _{\hel}^{0}|\geq |{\cal W} _{\hel}^{D}|$,  we express the RHELT efficiency as
\be
E_{\hel}=1-\frac{|{\cal W} _{\hel}^{D}|}{|{\cal W} _{\hel}^{0}|}\,;   E_{\hel}=\frac{|{\cal W} _{\hel}^{DA}|}{|{\cal W} _{\hel}^{0}|+|{\cal W} _{\hel}^{DA}|}=
\frac{1}{1+ [\frac{r}{R_{\hel}}]^6}, \,\,\,\,\label{effiH}
\ee
in terms of either  $|{\cal W} _{\hel}^{D}|$ or $|{\cal W} _{\hel}^{DA}|$. 

Equations (\ref{effiH})  yield
\be
|{\cal W}_{\hel}^{DA}|=|{\cal W}_{\hel}^{0}|\,[\,\frac{|{\cal W}_{\hel}^{0}|}{|{\cal W}_{\hel}^{D}|}\,-1\,]. \label{HrelatDAyD}
\ee
On the other hand, if  $|{\cal W} _{\hel}^{D}|> |{\cal W} _{\hel}^{0}|$, the second equation  (\ref{effiH}) of $E_{\hel}$ in terms of $|{\cal W} _{\hel}^{DA}|$ remains valid, but  $|{\cal W} _{\hel}^{DA}|$ should read
\be
|{\cal W}_{\hel}^{DA}|=|{\cal W}_{\hel}^{0}|\,[\,1-\,\frac{|{\cal W}_{\hel}^{0}|}{|{\cal W}_{\hel}^{D}|}\,]. \label{HrelatDAyDbis}
\ee

\begin{figure*}[t!]
\begin{centering}
\includegraphics[width=18cm]{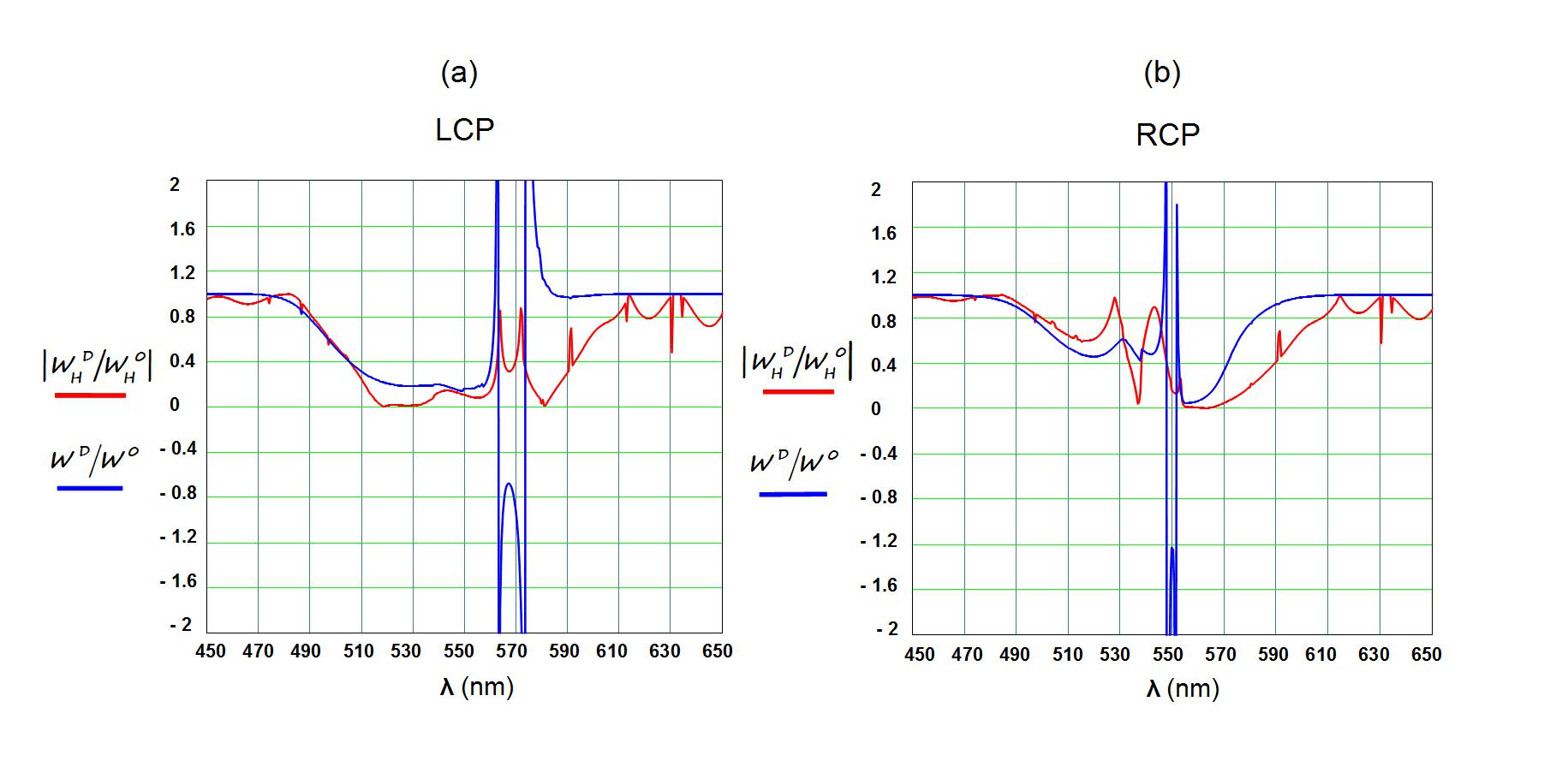}
\par\end{centering}
\caption{(Color online). Same configuration as in Fig. 5. The donor and the acceptor are chiral and magnetoelectric . Both ${\bf p}_D$ and ${\bf m}_D$, as well as ${\bf p}_A$ and ${\bf m}_A$, have well-defined helicity. (a) D and A have LCP  dipole moments. Ratios $|{\cal W}_{\hel}^{D}(\lambda)|/|{\cal W}_{\hel}^{0}(\lambda)|$   and ${\cal W}^{D}(\lambda)/{\cal W}^{0}(\lambda)$; ($k(\lambda)=n(\lambda)\frac{2\pi}{\lambda}$, $n(\lambda)=1$), with $a=15 nm$ being the radius of  both D and A. The distance $r$ between the centers of D and A is $r=35 nm$.  (b) Same as (a)  when both D and A have RCP dipole moments.}
\end{figure*}

So that elliminating $|{\cal W}_{\hel}^{DA}|$ between  (\ref{HrelatDAyDbis}) and  the second equation in (\ref{effiH}) we obtain $E_{\hel}$ in terms of $|{\cal W} _{\hel}^{D}|$:
\be
E_{\hel}=\frac{1}{1+\frac{|{\cal W}_{\hel}^{D}|}{|{\cal W}_{\hel}^{D}|-|{\cal W}_{\hel}^{0}|}}, \,\,\,\label{effiHbis}
\ee
when $|{\cal W} _{\hel}^{D}|> |{\cal W} _{\hel}^{0}|$.

Eqs. (\ref{HrelatDAyD}) and (\ref{HrelatDAyDbis}) show that the ellipsometric measurements of ${\cal W}_{\hel}^{0}$ and ${\cal W}_{\hel}^{D}$ yield $|{\cal W}_{\hel}^{DA}|$. Therefore the sign of the  transfer rate ${\cal W}_{\hel}^{DA}$ cannot be obtained from ${\cal W}_{\hel}^{D}$ and  ${\cal W}_{\hel}^{0}$ through the above equations since there is an evident difficulty in defining the RHELT efficiency from these latter quantities with their respective  signs, rather than from their moduli. This sign of ${\cal W}_{\hel}^{DA}$ might be determined by measuring the helicity emission from A, which is related to  the extinction of ${\cal W}_{\hel}^{DA}$ on interaction with A.  In fact, it is well-known that, for instance, in   a kind of FRET called {\it sensitized emission of the acceptor fluorescence} \cite{clegg} the intensity ${\cal W}^A \sim  {\cal W}^{DA}$, emitted from  the acceptor excited by RET from the donor, is detected, and this yields a measure of the RET rate  ${\cal W}^{DA}$. However this procedure might be more involved from the field helicity emitted by A, since we know that the  wavefield helicity transferred by RHELT   from D to A gives rise, on extinction by the acceptor, to an  helicity emitted by A plus a {\it converted} helicity in A which contains a rich variety of effects \cite{gutsche2, gutsche4}. Further experimental research is necessary to clarify this point.

\subsection{ Illustration: Donor illuminated by light of well-defined helicity}

Figure 6  depicts the spectra of  $|{\cal W}_{\hel}^{D}|/|{\cal W}_{\hel}^{0}|=\frac{|{\cal W}_{\hel}^{0 }|}{|{\cal W}_{\hel}^{DA}|+|{\cal W}_{\hel}^{0}|}$ and  ${\cal W}^{D}/{\cal W}^{0}=\frac{{\cal W}^{0 }}{{\cal W}^{DA}+{\cal W}^{0}}$ for the configuration  of  Example B above; namely, both particles D and A possessing well-defined helicity,  being circularly polarized dipoles either LCP [Fig. 6 (a)] or  RCP   [Fig.6 (b)]. As said above, in the case of quantum emitters  these ratios equal the {\it decay rate} of D in presence of A relative to to free-space decay rate of D: 
$|{\cal W}_{\hel}^{D}|/|{\cal W}_{\hel}^{0}|=|{\gamma _{\hel}^{D}}|/|{ \gamma _{\hel}^{0}}|$, and  ${\cal W}^{D}/{\cal W}^{0}={\gamma^{D}}/{ \gamma^{0}}$, (cf. eg. Eq. (8.142) of \cite{novotny}).  These spectra have been obtained from the RET and RHELT model by introducing Eqs. (\ref{topefret_2}), (\ref{tohelfret_2})  and  (\ref {W0}) into (\ref{WD}) and (\ref{HrelatDAyD})-(\ref{HrelatDAyDbis}). In particular, concerning the new phenomenon: RHELT,  put forward in this work, the spectra $|{\cal W}_{\hel}^{D}(\lambda)|/|{\cal W}_{\hel}^{0}(\lambda)|=|{\gamma _{\hel}^{D}}(\lambda)|/|{ \gamma _{\hel}^{0}}(\lambda)|$ constitute the signal which is possible to experimentally detect,  from either  polarizable particles or quantum emitters.

It is interersting in these figures the {\it  dramatic difference of the signal according to whether one employs  LCP  or RCP illumination},  given a chiral donor D. This is the essence of  the {\it circular dichroism} produced by the donor;  manifested in its helicity and energy emission  with, or without, the presence of the acceptor A.

 One also observes how the presence of A diminishes the decay rate of D, both in RHELT and RET, due to the transfer of  field helicity and energy  from D to A, and this decrease is more pronounced as   $|{\cal W}_{\hel}^{DA}|$ and  ${\cal W}^{DA}$ increase, [compare with the central zones of Figs.  5(a) and 5(c)]. In this way,  these relative values $|{ {\cal W}_{\hel}^{D}}|/|{ {\cal W} _{\hel}^{0}}|$ and ${{\cal W}^{D}}/{ {\cal W}^{0}}$ are less than 1; except in two regions: one is at  wavelengths  at both sides of the graphic window, where  this relative decay, both in energy and helicity, tends to $1$ because there one has $|{\cal W}_{\hel}^{DA }|<<|{\cal W}_{\hel}^{0 }|$ and   ${\cal W}^{DA }<<{\cal W}^{0 }$. Notice that the spikes that appear in these relative values correspond with those observed in $R _{\hel}$ and $R _{\cal E}$  in Figs.  5(b) and 5(d).

The other region where  ${{\cal W}^{D}}/{ {\cal W}^{0}}\ge 1$ is in wavelengths at which        ${\cal W}^{DA}(\lambda)<0$, as seen on comparison of Figs. 6(a) and 6(b) with Figs.  5(a) and 5(c). In this region, in the intervals about (562nm, 580nm) in  Fig. 6(a), and (547nm, 551nm) in Fig. 6(b), the relative emitted energy, or the  relative decay rate, of D asymptotically grows acquiring values larger than $1$ as  $|{\cal W}^{DA }|$ approaches ${\cal W}^{0 }$ in the denominator ${\cal W}^{DA }+{\cal W}^{0 }$ of ${\cal W}^{D}$. Nevertheless, this growth  will stop at the {\it saturation level  impossed by the excited donor lifetime}.  

The {\it sharp fall} of ${\gamma^{D}}/{ \gamma^{0}}$, or equivalently  of  ${\cal W}^{D}/{\cal W}^{0}$,  {\it to negative values}, shown in Figs. 6(a) and 6(b),  {\it is unphysical} since it is due to a negative  denominator ${\cal W}^{DA }+{\cal W}^{0 }$  in  the expression ${\cal W}^{D}/{\cal W}^{0}=\frac{{\cal W}^{0 }}{{\cal W}^{DA}+{\cal W}^{0}}$ when $|{\cal W}^{DA }|>{\cal W}^{0 }$. However as remarked above, negative values of both the decay rates and the emitted intensities   ${\cal W}^{D }$ are meaningless, and  $|{\cal W}^{DA }|<{\cal W}^{0 }$. As a consequence, the actual  values of  the  relative intensities emitted by D, or of its relative decay rates, in those  regions where  they appear negative should be those of saturation of D.

The same arguments concerning saturation apply to  $|{\gamma _{\hel}^{D}}|/|{ \gamma _{\hel}^{0}}|=|{\cal W}_{\hel}^{D}|/|{\cal W}_{\hel}^{0}|$ when $|{\cal W}_{\hel}^{DA}|$ approaches $|{\cal W}_{\hel}^{0}|$ in the denominator $|{\cal W}_{\hel}^{DA}|+|{\cal W}_{\hel}^{0}|$.

We stress that although, as said above,   these spectra of ${\cal W}^{D}$ and  ${\cal W}_{\hel}^{D}$ illustrated in Figs. 6(a) and 6(b) were simulated in consistency  with the RET and RHELT models of  the transfer rates ${\cal W}^{DA}$  and   ${\cal W}_{\hel}^{DA}$, {\it the  real procedure in experiments should work conversely}; i.e. the values of ${\cal W}^{D}$ and  ${\cal W}_{\hel}^{D}$, as well as those of ${\cal W}^{0}$ and   ${\cal W}_{\hel}^{0}$, are  what  the experimental measurements will yield, and then the transfer efficiencies, Eqs. (\ref{relatDAyD}), (\ref{HrelatDAyD}) and (\ref{HrelatDAyDbis}), interaction radii, and transfer rates will be derived.

\section{Conclusions}
It is remarkable that although  a vast majority of biological and pharmaceutical molecules are chiral, and this property immediately suggests to look at the helicity of electromagnetic wavefields on interaction with them, no theory or model of resonance helicity transfer between such "particles" exists to date;  and those existing on energy transfer rarely make use of their chiral characteristics. Thus the inclusion, as in this work,  of  recent advances  in {\it optical magnetism}, which involve the response of these bodies to the magnetic field of light, convey effects of both the electric and magnetic dipoles  even if only RET is addressed;  phenomenona which have so far been ruled out in standard FRET. Then if one also considers the theory of resonance helicity  transfer  put forward  in the previous pages, the new contributions of this paper may be summarized in the following main conclusions:

  1. {\it  The} classical electrodynamic {\it theory of  resonance  helicity transfer  (RHELT), established in this work  between two  generally magnetoelectric  bi-isotropic dipolar generic particles, chiral in particular}, that act as  donor and acceptor, respectively, and  both with angular momentum, {\it constitutes a new tool capable of adding a wealth of information contained in the helicity of the transferred twisted  fields, a quantity never used before in this context.}

2. {\it  Concerning the information conveyed by RHELT, there is the fact that}, as we have proven in our examples,  {\it its transfer rate is very sensitive to the states of polarization of generally elliptically rotating dipoles of the illuminated donor and the acceptor, as it contains four terms and four orientational factors}, rather than just one as in conventional  FRET with linearly polarized dipoles. {\it The same happens with its interaction radius}.

3. {\it Those four terms are discriminatory since they involve the chirality handedness of the donor D through its induced electric and magnetic dipole moments, while two of these terms explicitely exhibit the chirality of the acceptor A  through its cross electric-magnetic polarizability}.  In this way, the RHELT rate is  different when one changes the chiral symmetry of the particles, namely, on passing to the  arrangement with the particle  enantiomer. This effect may be envisaged as a sort of {\it REHLT dichroism}. Also, this is the reason behind {\it the high selectivity of the RHELT  rate and its  interaction radius to the  polarization of both the illumination and the response of D and A,  possessing a structural  symmetry}, under a given (generally elliptic, or particularly circular) polarization.  

4.  At the same time, we have formulated {\it the resonance energy transfer  (RET)  and its interaction radius   between two magnetoelectric chiral particles}. This  process again involves more orientational factors and terms than the well-known $\kappa^2$-factor and the single term of standard FRET. Like for RHELT, {\it  these  RET terms are discriminatory, and the RET rate is also very sensitive to the illumination, polarization states of D and A,  and to the  symmetry  of their structure.}

5. {\it An important  consequence} of the excitation of electric and magnetic dipoles in D,  and of the chirality of A, manifested by the presence of its cross electric-magnetic polarizability in the RET rate equation, {\it  is the possibility that this RET rate be negative} if A has a large enough cross-polarizability, thus being strongly chiral. {\it A negative RET rate means} the new effect in which {\it the emission by the donor is enhanced by the presence of the acceptor.} This phenomenon  does not exist in conventional FRET.

6.   We have  introduced the  {\it observables} and, as such, those {\it quantities measurable in experiments}.  They are  the {\it emitted  field energies an helicites, (or the energy and helicity  decay rates for quantum emitters), from D in presence of A, and their  respective free-space values}. In this way, we  established the {\it equations that allow one to derive from those observables} the RET and RHELT  rates, efficiencies, and interaction radii.   In particular, {\it  these equations  explicitely show  the emission enhancement from D in presence of A when the  RET rate is negative}, as remarked in  point 5 above.

7. An illustrative example plotting these {\it observables  relative  to their free-space values}, has provided an estimation of both  the RHELT and RET signals, showing how they are correlated with their respective tranfer rates.  This illustration has also highlighted  how the {\it  saturation level}, imposed by the excited  donor lifetime, {\it establishes a limit to these relative quantities},  and that {\it  the magnitudes of the energy and helicity transfer rates cannot surpass the donor  emission (or spontaneous  decay)  rates in free-space}.
 
 We should recall that after testing our equations with a known configuration, we have addressed particles bigger than those usually employed in standard FRET. Namely, we  assummed them to have a diameter of a few tens of nanometers, which yields greater polarizabilities and, hence, involve interaction distances larger than the FRET F\"{o}rster radius.  We expect that progress in synthesizing conjugates of fluorophore molecules with magnetoelectric chiral nanoparticles,  will allow experiments with such objects in which magnetoelectric effects occur and for which both helicity and energy  transfers  can be determined from their corresponding observables.

We also believe that further developments of our theory, as well as future experiments, may lead to applications that broaden the scope of  FRET techniques to strongly chiral particles,   adding to the  transfer of intensity  that of helicity of twisted fields, along with its potential information content, both on the induced electric and magnetic dipoles with angular momentum,  and  on the structural symmmetry of the interacting particles.

\section{Acknowledgments}
Work  supported by Ministerio de Ciencia y Tecnologia of Spain through grants   FIS2014-55563-REDC,  FIS2015-69295-C3-1-P and PGC2018-095777-B-C21. My special thanks to Drs. Alberto Berguer M.D. and Manuel de Pedro M.D. for removing a serious disease from my body, making it  possible the accomplishment of this work. I also thank Dr. Marcos W. Puga and two anonimous referees for helpful comments and suggestions.

\renewcommand{\theequation}{A1-\arabic{equation}}
  \setcounter{equation}{0} 
\renewcommand{\figurename}{Figure A1}
  \setcounter{figure}{0}
\appendix
\section{Appendix 1: Proof of Eqs. (6) and (7)  for the RHELT and RET  rates}
\subsection{A1.a. Proof of Eq. (6)}
Introducing in the helicity extinction, Eq. (2),  the acceptor dipole moments given by Eq. (4), one has
\be
{\cal W}_{\hel}^{DA}=2\pi c \mbox{Re} \{-\frac{1}{n^2}( \alpha_{e}^{A}s_{A\, i}^{p}s_{A\, j}^{p\,*}{E}_{D \, j} 
+\alpha_{em}^{A}s_{A\,i}^{p}s_{A\, j}^{m\,*} {B}_{D \, j}){B}^*_{D \, i} \nonumber \\
+(\alpha_{me}^{A}s_{A\, i}^{m}s_{A\, j}^{p\,*} {E}_{D \, j}
+\alpha_{m}^{A}s_{A\, i}^{m}s_{A\, j}^{m\,*} {B}_{D \, j}) {E}^*_{D \, i}
\}, \,(i,j=1,2,3).\,\,\,\,\nonumber
\ee
Making use of  Eqs. (3) for the near fields, the above equation becomes
\be
{\cal W}_{\hel}^{DA}=2\pi c \mbox{Re} \{-\frac{1}{n^2}[ \alpha_{e}^{A}s_{A\, i}^{p}s_{A\, j}^{p\,*} \frac{1}{\epsilon r^3} (3{ s}_{R\,j}({\bf p}_D\cdot  {\bf s}_R)-{ p}_{D\,j}) \nonumber \\
+\alpha_{em}^{A}s_{A\,i}^{p}s_{A\, j}^{m\,*} \frac{\mu}{ r^3} (3{ s}_{R\,j}({\bf m}_D\cdot  {\bf s}_R)-{ m}_{D\,j})] \nonumber \\
\times \frac{\mu}{ r^3} (3{s}_{R\,i}( {\bf m}^*_D\cdot {\bf s}_R)-{ m}^*_{D\,i}) \nonumber \\
+[\alpha_{me}^{A}s_{A\, i}^{m}s_{A\, j}^{p\,*}  \frac{1}{\epsilon r^3} (3{ s}_{R\,j}({\bf p}_D\cdot {\bf s}_R )-{ p}_{D\,j})  \nonumber \\
+\alpha_{m}^{A}s_{A\, i}^{m}s_{A\, j}^{m\,*} (3{ s}_{R\,j}({\bf m}_D\cdot  {\bf s}_R)-{ m}_{D\,j}) ]\nonumber \\
\times  \frac{1}{\epsilon r^3} (3{ s}_{R\,i}({\bf p}^*_D \cdot{\bf s}_R )-{ p}^*_{D\,i})  
\}\,\,\,\,\,\,\,\,\,\,\,\,\,\,\,\, \nonumber
\ee
On employing Eqs.  (5) for the donor dipole moments, the above equation reads
\be
{\cal W}_{\hel}^{DA}=2\pi c \mbox{Re} \{-\frac{1}{n^2}[ \alpha_{e}^{A}s_{A\, i}^{p}s_{A\, j}^{p\,*} \frac{p_D}{\epsilon r^3} (3{ s}_{R\,j}({\bf s}_{D}^{p}\cdot {\bf s}_R )-{ s}_{D\,j}^p) \nonumber \\
+\alpha_{em}^{A}s_{A\,i}^{p}s_{A\, j}^{m\,*}\, \frac{\mu m_D}{ r^3} (3{ s}_{R\,j}({\bf s}_{D}^{m}\cdot  {\bf s}_R) -{ s}_{D\,j}^m)] \nonumber \\
\times \frac{\mu m_D^* }{ r^3} (3{s}_{R\,i}( {\bf s}_{D}^{m *}\cdot {\bf s}_R)-{ s}_{D\,i}^{m *}) \nonumber \\
+[\alpha_{me}^{A}s_{A\, i}^{m}s_{A\, j}^{p\,*}  \frac{p_D}{\epsilon r^3} (3{ s}_{R\,j}({\bf s}_{D}^{p}\cdot {\bf s}_R )-{ s}_{D\,j}^p)  \nonumber \\
+\alpha_{m}^{A}s_{A\, i}^{m}s_{A\, j}^{m\,*}\, \frac{\mu m_D}{ r^3} (3{ s}_{R\,j}({\bf s}_{D}^m\cdot {\bf s}_R )-{ s}_{D\,j}^m) ]\nonumber \\
\times  \frac{p_D^*}{\epsilon r^3} (3{ s}_{R\,i}({\bf s}_{D}^{p *}\cdot {\bf s}_R )-{ s}_{D\,i}^{p\,*})  
\}\,\,\,\,\,\,\,\,\,\,\,\,\,\,\,\, \nonumber
\ee
Rearranging terms, we write the last equation as
\be
{\cal W}_{\hel}^{DA}=\frac{2\pi c}{\epsilon r^6} \mbox{Re} \{-\frac{\alpha_{e}^{A}} {\epsilon}p_D m_D^*[ 3 ( {\bf s}_{A}^{p}\cdot {\bf s}_{R})({\bf s}_{R}\cdot {\bf s}_{D}^{m}) \nonumber \\
-( {\bf s}_{A}^{p}\cdot{\bf s}_D^{m})][ 3 ( {\bf s}_{R}\cdot {\bf s}_{A}^{p})({\bf s}_{D}^{p}\cdot {\bf s}_{R} )-( {\bf s}_{A}^{p\,*}\cdot{\bf s}_{D}^{p\,*})]\nonumber \\
+\mu\alpha_{m}^{A}m_D p_D^* [ 3 ( {\bf s}_{A}^{m}\cdot{\bf s}_{R} )({\bf s}_{R}\cdot {\bf s}_{D}^{p}) \nonumber \\
-( {\bf s}_{A}^{m}\cdot{\bf s}_D^{p})][ 3 ( {\bf s}_{R}\cdot {\bf s}_{A}^{m})({\bf s}_{D}^{m}\cdot{\bf s}_{R} )-( {\bf s}_{A}^{m\,*}\cdot{\bf s}_{D}^{m\,*})]\nonumber \\
 -\mu\alpha_{em}^{A}|m_D|^2 [ 3 ( {\bf s}_{A}^{p}\cdot{\bf s}_{R} )({\bf s}_{R}\cdot {\bf s}_{D}^{m}) \nonumber \\
-( {\bf s}_{A}^{p}\cdot{\bf s}_D^{m})][ 3 ( {\bf s}_{R}\cdot {\bf s}_{A}^{m})({\bf s}_{D}^{m}\cdot{\bf s}_{R} )-( {\bf s}_{A}^{m\,*}\cdot{\bf s}_{D}^{m\,*})]\nonumber \\
+\frac{\alpha_{me}^{A}}{\epsilon}|p_D|^2 [ 3 ( {\bf s}_{A}^{m}\cdot {\bf s}_{R})({\bf s}_{R}\cdot {\bf s}_{D}^{p}) \nonumber \\
-( {\bf s}_{A}^{m}\cdot{\bf s}_D^{p})][ 3 ( {\bf s}_{R}\cdot {\bf s}_{A}^{p})({\bf s}_{D}^{p}\cdot{\bf s}_{R} )-( {\bf s}_{A}^{p\,*}\cdot{\bf s}_{D}^{p\,*})]\}.\,\,\,\,
\ee
The notation for the  scalar product of two complex vectors ${\bf a}$ and ${\bf b}$ employed  here is: ${\bf a}\cdot{\bf b}=a_i b_i^*$. 
The above expression is Eq. (6) with the orientational factors (8)-(11).

\subsection{A1.b. Proof of Eq. (7)}
By substituting in the energy extinction, Eq. (1),  the acceptor dipole moments by their  Eqs. (4), we get
\be
{\cal W}^{DA}=\frac{\omega}{2}\mbox{Im}\{( \alpha_{e}^{A}s_{A\, i}^{p}s_{A\, j}^{p\,*}{E}_{D \, j} +\alpha_{em}^{A}s_{A\,i}^{p}s_{A\, j}^{m\,*} {B}_{D \, j}){E}^*_{D \, i} \,\,\,\,\,\,\nonumber \\
+(\alpha_{me}^{A}s_{A\, i}^{m}s_{A\, j}^{p\,*} {E}_{D \, j}
+\alpha_{m}^{A}s_{A\, i}^{m}s_{A\, j}^{m\,*} {B}_{D \, j}) {B}^*_{D \, i}
\}, \,(i,j=1,2,3).\,\,\,\,\nonumber
\ee
Using  Eqs. (3)  for the near fields, the above equation reads
\be
{\cal W}^{DA}=\frac{\omega}{2}\mbox{Im}\{[ \alpha_{e}^{A}s_{A\, i}^{p}s_{A\, j}^{p\,*} \frac{1}{\epsilon r^3} (3{ s}_{R\,j}({\bf p}_D\cdot {\bf s}_R )-{ p}_{D\,j}) \nonumber \\
+\alpha_{em}^{A}s_{A\,i}^{p}s_{A\, j}^{m\,*} \frac{\mu}{ r^3} (3{ s}_{R\,j}({\bf m}_D\cdot {\bf s}_R )-{ m}_{D\,j})] \nonumber \\
\times \frac{1}{\epsilon  r^3} (3{s}_{R\,i}({\bf p}^*_D\cdot {\bf s}_R )-{ p}^*_{D\,i}) \nonumber \\
+[\alpha_{me}^{A}s_{A\, i}^{m}s_{A\, j}^{p\,*}  \frac{1}{\epsilon r^3} (3{ s}_{R\,j}({\bf p}_D\cdot {\bf s}_R )-{ p}_{D\,j})  \nonumber \\
+\alpha_{m}^{A}s_{A\, i}^{m}s_{A\, j}^{m\,*} (3{ s}_{R\,j}({\bf m}_D\cdot {\bf s}_R )-{ m}_{D\,j}) ]\nonumber \\
\times  \frac{\mu}{r^3} (3{ s}_{R\,i}({\bf m}^*_D\cdot {\bf s}_R )-{ m}^*_{D\,i})  
\}\,\,\,\,\,\,\,\,\,\,\,\,\,\,\,\, \nonumber
\ee
Introducing in this equation Eqs.  (5)  for the donor dipole moments,  one obtains
\be
{\cal W}^{DA}=\frac{\omega}{2}\mbox{Im} \{[ \alpha_{e}^{A}s_{A\, i}^{p}s_{A\, j}^{p\,*} \frac{p_D}{\epsilon r^3} (3{ s}_{R\,j}({\bf s}_{D}^{p}\cdot {\bf s}_R )-{ s}_{D\,j}^p) \nonumber \\
+\alpha_{em}^{A}s_{A\,i}^{p}s_{A\, j}^{m\,*}\, \frac{\mu m_D}{ r^3} (3{ s}_{R\,j}({\bf s}_{D}^{m}\cdot  {\bf s}_R) -{ s}_{D\,j}^m)] \nonumber \\
\times \frac{ p_D^* }{\epsilon r^3} (3{s}_{R\,i}({\bf s}_{D}^{p *}\cdot {\bf s}_R )-{ s}_{D\,i}^{p *}) \nonumber \\
+[\alpha_{me}^{A}s_{A\, i}^{m}s_{A\, j}^{p\,*}  \frac{p_D}{\epsilon r^3} (3{ s}_{R\,j}( {\bf s}_{D}^{p}\cdot {\bf s}_R)-{ s}_{D\,j}^p)  \nonumber \\
+\alpha_{m}^{A}s_{A\, i}^{m}s_{A\, j}^{m\,*}\, \frac{\mu m_D}{ r^3} (3{ s}_{R\,j}({\bf s}_{D}^m\cdot {\bf s}_R )-{ s}_{D\,j}^m) ]\nonumber \\
\times  \frac{\mu m_D^*}{ r^3} (3{ s}_{R\,i}({\bf s}_{D}^{m *}\cdot {\bf s}_R )-{ s}_{D\,i}^{m\,*})  
\}\,\,\,\,\,\,\,\,\,\,\,\,\,\,\,\, \nonumber
\ee
Regrouping terms, and recalling that ${\bf a}\cdot{\bf b}=a_i b_i^*$,  the last equation is written as
\be
{\cal W}^{DA}=\frac{\omega}{2 r^6}\mbox{Im} \{\frac{\alpha_{e}^{A}} {\epsilon^2}|p_D|^2 [ 3 ( {\bf s}_{R}\cdot {\bf s}_{A}^{p})({\bf s}_{D}^{p}\cdot {\bf s}_{R}) \nonumber \\
-({\bf s}_D^{p}\cdot {\bf s}_{A}^{p})][ 3 ( {\bf s}_{R}\cdot{\bf s}_{A}^{p\,*} )({\bf s}_{D}^{p\,*}\cdot{\bf s}_{R} )-( {\bf s}_{D}^{p\,*}\cdot{\bf s}_{A}^{p\,*})]\nonumber \\
+\mu^2 \alpha_{m}^{A}|m_D|^2[ 3 ( {\bf s}_{R}\cdot {\bf s}_{A}^{m})({\bf s}_{D}^{m}\cdot{\bf s}_{R} ) \nonumber \\
-({\bf s}_D^{m}\cdot {\bf s}_{A}^{m})][ 3 ({\bf s}_{R}\cdot {\bf s}_{A}^{m\,*} )({\bf s}_{D}^{m\,*}\cdot {\bf s}_{R})-( {\bf s}_{D}^{m\,*}\cdot {\bf s}_{A}^{m\,*})]\nonumber \\
 +\frac{\mu}{\epsilon}\alpha_{em}^{A} p_D^* m_D[ 3 ({\bf s}_{A}^{p}\cdot  {\bf s}_{R})({\bf s}_{R}\cdot {\bf s}_{D}^{p}) \nonumber \\
-( {\bf s}_{A}^{p}\cdot{\bf s}_D^{p})] [ 3 ( {\bf s}_{R}\cdot {\bf s}_{A}^{m})({\bf s}_{D}^{m}\cdot {\bf s}_{R}) \nonumber \\
-( {\bf s}_{D}^{m}\cdot{\bf s}_{A}^{m})]
+\frac{\mu}{\epsilon} \alpha_{me}^{A}p_D m_D^* [ 3 ( {\bf s}_{A}^{p\,*}\cdot {\bf s}_{R})({\bf s}_{R}\cdot {\bf s}_{D}^{p\,*}) \nonumber \\
-({\bf s}_{A}^{p\,*}\cdot {\bf s}_D^{p\,*} )][ 3 ( {\bf s}_{R}\cdot {\bf s}_{A}^{m\,*})({\bf s}_{D}^{m\,*}\cdot {\bf s}_{R} )-( {\bf s}_{D}^{m\,*}\cdot {\bf s}_{A}^{m\,*})]\}. \,\,\,\,\,\,\,\,\,\,\,\,
\ee
Which is Eq. (7) with the orientational factors (12)-(14).

\renewcommand{\theequation}{A2-\arabic{equation}}
  \setcounter{equation}{0} 
\renewcommand{\figurename}{Figure A2}
  \setcounter{figure}{0}
\appendix

\section{Appendix 2:  Emission, absorption and extinction spectra  of donor and acceptor }
In standard FRET   the normalized emission spectrum $f_{e}^{D}(\omega)$ of the donor D electric dipole and the absorption spectrum $\sigma_{e\, A}^{a}(\omega)$ of the acceptor A  electric dipole are defined in terms of  $\mbox{Im}\{\alpha_{e}^{D}(\omega)\}$  and   $\mbox{Im}\{\alpha_{e}^{A}(\omega)\}$, respectively \cite{novotny,clegg}. However, when D and/or A is chiral, one will also need to link  the magnetic and cross electric-magnetic polarizabilities with the respective emission or absorption spectra accounting for the excitation of the magnetic dipole and the electric-magnetic interaction between both dipoles, respectively. 

In addition, if there is also  scattering by D and/or A, one needs to introduce   the extinction (rather than just the absorption)  cross-section linked to the polarizabilities. This is done by using the optical theorem of energy, expressed in terms of the donor or the  acceptor polarizabilities, which according to Eq. (30) of \cite{nieto2} for a chiral particle on illumination with an elliptically polarized plane wave, [see also Eq. (1)], reads  (we shall drop the scripts $D$ and $A$, understanding that  the polarizabilities dealt with now apply to  either donor and/or acceptor):
\be
{\cal W}^{a}+\frac{2 k^3}{3 }\{[\epsilon^{-1} |\alpha_{e}|^{2}+ n^{2}\mu|\alpha_{m}|^2    
+( \epsilon^{-1} n^{2}+ \mu) \,\,\,\,\,\,\nonumber \\
 \times|\alpha_{me}|^{2} ]|e_{i}|^2    
 +4k \sqrt{\frac{\mu}{\epsilon}}\mbox{Im} [ \epsilon^{-1} \alpha_{me}^{*}   \alpha_{e}  - \mu \alpha_{me} \alpha_{m}^{*}]{\hel^{i}}\}=
  \nonumber \\  \,\,\,\,\,\, 
4k \sqrt{\frac{\mu}{\epsilon}}\alpha_{me}^{R} {\hel^{i}}+ (\alpha_{e}^{I} +n^2  \alpha_{m}^{I}) |e_i|^2   .  \,\,\,\,\,\,\,\,\,\, \,\,\,\,\label{top2}
\ee
The superscripts $I$ and $R$ denote  imaginary and real part, respectively. $ {\hel^{i}}$  is the helicity density, [cf.  Eq. (26)], of the  field incident on the particle, which we shall now consider  to be  circularly polarized (CPL), so that \cite{nieto2}:  $ {\hel^{i}}= \pm\frac{\epsilon}{2k}|e_{i}|^{2}$.

Dividing (\ref{top2}) by the magnitude of the time-averaged incident Poynting vector $<S_i>= (c\epsilon/4 \pi n)|{\bf E}_{i}|^{2}$, and using the above expression of ${\hel^{i}}$, we obtain
\be
{\sigma}^{a}+\frac{8\pi  k^4}{3 }\{[\epsilon^{-2} |\alpha_{e}|^{2}+ \mu^{2}|\alpha_{m}|^2 +2 \frac{\mu}{\epsilon}|\alpha_{me}|^{2}]    \nonumber  \\
  \pm 2 \sqrt{\frac{\mu}{\epsilon}}   \mbox{Im} [ \epsilon^{-1} \alpha_{me}^{*}   \alpha_{e}  - \mu \alpha_{me} \alpha_{m}^{*}]\}=    \nonumber  \\ \,\,\,\,\,\,
4\pi k  (\epsilon^{-1}\alpha_{e}^{I} +\mu   \alpha_{m}^{I} \pm 2 \sqrt{\frac{\mu}{\epsilon}}\alpha_{me}^{R})    .   \label{top2b}
\ee
The left side of (\ref{top2b}) is the absorption cross-section $\sigma^{a}$ plus a term which represents the scattering cross-section $\sigma^{s}$ of the particle, (either $D$ or $A$).  This sum  is  the extinction cross-section:  $\sigma^{ext}=\sigma^{a}+\sigma^{s}$. Notice that $\sigma^{s}$ contains terms with the cross-polarizability $\alpha_{me}$ added to the well-known quadratic terms in $|\alpha_{e}|$ and $|\alpha_{m}|$ of  non bi-isotropic particles \cite{opex2010}. On the other hand, the right side of  (\ref{top2b}), which expresses  $\sigma^{ext}$,  has a  term with $\alpha_{me}^{R}$  added to the well known extinction terms: $\epsilon^{-1}\alpha_{e}^{I} +\mu   \alpha_{m}^{I}$ corresponding to  an achiral particle \cite{opex2010}.

Addressing the operation: $\pm$ in the left side of     (\ref{top2b}), if we substract this equation taking  the sign - from that taking the sign +,   we obtain 
\be
(\sigma_{LCP}^{a}-\sigma_{RCP}^{a})+\frac{32\pi  k^4}{3 } \sqrt{\frac{\mu}{\epsilon}}\mbox{Im} [ \epsilon^{-1} \alpha_{me}^{*}   \alpha_{e}  - \mu \alpha_{me} \alpha_{m}^{*} ]\}\nonumber \,\,\,\,\,\, \\
=16\pi k  \sqrt{\frac{\mu}{\epsilon}} \alpha_{me}^{R}. \,\,\,\,\,\,\, \,\,\,\,\, \label{top3}
\ee
The quantity $(\sigma_{LCP}^{a}-\sigma_{RCP}^{a})$ and   the second term of the left side of (\ref{top3}) are  the difference between  the particle absorption cross-sections and between its scattering cross-sections $(\sigma_{LCP}^{s}-\sigma_{RCP}^{s})$ with  LCP and RCP plane wave incidence, respectively.
Thus the whole left side of  (\ref{top3}) is the difference of the particle extinction cross sections $(\sigma_{LCP}^{ext}-\sigma_{RCP}^{ext})$.  Hence this latter difference constitutes the meaning of the right side of (\ref{top3})  which thereby represents the {\it circular dichroism (CD) cross-section}: $\sigma_{me}^{CD}=\sigma_{LCP}^{ext}-\sigma_{RCP}^{ext}$ of the chiral particle \cite{nieto3, tang1},  characterized by $\alpha_{me}^{R}$ as (\ref{top3}) shows. Namely, we write Eq. (\ref{top3}) as
\be
\alpha_{me}^{R}=\frac{1}{16\pi k}  \sqrt{\frac{\epsilon}{\mu}} \sigma_{me}^{CD} ,
  \,\,\,\, \sigma_{me}^{CD}=\sigma_{LCP}^{ext}-\sigma_{RCP}^{ext}.  \,\, \,\,\,\,\,\,\,\,\label{top4}
\ee

Notice that, on the other hand,   adding  Eq.(\ref{top2b}) with the sign $-$ to that with the sign $+$, one gets  the usual relationship  between $\sigma_{LCP}^{ext}+\sigma_{RCP}^{ext}$ and the modulus and the imaginary parts of $\alpha_e$ and $\alpha_m$ in which there is the additional term $2 \frac{\mu}{\epsilon}|\alpha_{me}|^{2}$. In this connection,  the ratio  $(\sigma_{LCP}^{ext}-\sigma_{RCP}^{ext})/(\sigma_{LCP}^{ext}+\sigma_{RCP}^{ext})$ is the well-known {\it dissymmetry factor} of CD \cite{tang1,schellman}. However, as shown in  (\ref{top3}), the dichroism signal is generally not only described by the difference of absorption cross-sections, as usually formulated \cite{schellman, tang1}; but it  contains an additional second term $\sigma_{LCP}^{s}-\sigma_{RCP}^{s}$ which accounts for scattering.
On the other hand, $\alpha_{me}^{I}$ is proportional to the {\em optical rotation}  (OR) {\em cross-section} $\sigma_{me}^{OR}$:
\be
 \sigma_{me}^{OR}=16\pi k\sqrt{\frac{\mu}{\epsilon}}\alpha_{me}^{I}. \label{sigOR}
\ee

 In many instances  $ {\sigma}^{ext}\approx {\sigma}^{a}$, hence there being no strong coupling, or multiple feedback, between D and A, However, for a magnetoelectric nanoparticle with little absorption \cite{g-etxarri,geffrin,kuznetsov,staude,kivshar_reviews},   $ {\sigma}^{s}$ will dominate and, after normalizing it to $1$,  it  constitutes by itself the  emission spectrum $f(\omega)$. 

Summarizing, for a donor lossless nanoparticle:  $ \sigma_{me}^{CD} \approx \sigma_{LCP}^{s}-\sigma_{RCP}^{s}$, whereas for a donor, or acceptor, absorbing molecule:  $ \sigma_{me}^{CD} \approx\sigma_{LCP}^{a}-\sigma_{RCP}^{a}$.

 In those cases in which  one can separately associate  the imaginary part of  the  electric and magnetic  polarizabilities to  the particle electric and  magnetic extinction cross sections:  $\sigma_{e}^{ext}(\omega)$, $\sigma_{m}^{ext}(\omega)$, respectively, [cf. Eq.(\ref{top2b})], averaging over the three orientations of the particle, which yields  $1/3$  times  the polarizabilities, one  has from (\ref{top2b}) and  (\ref{top3}) for an acceptor molecule, (if  scattering is neglected):
\be
\alpha_{e}^{A\,I}(\omega)=\frac{3\epsilon}{4\pi k}\sigma_{e\,A}^{ext}(\omega)\approx\frac{3\epsilon}{4\pi k}\sigma_{e\,A}^{a}(\omega) , \,\,\,\nonumber \\
\alpha_{m}^{A\,I}(\omega)=\frac{3}{4\pi k\mu}\sigma_{m\,A}^{ext}(\omega)\approx\frac{3}{4\pi k\mu}\sigma_{m\,A}^{a}(\omega).  \label{sigmaseem} \\ 
\alpha_{me}^{A\,R}(\omega)=\frac{3}{16\pi k}\sqrt{\frac{\epsilon}{ \mu}}\sigma_{me\,A}^{CD}(\omega) , \nonumber\,\,\,\,\, \\
 \alpha_{me}^{A\,I}(\omega)=\frac{3}{16\pi k}\sqrt{\frac{\epsilon}{\mu}}\sigma_{me\,A}^{OR}(\omega).\,\,\, \,\,\label{sigORR} 
\ee
And for a donor nanoparticle, (sometimes  either absorption or scattering is neglected):
\be
\alpha_{e}^{D\,I}(\omega)=\frac{3\epsilon}{4\pi k}\sigma_{e\,D}^{ext}(\omega)\propto\frac{3\epsilon}{4\pi k}f_{e}^{D}(\omega), \,\,\,\,\,\nonumber \\
\alpha_{m}^{D\,I}(\omega)=\frac{3}{4\pi k\mu}\sigma_{m\,D}^{ext}(\omega)\propto\frac{3}{4\pi k\mu}f_{m}^{D}(\omega),  \label{sigmasoo}
\\
\alpha_{me}^{D\,R}(\omega)\propto \frac{3}{16\pi k}\sqrt{\frac{\epsilon}{ \mu}}f_{me\,D}^{CD}(\omega) ,
 \nonumber \\
 \alpha_{me}^{D\,I}(\omega)\propto\frac{3}{16\pi k}\sqrt{\frac{\epsilon}{\mu}}f_{me\,D}^{OR}(\omega). \,\,\,\, \label{sigOIR}
\ee
Defining the $f$-emission spectra of D  in $fs$, convey  the  above  proportionality  factor in $nm^2 fs^{-1}$.

These expressions are complemented with  the dispersion relations \cite{barron1} (dropping again the superscripts $A$ and $D$):
\be
\alpha_{e,m,me}^{R}(\omega)=\frac{2}{\pi}{\cal P}\int_{0}^{\infty}d\omega'  \, \frac{\omega'\,\alpha_{e,m,me}^{I}(\omega')}{\omega'^{2} -\omega^{2}}, \label{HT1} \\
\alpha_{e,m,me}^{I}(\omega)=-\frac{2\omega}{\pi}{\cal P}\int_{0}^{\infty}d\omega'  \, \frac{\alpha_{e,m,me}^{R}(\omega')}{\omega'^{2} -\omega^{2}}. \label{HT12}
\ee
${\cal P}$ denoting principal value. From $k(\omega)=2n(\lambda)\pi/\lambda=n(\omega) \omega/c$, we may derive the real part of the polarizabilities from their imaginary parts  as:
\be
\alpha_{e,m,me}^{R}(\lambda)=-\frac{2\lambda^2}{\pi}{\cal P}\int_{0}^{\infty}d\lambda'  \, \frac{ \,\alpha_{e,m,me}^{I}(\lambda')}{\lambda' \,  (\lambda'^{2} -\lambda^{2})}. \,\,\,\,\label{HT2}\\
\alpha_{e,m,me}^{I}(\lambda)=\frac{2\lambda}{\pi}{\cal P}\int_{0}^{\infty}d\lambda'  \, \frac{\alpha_{e,m,me}^{R}(\lambda')}{\lambda'^{2} -\lambda^{2}}.\,\,\,\, \label{HT22}
\ee
In passing, we note that Eqs.  (\ref{sigORR}), (\ref{sigmasoo}) and (\ref{HT22}) lead to the dispersion relation between the CD  and OR cross-sections \cite{schellman}:
\be
\sigma_{me}^{OR}(\lambda)=\frac{2}{\pi}{\cal P}\int_{0}^{\infty}d\lambda'  \, \frac{\lambda'\,\sigma_{me}^{CD}(\lambda')}{\lambda'^{2} -\lambda^{2}}. \label{OT22} \\
\sigma_{me}^{CD}(\lambda)=-\frac{2\lambda}{\pi}{\cal P}\int_{0}^{\infty}d\lambda'  \, \frac{\,\sigma_{me}^{OR}(\lambda')}{\lambda'^{2} -\lambda^{2}}. \label{OT23} 
\ee 
For a distribution  of  donors and acceptors which emit and absorb over a range of frequencies, one should  generalize  (\ref{sigmaseem}) - (\ref{sigOIR}) to  the imaginary part of    {\it effective} $\alpha$'s in terms of  overlapping integrals of the  emission spectra of D, $f_{e,m,me}^{D}(\omega)$ and  absorption espectra of A, $\sigma_{e,m,me\,A}^{a}(\omega)$; so that  we have 
\be
\alpha_{e}^{I\,eff}=\frac{3}{4\pi}    \int_{0}^{\infty}d\omega \, \frac{\epsilon(\omega) f_{e}^{D}(\omega)\sigma_{e\,A}^{a}(\omega)}{ k(\omega)}. \label{sigmaso1}
\ee
\be
\alpha_{m}^{I\, eff}=\frac{3}{4\pi}    \int_{0}^{\infty}d\omega \, \frac{ f_{m}^{D}(\omega)\sigma_{m\,A}^{a}(\omega)}{\mu(\omega) k(\omega)}. \,\, \, \,\, \,\,\, \,\,\,\,\, \label{sigmaso2} \\
\alpha_{me}^{R\, eff}=\frac{3}{16\pi}\int_{0}^{\infty}d\omega \, \sqrt{\frac{\epsilon(\omega)}{\mu(\omega)}} \frac{f_{me\,D}^{CD}(\omega)\,\sigma_{me\,A}^{CD}(\omega) }{k(\omega)}. \,\,\, \,\,\,\,\, \,\, \,\,\,  \,\,\,\, \label{sigmaso3} \\
 k(\omega)=n(\omega) \frac{\omega}{c}. \,\, \,  \nonumber
\ee
The $f_D$s fulfilling: $ \int_{0}^{\infty}d\omega \, f_D(\omega) = 1 $.
And again from $k(\omega)=2n(\lambda)\pi/\lambda=n(\omega) \omega/c$, the above expressions also read:
\be
\alpha_{e}^{I\, eff}=\frac{3c}{4\pi}    \int_{0}^{\infty}d\lambda \, \frac{\epsilon(\lambda) f_{e}^{D} (\lambda)\sigma_{e\,A}^{a}(\lambda)}{ n(\lambda) \lambda}. \label{sigmaso11}
\ee
\be
\alpha_{m}^{I\, eff}=\frac{3c}{4\pi}    \int_{0}^{\infty}d\lambda \, \frac{ f_{m}^{D}(\lambda)\sigma_{m\,A}^{a}(\lambda)}{ n(\lambda)\mu(\lambda) \lambda}
. \,\, \, \,\,\,\,\,\, \,\,\, \,\label{sigmaso22} \\
\alpha_{me}^{R\, eff}=\frac{3c}{16\pi}    \int_{0}^{\infty}d\lambda\, \sqrt{\frac{\epsilon(\lambda)}{\mu(\lambda)}}\frac{f_{me\,D}^{CD}(\lambda)\sigma_{me\,A}^{CD}(\lambda)} {n(\lambda) \lambda}. \,\,\, \, \,\, \,\,\, \,\,\,\,  \label{sigmaso33} 
\ee
 Where now the normalization of  the  $f_D $s   is:  $ 2\pi c\int_{0}^{\infty}d\lambda\, f_D(\lambda)/\lambda^2 = 1$.

\renewcommand{\theequation}{A3-\arabic{equation}}
  \setcounter{equation}{0} 
\renewcommand{\figurename}{Figure A3 -}
  \setcounter{figure}{0}
\appendix

\section{Appendix 3: Test and calibration of  RHELT and RET formulations}
 In this section we calibrate our helicity and energy transfer equations [cf. Eqs. (15) and (16)] with a known configuration. This will also serve to calibrate the range of  the RHELT and RET radii, [see Eqs.  (20) and (21)], as well as the sensitivity of RHELT to the chirality of D and/or A.

\begin{figure*}[t]
\begin{centering}
\includegraphics[width=18cm]{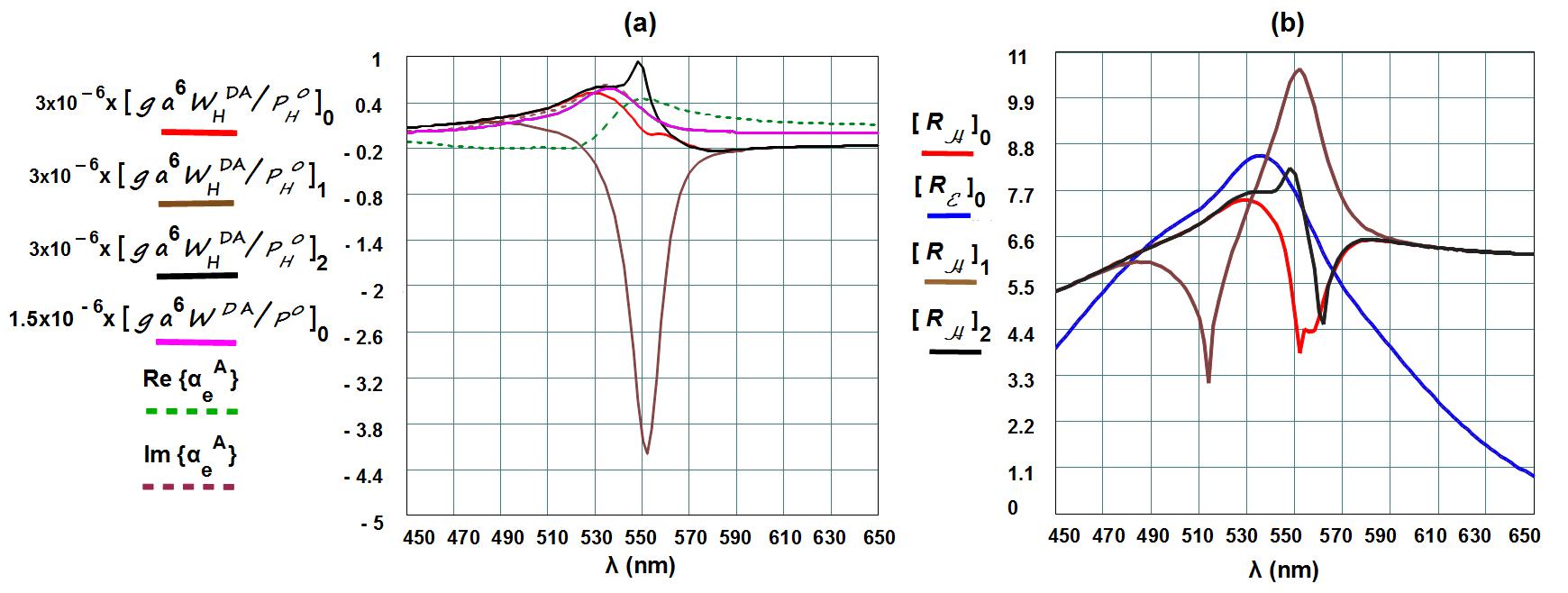}
\par\end{centering}
\caption{(Color online). The electric dipole polarizabilities of D and  A are equal to those of \cite{novotny}. Their cross electric-magnetic and magnetic polarizabilities are six and seven orders of magnitude smaller, respectively, than the electric polarizability.  ${\bf p}_D$ and ${\bf m}_D$ are excited by  circularly polarized light (CPL), either with $e^{+}=6$,  $e^{-}=0$, (left circular, LCP), or  $e^{+}=0$,  $e^{-}=6$ (right circular, RCP), [see Eq. (25)].  ${\bf p}_A$ and ${\bf m}_A$ are also CPL with the same polarization as ${\bf p}_D$ and ${\bf m}_D$.  The response of both the acceptor and its enantiomer, (i.e. the molecule  with $\alpha_{me}^{A}$ of opposite sign), are shown. Quantities in brackets with subindex  $0$ correspond to LCP illumination, the acceptor having  cross-polarizability: $\alpha_{me}^{A}$. Quantities  in brackets with subindex $1$   correspond to LCP illumination, the acceptor being the enantiomer, (i.e.  with cross-polarizability: $-\alpha_{me}^{A}$).   Quantities in brackets with subindex $2 $ correspond to RCP illumination, the acceptor having  cross-polarizability: $\alpha_{me}^{A}$.   
 (a) $[ga^6{\cal W}_{\hel}^{DA}(\lambda)/{\cal P}_{\hel}^{0}]_j$, ($j=0,1,2$), where   ${\cal P}_{\hel}^{0}(\lambda)=\mbox{Im} [ p_{D} m^*_{D}]$;   $[ga^6{\cal W}^{DA}(\lambda)/{\cal P}^{0}(\lambda)]_0$, (only this quantity is plotted because its line coincides with those of subindex 1 and 2), where  ${\cal P}^{0}(\lambda)=(|p_{D}|^{2}/\epsilon+\mu |m_{D}|^{2})$.  [$k(\lambda)=n(\lambda)(2\pi/\lambda)$, $n(\lambda)=1$,  $g=\frac{3r^6}{4a^6 k^3(\lambda)}$].  All these quantities are plotted in arbitrary units. Also shown are  $\mbox{Re}\{\alpha_{e}^{A}(\lambda)\}$ and  $\mbox{Im}\{\alpha_{e}^{A}(\lambda)\}$ in $nm^3$.  (b)   $[R _{\hel}(\lambda)]_j$  in $nm$, ($j=0,1,2$),  and $[R _{\cal E}(\lambda)]_0$   in $nm$,  (only this quantity is plotted because its line coincides with those of subindex 1 and 2). All  ${\cal K}$-factors are  averaged  to  $2/3$. Notice   that since the RET rates and radii coincide in these three cases, for this molecule  these quantities are  not  affected neither by the chirality of the illumination nor by that of A. As seen, this is in contrast with the higher sensitivity of the RHELT  rates and radii.}
\end{figure*}

We consider D illuminated by a left-handed circularly polarized (LCP) plane wave, - see Eq. (25) -. We employ the D and A electric dipole  lineshapes and parameters  of  Eq.(8.172) of \cite{novotny}, Section 8.6.2. They are reproduced in Section A3.a of Appendix 3, Eqs. (A3-1) and (A3-2); also assuming both D and A with a very small chirality and  magnetic dipole moment, so that there are weak electric-magnetic, as well as  weak magnetic, interactions.

The quantities  ${\cal W} _{\hel}^{DA}(\lambda)$, ${\cal W}^{DA}(\lambda)$,  $R _{\hel}(\lambda)$ and $R _{\cal E}(\lambda)$, as well as  $\mbox{Im}\{\alpha_{e}^{A}(\lambda)\}$ and  $\mbox{Re}\{\alpha_{e}^{A}(\lambda)\}$, are obtained  from Eqs. (15), (16), (20) and (21), along with Eqs.(A2-6) and (A2-10) of Appendix 2, using the lineshapes (\ref{f_D}) and (\ref{sigma_A}) and parameters of Section A3.a of this Appendix 3.
The only significant contribution to ${\cal W}^{DA}$ and $R _{\cal E}$ comes from the first term of (16) and (21) , namely of the electric dipole moments of D and A,  which are equal to those  of \cite{novotny} thus yielding a result akin to that of standard FRET. As seen in Figs. A 3-1(a) and A3-1(b) of this Appendix 3,  $N{W}^{DA}(\lambda)$ and $R _{\cal E}(\lambda)$ do not change with the choice of chirality of the illumination, (i.e. according to whether it is LCP or RCP), neither with that of the acceptor, (namely, the handedness of A, characterized by the sign of  $\alpha_{me}^{A}$).

Like in \cite{novotny}, (where D is  a {\it fluorescein} molecule whose estimated average diameter is $0.69 \pm 0.02$ nm), a  F\"{o}rster radius  of  $7.6 nm$ is obtained from Eq. (21)  using  all orientational factors averaged to $2/3$. Also, an effective  $\mbox{Im}\{\alpha_{e}^{A\,eff}\}=0.32 nm^3$, given by the overlapping integral (\ref{sigmaso1}) of Appendix 2, is obtained. This latter  value, $0.32 nm^3$, is compared with those of  $\mbox{Im}\{\alpha_{e}^{A}(\lambda)\}$ derived from Eq. (\ref{sigmaseem}) of  Appendix 2, which as seen in Fig.A 3-1(a), has a maximum of $0.6 nm^3$ at $\lambda=538 nm$, and acquires that value $0.32 nm^3$  at $\lambda$ close to $550 nm$. As shown in Fig.A 3-1(b), and consistently with this latter value  $\mbox{Im}\{\alpha_{e}^{A}(550)\}=0.32 nm^3$,  the above quoted  F\"{o}rster  radius $R _{\cal E}=7.6 nm$ occurs at $\lambda=550 nm$. 

Hence Figs. A3-1(a) and A3-1(b) constitute a confirmation of the adequacy of our formulation since  Fig.A3-1(b) exhibits values of $R _{\cal E}$ between $4 nm$ and $8.5 nm$ in the interval of wavelengths: $[450,590]$ $ nm$. On the other hand, the resonant  $\mbox{Im}\{\alpha_{e}^{A}(\lambda)\}$ yields the peak  $R _{\cal E}(538)=8.5 nm$. Moreover, $\lambda=550 nm$ is approximately the wavelength at which $ f_{e}^{D} (\lambda)$ and $\sigma_{e\,A}^{a}(\lambda)$ cross each other, (cf. Fig. 8.14 of \cite{novotny}). Illumination of D with elliptically polarized light does not appreciably change the values of  $R _{\hel}$ and $R _{\cal E}.$

It is surprising, notwithstanding, that such small (but not zero)  cross electric-magnetic  dipole and magnetic dipole parameters, and hence polarizabilities,  as seen in Section A3.a below, (which  are respectively six and seven orders of magnitude smaller than the electric  dipole one), yield non-negligible values of the helicity transfer distance $R _{\hel}(\lambda)$ and  normalized  helicity transfer    $ga^6{\cal W}_{\hel}^{DA}(\lambda)/{\cal P}_{\hel}^{0}$, as shown in  the above Figs. A3-1(a) and A3-1(b). The cause is the denominator $k^3(\lambda)  {\cal P} ^{0}_{\hel}(\lambda)=k^3\mbox{Im} [ p_{D} m^*_{D}]$, which still is six orders of magnitude smaller than the numerator ${\cal W}_{\hel}^{DA}(\lambda)$,   (to which only the first and fourth terms contribute in  Eq.  (15); the second and third terms being much smaller than this denominator), and it is of the same order  of magnitude as the numerator in (20);  thus resulting in a large ratio   $R _{\hel}^{6}(\lambda)=ga^6{\cal W}_{\hel}^{DA}(\lambda)/{\cal P}_{\hel}^{0}$,  and hence in a  $R _{\hel}(\lambda)$ comparable to $R _{\cal E}$, as shown in Fig. A3-1(b). Of course were zero the cross electric-magnetic and magnetic polarizabilities, both $ga^6{\cal W}_{\hel}^{DA}(\lambda)/{\cal P}_{\hel}^{0}(\lambda)$ and $R _{\hel}(\lambda)$ would become zero. This non-negligible value of  $ga^6{\cal W}_{\hel}^{DA}(\lambda)/{\cal P}_{\hel}^{0}(\lambda)$ and    $R _{\hel}(\lambda)$ for  very small values of the electric-magnetic and magnetic polarizabilities versus the electric ones, is a  remarkable feature of  the  RHELT  equations.

Linked to this latter fact is that  {\it both ${\cal W}_{\hel}^{DA}$ and   $R _{\hel}$  are very sensitive to variations in either the incident polarization}, (e.g. changes from LCP to RCP illumination of D), {\it and in the sign of } $\alpha_{me}^{A}$, (namely, on  the  handedness of  the acceptor particle), {\it while  ${\cal W}^{DA}$ and   $R _{\cal E}$ were not altered by  these changes}.  This is seen in  Figs. A3-1(a) and A3-1(b). {\it The large minimum in ${\cal W}_{\hel}^{DA}$ for   LCP light incident on the enantiomer of D, and the non-zero values of $R _{\hel}$ for such small magnetoelectric response of D and/or  A,  are manifestations of the  high sensitivity of the transfer of helicity to these chirality and magnetic properties of D on comparison with that of energy transfer. }

{\it The electric polarizability of a particle is in the same range of values as its volume}. Accordingly, we have obtained  (not shown for brevity) that other doped acceptor molecules with an   order of magnitude in their size similar to that of the example of  the above Figs. A3-1(a) and A3-1(b),  like a functionalized exahelicene, (average radius: $a=0.242 nm$;  $\alpha_{e}^{A\, ,R}=0.0104 nm^3$,  $\alpha_{me}^{A\, ,I}=- 0.62 \times 10^{-5} nm^3$  at $\lambda=589 nm$),  \cite{barron1} yield  RET radii which vary with $\lambda$ in the same range as in the above example, namely: $4-8 nm$  at wavelengths akin to those  of  Figs. A3-1(a) and A3-1(b).  $R _{\hel}(\lambda)$  is in the same range of values as $R _{\cal E}$.

\subsection{A3.a.  Data for test and calibration of  the {RHELT} and {RET} equations. Donor and acceptor  are molecules whose electric dipole  lineshapes and parameters, are  those of Eq. (8.172) of  \cite{novotny}}
We use electric dipole parameters for both donor, D, and acceptor, A, close to those of \cite{novotny}, (cf. Eq.(8.172) of  Section 8.6.2 of \cite{novotny}),  and data therein. However, we also use lineshapes for the electric-magnetic and magnetic dipole interactions, even though the polarizabilities of D and A associated to these $e-m$ and $m$ interactions, are six and seven orders of  magnitude smaller, respectively, than those of the electric dipole interaction. With the notation of Appendix 2, one has:

For the donor:
\be
f_{e,\,m;\,me}^{D;\,CD}(\lambda)=D_{e,\,m,\,me}^{(1)} e^{-[(\lambda-\lambda_{e,\,m,\,me}^{(1)\,D})/\Delta\lambda_{e,\,m,\,me}^{(1)\,D}]^2 } \nonumber \\
+D_{e,\,m,\,me}^{(2)} e^{-[(\lambda-\lambda_{e,\,m,\,me}^{(2)\,D})/\Delta\lambda_{e,\,m,\,me}^{(2) \,D}]^2 }. \,\, \,\,\,\,\label{f_D} 
\ee
 With the normalization of  these  $f^{D}$'s:  $ 2\pi c\int_{0}^{\infty}d\lambda\, f_{e,\,m;\,me}^{D;\,CD}(\lambda))/\lambda^2 = 1$, and with $n(\lambda)=1$, $c=300\, nm\times fs^{-1}$.  

 $D_{e}^{(1)}=2.52 fs$,  $D_{e}^{(2)}=1.15 fs$, $\lambda_{e}^{(1)\,D}= 512.3 nm$, $\lambda_{e}^{(2)\,D}= 541.7 nm$,  $\Delta\lambda_{e}^{(1)\,D}= 16.5 nm$, $\Delta\lambda_{e}^{(2)\,D}= 35.6 nm$,
 $D_{m}^{(1)}=0.8 \times 10^{-8} fs$,  $D_{m}^{(2)}=0.11fs$, $\lambda_{m}^{(1)\,D}= 539 nm$, $\lambda_{m}^{(2)\,D}= 561 nm$,  $\Delta\lambda_{m}^{(1)\,D}=11 nm$, $\Delta\lambda_{m}^{(2)\,D}=29 nm$,
 $D_{me}^{(1)}=3\times 10^{-8} fs$,  $D_{me}^{(2)}=2.3 \times10^{-8} nm^2$, $\lambda_{me}^{(1)\,D}= 558.1 nm$, $\lambda_{me}^{(2)\,D}= 534.3 nm$,  $\Delta\lambda_{me}^{(1)\,D}=11.7 nm$, $\Delta\lambda_{me}^{(2)\,D}=28.5 nm$.

And for the acceptor:
\be
 \sigma_{e,\,m;\,me}^{ext;\,CD}(\lambda)=A_{e,\,m,\,me}^{(1)} e^{-[(\lambda-\lambda_{e,\,m,\,me}^{(1)\,A})/\Delta\lambda_{e,\,m,\,me}^{(1)\,A}]^2 } \nonumber \\
+A_{e,\,m,\,me}^{(2)} e^{-[(\lambda-\lambda_{e,\,m,\,me}^{(2)\,A})/\Delta\lambda_{e,\,m,\,me}^{(2)\,A}]^2 }. \,\,\,\,\, \label{sigma_A}
\ee
With   $A_{e}^{(1)}=0.021nm^2$,  $A_{e}^{(2)}=0.013nm^2$, $\lambda_{e}^{(1)\,A}= 535.8 nm$, $\lambda_{e}^{(2)\,A}= 514.9 nm$,  $\Delta\lambda_{e}^{(1)\,A}=15.4 nm$, $\Delta\lambda_{e}^{(2)\,A}=36.9 nm$
 $A_{m}^{(1)}=1\times 10^{-9}nm^2$,  $A_{m}^{(2)}=4.4 \times 10^{-10} nm^2$, $\lambda_{m}^{(1)\,A}= 553.1 nm$, $\lambda_{m}^{(2)\,A}= 533.3 nm$,  $\Delta\lambda_{m}^{(1)\,A}=10.1 nm$, $\Delta\lambda_{m}^{(2)\,A}=20.5 nm$,
 $A_{me}^{(1)}=4.1 \times 10^{-8} nm^2$,  $A_{me}^{(2)}=23 nm^2$, $\lambda_{me}^{(1)\,A}= 558.1 nm$, $\lambda_{me}^{(2)\,A}= 534.3 nm$,  $\Delta\lambda_{me}^{(1)\,A}=11.7 nm$, $\Delta\lambda_{me}^{(2)\,A}=28.5 nm$. 

\renewcommand{\theequation}{A4-\arabic{equation}}
  \setcounter{equation}{0} 
\renewcommand{\figurename}{Figure A4}
  \setcounter{figure}{0}
\appendix

\section{Appendix  4:  Calculation of  orientational averages of the ${\cal K}$-factors}
To illustrate how the orientational averages of the ${\cal K}$-factors are obtained, we show here  the calculation of the term of $<{\cal K} ^{(3)}>$:
\be
<{\cal K}^{(3)}_{2}>=-3<({\bf s}_{A}^{p} \cdot{\bf s}_{R}) ({\bf s}_{R}\cdot {\bf s}_{D}^{p}) ({\bf s}_{D}^{m}\cdot{\bf s}_{A}^{m})>.  \nonumber  \,\,\,\,\,\, 
\ee
From Eqs. (30), (36), (37), (40) and (41),  and Fig. 2 we get:
\be
<{\cal K}^{(3)}_{2}>=-\frac{3}{(4\pi)^2}\int_{0}^{2\pi}d\beta \int_{0}^{2\pi}d\phi\int_{0}^{\pi}d\alpha\sin \alpha \int_{0}^{\pi}d\theta \sin\theta \,\,\,\,\,\,\,\,\,\,\,\,\,\,\,\,\,\,\,\,   \nonumber \\ 
\{ [ ({ s}_{A\,\perp}^{p}\sin \phi +{ s}_{A\,\parallel}^{p}\cos\theta\cos\phi)\sin\alpha\cos\beta \,\,\,\,\,\,\,\,\,\,\,\,\,\,\,\,\,\,\,\,  \nonumber \\ 
+ (-{ s}_{A\,\perp}^{p}\cos \phi +{ s}_{A\,\parallel}^{p}\cos\theta\sin\phi)\sin\alpha\sin\beta
-{ s}_{A\,\parallel}^{p}\sin\theta\cos\alpha]\,\,\,\,\,\,\,\,\,\,\,\,\,\,\,\,\,\,\,\,  \nonumber \\ 
\times (s_{D\, x}^{p\, *}\sin\alpha\cos\beta  +s_{D\, y}^{p\, *}\sin\alpha\sin\beta) \,\,\,\,\,\,\,\,\,\,\,\,\,\,\,\,\,\,\,\,  \nonumber \\
 \times [\xi_{x} s_{D\, x}^{p}(\zeta_{\perp}^{\,*}{ s}_{A\,\perp}^{p\,*}\sin \phi +\zeta_{\parallel}^{\,*}{ s}_{A\,\parallel}^{p\,*}\cos\theta\cos\phi)  \,\,\,\,\,\,\,\,\,\,\,\,\,\,\,\,\,\,\,\,  \nonumber \\ 
+\xi_{y}s_{D\, y}^{p}(-\zeta_{\perp}^{\,*}{ s}_{A\,\perp}^{p\,*}\cos \phi +\zeta_{\parallel}^{\,*}s_{A\,\parallel}^{p\,*}\cos\theta\sin\phi)]\}.   \,\,\,\,\,  \,\,\,\,\,\,\,\,\,\,\,     \nonumber
\ee
The terms that will not vanish on integration in $\alpha$ and $\beta$ yield
\be
<{\cal K}^{(3)}_{2}>=-\frac{3}{(4\pi)^2}\int_{0}^{2\pi}d\beta \int_{0}^{2\pi}d\phi\int_{0}^{\pi}d\alpha\sin \alpha \int_{0}^{\pi}d\theta \sin\theta \,\,\,\,\,\,\,\,\,\,\,\,\,\,\,\,\,\,\,\,   \nonumber \\ 
\{ | { s}_{A\,\perp}^{p}|^2 | s_{D\, x}^{p}|^{2}\zeta_{\perp}^{\,*}\xi_x\sin^{2} \phi \sin^{2}\alpha \cos^{2} \beta
\,\,\,\,\,\,\,\,\,\,\,\,\,\,\,\,\,\,\,\,  \nonumber \\ 
+ |{ s}_{A\,\parallel}^{p}|^2 | s_{D\, x}^{p}|^{2}\zeta_{\parallel}^{\,*}\xi_x \cos^{2}\theta\cos^{2} \phi \sin^{2}\alpha \cos^{2} \beta \,\,\,\,\,\,\,\,\,\,\,\,\,\,\,\,\,\,\,\,   \nonumber \\ 
+ | { s}_{A\,\perp}^{p}|^2 | s_{D\, y}^{p}|^{2}\zeta_{\perp}^{\,*}\xi_y\cos^{2} \phi \sin^{2}\alpha \sin^{2} \beta
\,\,\,\,\,\,\,\,\,\,\,\,\,\,\,\,\,\,\,\,  \nonumber \\ 
+ | { s}_{A\,\parallel}^{p}|^2 | s_{D\, y}^{p}|^{2}\zeta_{\parallel}^{\,*}\xi_y \cos^{2}\theta\sin^{2} \phi \sin^{2}\alpha \sin^{2} \beta. \,\,\,\,\,\,\,\,\,\,\,\,\,\,\,\,\,    \nonumber
\ee
After integration it is straightforward to obtain:
\be
<{\cal K}^{(3)}_{2}>=-\frac{1}{2}(| { s}_{A\,\perp}^{p}|^2 \zeta_{\perp}^{\,*}+\frac{1}{3}| { s}_{A\,\parallel}^{p}|^2 \zeta_{\parallel}^{\,*})
(| s_{D\, x}^{p}|^{2}\xi_x + | s_{D\, y}^{p}|^{2}\xi_y). \,\,\,  \,\,\,\,\,\,\,\,\,\,\,    \nonumber
\ee
All other terms of  orientational averages of  the ${\cal K}$-factors  are derived in similar fashion.
\renewcommand{\theequation}{A5-\arabic{equation}}
  \setcounter{equation}{0} 
\renewcommand{\figurename}{Figure A5 -}
  \setcounter{figure}{0}
\begin{figure*}[t]
\begin{centering}
\includegraphics[width=18cm]{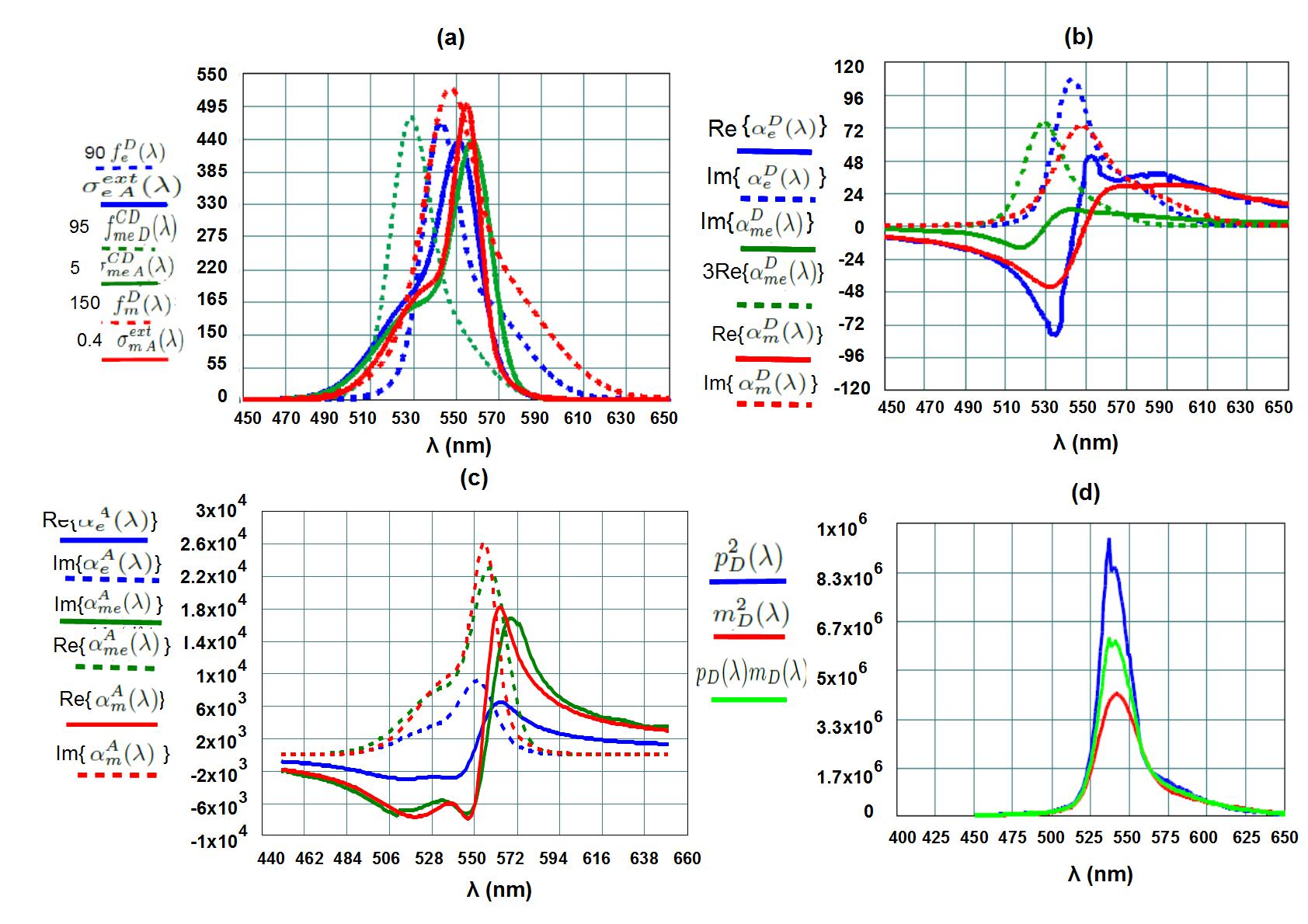}
\par\end{centering}
\caption{(Color online). Data of examples A and B:  Both donor  and acceptor are chiral and magnetoelectric. $n(\lambda)=1$. (a) Emission spectra in $fs$: $f_{e}^{D}(\lambda)$,   $f_{m}^{D}(\lambda)$,  $f_{me\,D}^{CD}(\lambda)$ of D, and extinction and dichroism cros-sections in $nm^2$:  $\sigma_{e\,A}^{ext}(\lambda)$, $\sigma_{m\,A}^{ext}(\lambda)$, $\sigma_{me\,A}^{CD}(\lambda)$ of A. (b) Real and imaginary parts of the polarizabilities of D in $nm^3$:  $\alpha_{e}^{D}(\lambda)$,   $\alpha_{m}^{D}(\lambda)$ and  $\alpha_{me}^{D}(\lambda)$. (c)  Real and imaginary parts of the polarizabilities of A in $nm^3$: $\alpha_{e}^{A}(\lambda)$,   $\alpha_{m}^{A}(\lambda)$ and  $\alpha_{me}^{A}(\lambda)$. (d) Donor distributions: $p_D^2(\lambda)$, $m_D^2(\lambda)$ and $p_D(\lambda)m_D(\lambda)$ in arbitrary units  for incident elliptic polarization with  $e_i^+= 7$,  $e_i^-= 3$  on D.  The wavelength $\lambda$ is in $nm$.} 
\end{figure*}

\appendix

\section{Appendix 5:  Data for examples A and B:   RHELT and RET when both donor and acceptor are chiral and magnetoelectric}

Fig. A5-1(a) shows the spectra of the emission distributions of D, as well as  extinction and CD cross-sections of A according to Eqs. (\ref{f_D}) and  (\ref{sigma_A}). $n(\lambda)=1$ and $c=300 nm \times fs^{-1}$. The polarizabilities of D and A,  along with the products of the induced donor dipoles $p_D^2$, $m_D^2$ and $p_D m_D$  for incident elliptic polarization with  $e_i^+= 7$,  $e_i^-= 3$  on D, are shown in Fig. A5-1(b), (c) and (d), respectively.

We have chosen the lineshapes  for the donor emission distributions of  Eq. (\ref{f_D}) with

 $D_{e}^{(1)}=3.91 fs$,  $D_{e}^{(2)}= 1.86 fs$, $\lambda_{e}^{(1)\,D}= 542 nm$, $\lambda_{e}^{(2)\,D}= 560 nm$,  $\Delta\lambda_{e}^{(1)\,D}=11 nm$, $\Delta\lambda_{e}^{(2)\,D}=29 nm$,
 $D_{m}^{(1)}=2.18 fs$,  $D_{m}^{(2)}=1.52fs$, $\lambda_{m}^{(1)\,D}= 546 nm$, $\lambda_{m}^{(2)\,D}= 561 nm$,  $\Delta\lambda_{m}^{(1)\,D}=15 nm$, $\Delta\lambda_{m}^{(2)\,D}=39 nm$,
 $D_{me}^{(1)}=3.75 fs$,  $D_{me}^{(2)}=1.54 fs$, $\lambda_{me}^{(1)\,D}= 528 nm$, $\lambda_{me}^{(2)\,D}= 562 nm$,  $\Delta\lambda_{me}^{(1)\,D}=12 nm$, $\Delta\lambda_{me}^{(2)\,D}=26 nm$.

Whereas for the magnetoelectric acceptor,[cf. Eq.(\ref{sigma_A})], the parameters are:   

$A_{e}^{(1)}=310 nm^2$,  $A_{e}^{(2)}=180 nm^2$, $\lambda_{e}^{(1)\,A}= 552.1 nm$, $\lambda_{e}^{(2)\,A}= 534.3 nm$,  $\Delta\lambda_{e}^{(1)\,A}=11.7 nm$, $\Delta\lambda_{e}^{(2)\,A}=28.5 nm$
 $A_{m}^{(1)}=910 nm^2$,  $A_{m}^{(2)}=480 nm^2$, $\lambda_{m}^{(1)\,A}= 555.2 nm$, $\lambda_{m}^{(2)\,A}= 539.1 nm$,  $\Delta\lambda_{m}^{(1)\,A}=8.1 nm$, $\Delta\lambda_{m}^{(2)\,A}=26.2 nm$,
 $A_{me}^{(1)}=71 nm^2$,  $A_{me}^{(2)}=32 nm^2$, $\lambda_{me}^{(1)\,A}= 558.1 nm$, $\lambda_{me}^{(2)\,A}= 534.3 nm$,  $\Delta\lambda_{me}^{(1)\,A}=11.7 nm$, $\Delta\lambda_{me}^{(2)\,A}=28.5 nm$.


\begin{thebibliography}{99}
\bibitem{allen1}   L. Allen, S. M.  Barnett and M. J.  Padgett, eds, {\it Optical Angular Momentum}, (IOP Publishing, Bristol, UK, 2003).

\bibitem{andrews1} D. L. Andrews and M. Babiker, eds., {\it The Angular Momentum of  Light} (Cambridge University press, Cambridge, 2013).

\bibitem{yao} M. Yao and M. Padgett, Adv. Opt. Photon. {\bf 3}, 161 (2011).

\bibitem{andrews2} D. L. Andrews,  M. M.  Coles, M. D. Williams,  and D. S. Bradshaw,   Proc. SPIE {\bf 8813}, 88130Y (2013).

\bibitem{boyd}M. N. O'Sullivan,  M. Mirhosseini, M. Malik,  and R. W. Boyd, , Opt. Express {\bf 20}, 24444 (2012).

\bibitem{schellman}J. A. Schellman,  Chem. Rev. {\bf 75}, 323 (1975).

\bibitem{richard} F. S. Richardson,  and J. P. Riehl,   Chem. Rev.{\bf 77}, 773 (1977).

\bibitem{vuong}L. Vuong, A.  Adam, J.  Brok, P.  Planken,  and  H. Urbach, Phys. Rev. Lett. {\bf 104}, 083903 (2010).

\bibitem{bliokh1}K. Y. Bliokh,  F. Rodriguez-Fortu\~{n}o, F. Nori, and  A.  V. Zayats,  Nat. Photonics {\bf 9}, 796 (2015).

\bibitem{schukov1} S. Sukhov,  V. Kajorndejnukul,  R. R. Naraghi,  and A. Dogariu,  Nat. Photonics {\bf 9}, 809 (2015).

\bibitem{brasse} D. Hakobyan,  and E.  Brasselet,  Opt. Express {\bf 23}, 31230 (2015).

\bibitem{bliokh2}K. Y.  Bliokh,  D. Smirnova,  and  F. Nori,   Science {\bf 348}, 1448 (2015).

\bibitem{andrews3} D. S. Bradshaw,  J. M. Leeder,  M. M.  Coles,  and D.L. Andrews,  Chem. Phys. Lett. {\bf 626}, 106  (2015).

\bibitem{nieto1} M. Nieto-Vesperinas,  J. Opt. {\bf 19}, 065402 (2017).

\bibitem{bliokh3} K. Y. Bliokh,  Y. S.  Kivshar, and  F. Nori,  Phys. Rev. Lett. {\bf 113}, 033601 (2014).

\bibitem{kivshar} A. Krasnok,  S.  Glybovski,  M. Petrov,  S. Makarov,  R. Savelev, P.  Belov,  C. Simovski and  Y.S. Kivshar,  Appl. Phys. Lett. {\bf 108}, 211  (2016.). (doi:10.1063/1.4952740).

\bibitem{tang1} Y. Tang,  and A. E. Cohen,  Phys. Rev. Lett. {\bf 104}, 163901 (2010).

\bibitem{bliokh4} K.Y. Bliokh and  F.  Nori,   Phys. Rev. A {\bf 83}, 021803 (2011).

\bibitem{bliokh5} K. Y.  Bliokh, A. Y. Bekshaev and F. Nori,  New. J. Phys. {\bf 15} 033026 (2013).

\bibitem{cameron1}R. P.  Cameron,   S. M. Barnett  and  A. M.  Yao,  New J. Phys. {\bf 14}, 053050 (2012).

\bibitem{cameron2} R. P. Cameron and S. M. Barnett, New J. Phys. {\bf 14}, 123019 (2012).

\bibitem{nieto2}M.  Nieto-Vesperinas, Phys. Rev. A {\bf 92}, 023813 (2015).

\bibitem{nieto3} M. Nieto-Vesperinas,  Phil. Trans. R. Soc. A {\bf 375}, 20160314 (2017).

\bibitem{gutsche1}P. Gutsche,  P. I. Schneider,  S. Burger and  M. Nieto-Vesperinas,  IOP Conf. Series: Journal of Physics {\bf 963}, 012004 (2018).  arXiv:1712.07091 (2018).

\bibitem{gutsche2} P. Gutsche,  L. V.  Poulikakos,  M. Hammerschmidt, S.  Burger. and F. Schmidt,  Proc. SPIE {\bf 9756}, 97560X arXiv:1603.05011 (2016).

\bibitem{gutsche3} L. V. Poulikakos, P.  Gutsche, K. M. McPeak, S.  Burger, J. Niegemann,  C.  Hafner and  D. J. Norris,  ACS Photonics {\bf 3} 1619–25 (2016).

\bibitem{corbato1}I.  Fernandez-Corbaton and G. Molina-Terriza,  Phys. Rev. B {\bf 88}, 085111 (2013).

\bibitem{zambra1} X. Zambrana-Puyalto and  N. Bonod,  Nanoscale {\bf 8}, 10441 (2016).

\bibitem{gutsche4} P. Gutsche  and M. Nieto-Vesperinas,  Sci. Reps. {\bf  8} 9416  (2018). DOI:10.1038/s41598-018-27496-w.

\bibitem{tang2} Y. Tang, Y. and A. E. Cohen,   Science {\bf 332}, 333 (2011).

\bibitem{tang3} N. Yang, Y. Tang, A. E. Cohen, Nano Today {\bf 4}, 269,  (2009)

\bibitem{choi} J. S. Choi and M.  Cho, Phys. Rev. A {\bf 86}, 063834 (2012).

\bibitem{klimov} D. V. Guzatov and V. V.  Klimov,  New J. Phys. {\bf 14}, 123009 (2012).

\bibitem{dionne1} H.  Alaeian, and  J. A. Dionne,  Phys. Rev. B {\bf 91}, 245108 (2015).

\bibitem{schafer1}  M. Sch\"{a}ferling,  D. Dregely,  M. Hentschel and  H. Giessen, Phys. Rev. X {\bf 2}, 031010 (2012).

\bibitem{giessen1} M. Hentschel, M. Schäferling,  X. Duan,  H. Giessen, and  N. Liu, Chiral plasmonics. Sci. Adv. {\bf 3}, e1602735 (2017).

\bibitem{schafer2}C.  Kramer,  M. Schäferling, T.  Weiss,  H. Giessen, and  M. Brixner,  ACS Photonics {\bf 4}, 396 (2017).

\bibitem{dionne2} A. Garcia-Etxarri, A. and J. A. Dionne,  Phys. Rev. B {\bf 87}, 235409 (2013).

\bibitem{wang} H. Wang, Z.  Li,  H.  Zhang,  P. Wang  and S. Wen,   Sci. Rep. {\bf 5}, 8207 (2015).

\bibitem{carminati} R. Vincent, and R. Carminati,  Phys. Rev. B {\bf 83}, 165426 (2011).

\bibitem{hu} L. Hu,  X. Tian,  Y.Huang,  L. Fang and T. Fang,  Nanoscale {\bf  8}, 3720 (2016).

\bibitem {fret1} T. F\"{o}rster, in {\it Modern Quantum Chemistry}, ed. O. Sinanoglu,  (Academic P., New York 1965),  pp. 93-137.

\bibitem{fret2}  L. Stryer  and R.P  Haugland,  Proc. Natl. Acad. Sci. USA {\bf  58} 719 (1967).

\bibitem{circpollibro1} J, P. Riehl and G. Muller, Circularly polarized luminescence spectroscopy and emission detected dichroism  Chapt. 3, pp 64 of: Comprehensive spectroscopy. Vol. 1, 
N. Berova,  P. L. Polavarapu,  K. Nakanishi and R. W. Woody.  J. Wiley, Hoboken, New Jersey 2012.

\bibitem{circpolemission}  F. S. Richardson and J. P. Riehl, Circularly Polarized Luminescence Spectroscopy, Chem. Rev. {\bf 77} 773 (1977).

\bibitem{kagan} C. R. Kagan, C. B. Murray, M. Nirmal, and M. G. Bawendi, Phys. Rev. Lett. {\bf 76},  1517 (1996).

\bibitem{clegg} R. M. Clegg, F\"{o}rster  resonance energy transfer—FRET. 
what is it, why do it, and how it’s done, Ch.1 of  "FRET and FLIM techniques", T.W.J. Gadella, ed., Laboratory Techniques in Biochemistry and Molecular Biology {\bf 33}, 1, (Academic Press, 2009).

\bibitem{novotny} L. Novotny L Hand B. Hecht,  Principles of nano-optics, 2nd edn. Cambridge, UK: Cambridge University Press (2012).

\bibitem{craig1} D. P. Craig and T. Thirunamachandran, Chem. Phys. {\bf 167},  229 (1992)

\bibitem{salam1} A. Salam, Mol. Phys. {\bf 87} 919 (1996).

\bibitem{craig2}  D. P. Craig and T. Thirunamachandran, Theor. Chem. Acc.  {\bf 102} 112 (1999). DOI 10.1007/s00214980m157.

\bibitem{salam2} A. Salam, AIP Conference Proceedings {\bf 1642},  90 (2015); doi: 10.1063/1.4906634.

\bibitem{silas} S. J. Leavesley and T. C. Rich, Cytometry A {\bf 89}, 325 (2016).

\bibitem{nietoJOSA}  M. Nieto-Vesperinas, R. Gomez-Medina, and J. J. Saenz, 
 J. Opt. Soc. Am. A {\bf 28}   54 (2011).

\bibitem{g-etxarri} A. Garcia-Etxarri,  R. Gomez-Medina, L.S. Froufe-Perez, C. Lopez, L. Chantada, F. Scheffold, J. Aizpurua, M. Nieto-Vesperinas and J.J. Saenz, Opt. Express {\bf 19}, 4815 (2011).

\bibitem{geffrin} J. M. Geffrin, B. Garcia-Camara, R. Gomez-Medina, P. Albella, L. S. Froufe-Perez, C. Eyraud, A. Litman, R. Vaillon, F. Gonzalez, M. Nieto-Vesperinas, J. J. Saenz and F. Moreno, Nat. Commun. {\bf 3} 1171 (2012).

\bibitem{kuznetsov} A. I. Kuznetsov, A. E. Miroshnichenko, Y. H. Fu, J. Zhang  and B. Luk’yanchuk,  Sci. Reps. {\bf 2}, 492 (2012).

\bibitem{staude}  M. Decker and I. Staude,  J. Opt. {\bf 18}, 103001 (2016) .

\bibitem{kivshar_reviews} A. I. Kuznetsov, A. E. Miroshnichenko, M. L. Brongersma,Y. S. Kivshar,  B. Luk’yanchuk,  Science  {\bf  354} (2016)  2472. 

\bibitem{nieto4}  M. Nieto-Vesperinas,  Opt. Lett. {\bf 40},  3021 (2015).

\bibitem{madrazo} A. Madrazo, M.Nieto-Vesperinas and N. Garcia, Phys. Rev. B {\bf 53}, 3654 (1996).

\bibitem{garcia} M. Nieto-Vesperinas and  N. Garcia, (eds), {\it Optics at the Nanometer Scale:
Imaging and Storing with Photonic Near Fields}, NATO ASI Series, E-319, (Springer, 1996. Reprinted: 2012).

\bibitem{kumar} J. Kumar, T. Nakashima and T. Kawai,  J.  Phys. Chem. Lett.   {\bf 6}, 3445  (2015).

\bibitem{pyramids} W.  Yan, L. Xu, Ch. Xu, W. Ma, H. Kuang, L. Wang and N. A. Kotov,
 J. Am. Chem. Soc.  {\bf 134}, 15114  (2012).

\bibitem{jaque}  P. Haro-Gonzalez, B. del Rosal a L. M. Maestro, E. Martin Rodriguez, R. Naccache,
J. A. Capobianco, K. Dholakia, J. Garciıa Solea and D. Jaque, Nanoscale {\bf 5}, 12192 (2013). 

\bibitem{dionne} A.  Lay, D. S. Wang, M. D. Wisser, R. D. Mehlenbacher, Y. Lin,
M. B. Goodman, W. L. Mao and J. A. Dionne,
Nano Lett. {\bf 17},  4172 (2017).

\bibitem{chinos}S. Wen, J. Zhou, K.  Zheng, A. Bednarkiewicz, X. Liu and  D. Jin, 
 Nat. Comm. {\bf 9}, 2415  (2018).

\bibitem{shivola} I.V. Lindell and A. Shivola, Electromagnetic waves in chiral and bi-isotropic media, Artech House, London, 1994.

\bibitem{berney} C. Berney and G. Danuser,  Biophys. J.  {\bf 84}, 3992 (2003).

\bibitem{cana} A. Canaguier-Durand, J. A Hutchison, C. Genet and T. W. Ebbesen,
New J. Phys. {\bf 15},  123037  (2013).

\bibitem{born} M. Born and E. Wolf, {\it Principles of Optics}, 7 th edition, Cambridge U.P., Cambridge, 1999. 

\bibitem{banzer} S. Nechayev, S. Eismann, G. Leuchs and P. Banzer, Phys Rev. B {\bf 99}, 075155 (2019);
S. Nechayev and P. Banzer, Phys. Rev. B {\bf 99},  241101(R)  (2019).

\bibitem{opex2010} M. Nieto-Vesperinas, J. J. Saenz,  R. Gomez-Medina  and L. Chantada,
  Opt. Exp. {\bf  18}, 11428 (2010).

\bibitem{barron1} L. D. Barron,{\it Molecular Light Scattering and Optical Activity},  (Cambridge U.P., Cambridge, 2004).

\end{thebibliography}
\end{document}